\documentclass[preprint,12pt,authoryear]{elsarticle}
\usepackage{setspace}

\usepackage{amssymb}
\usepackage{epstopdf}
\usepackage{xcolor}
\usepackage{lineno}

\journal{Planetary and Space Science}
\begin{document}
%%\linenumbers
\begin{frontmatter}



  \title{New phase-magnitude curves for some Main Belt asteroids, fit of
    different photometric systems and calibration of the albedo - photometry
    relation}


\author[1]{A. \, Carbognani}
\ead{albino.carbognani@gmail.com}
\author[2]{A. \, Cellino}
\author[1,3]{S. \,Caminiti}

\address[1]{Astronomical Observatory of the Autonomous Region of the Aosta
  Valley (OAVdA), Lignan 39, 11020 Nus, Aosta, Italy}
\address[2]{INAF - Osservatorio Astrofisico di Torino, Strada Osservatorio 20,
  10025, Pino Torinese, Torino, Italy}
\address[3]{Universit\`a degli studi di Pavia, Dipartimento di Fisica,
  Via Bassi 6, 27100 Pavia, Italy} 

\begin{abstract}
  Results of photometric observations of a small sample of selected Main Belt
  asteroids are presented. The obtained measurements can be used to achieve a 
  better calibration of the asteroid
  photometric system ($H$, $G_1$, $G_2$) adopted by the IAU, and to make
  comparisons with best-fit curves that can be obtained using different
  photometric systems. The new data have been obtained as a first feasibility study of a
  more extensive project planned for the future, aimed at obtaining a reliable 
  calibration of possible relations between some parameters characterizing the phase-magnitude curves
  and the geometric albedo of asteroids. This has important potential
  applications to the analysis of asteroid photometric data obtained by the
  Gaia space mission.

\end{abstract}


\end{frontmatter}


\section{Introduction}
\label{1}
The phase - magnitude curve (hereinafter, phase - mag curve) of an asteroid
describes the variation of brightness, expressed in magnitudes and
normalized to unit distance from Sun and observer, as a function of
varying phase angle. The latter is the angle between the directions to the
observer and to the Sun as seen from the observed body.

It is well known that the magnitude (by definition, the brightness expressed
in a logarithmic scale) of small bodies of the Solar system tends to increase nearly 
linearly (the objects becoming much fainter) for increasing phase angle. In most cases, a
so-called {\em opposition effect} is also observed, namely a non-linear magnitude
surge occurring when the object is seen close to solar opposition, at phase
angles generally below $6^{\circ}$. 

The photometric behaviour of atmosphereless Solar system bodies is determined
by their macroscopic shapes and by the light scattering properties of their
surfaces, related to composition, texture and roughness. In particular, 
%Roughness quantifies
%an intrinsic low-scale disorder in the distribution of the orientation of the
%vectors normal to different points of a given surface, small roughness
%meaning a smooth surface.
%At larger scales,
macroscopic roughness and local topographic features of the surface produce
shadowing effects, depending upon illumination conditions.
Shadows tend to disappear when the object is viewed from nearly the same direction as
the illumination, as an object approaches solar opposition.
Shadowing effects contribute therefore to the existence of the above-mentioned
opposition effect. 
%As the phase angle increases, there is a the decrease in brightness, which is
%due to the interplay of the scattering properties of the surface, by the
%increasing importance of shadowing (mainly for surfaces for which multiple
%scattering is less important), and by the presence of a defect of
%illumination, corresponding to an increasingly large fraction of the surface
%of the object facing the observer being not illuminated by the Sun. 

It has been known since several decades, however, that another mechanism,
named {\em coherent backscattering}, plays a more fundamental role in
determining the opposition effect. Coherent backscattering is a
phenomenon of constructive interference of light beams following different
optical paths to reach the observer. It starts to be particularly effective
when the body is seen close to solar opposition.
Coherent backscattering is enhanced by multiple light scattering, and
for this reason it tends to be stronger for asteroids having higher albedo
\citep{MuinonenACM93, Muinonenetal12, DlugachMishchenko13}.

Surface scattering properties
are responsible not only for the variations of brightness that are
measured at different epochs and in different illumination conditions, but also
of some corresponding variations in the state of linear polarization of
the sunlight scattered by the surfaces. For this reason, the
phase - mag curves of asteroids and the corresponding phase - linear polarization
curves are fundamental sources of information that in
principle can be used to infer hints about important surface properties,
including the geometric albedo, the texture and roughness of the regolith,
whose determination is in general difficult by using remote observation
techniques \citep{MNRAS2}.

So far, however,
the number of asteroids for which we have both good-quality
phase - mag {\em and} phase - polarization curves, is surprisingly
limited.
%For what concerns asteroid polarimetry, there have been since some
%years systematic attempts by some authors to increase the still very
%limited data-base of good-quality phase - polarization curves.
%[{\bf Add references}].
%This effort
%deserves to be complemented on the side of
This means that it is desirable to set up observing programs aimed at obtaining
new phase - mag and phase - polarization curves of the same targets.

On the side of photometry, a most notable effort has been carried out by \citet{Shevchenkoetal2016}.
The present paper presents the results of
a pilot program originally conceived as a feasibility check of a larger
project to be carried out at OAVdA (Astronomical Observatory of the Autonomous 
Region of Aosta Valley). After the end of this program, we were forced to interrupt our photometric activities due to
insufficient staffing. In turn, this was also the consequence of a big effort carried 
out by OAVdA in order to put into operation an array of five new 40-cm telescopes 
necessary to participate in an European APACHE
(A PAthway toward the Characterization of Habitable Earths)
project designed for the purpose of the discovery of extrasolar planets, by means
of the detection of photometric transit events. Now, after the end of this program 
in late 2017, we are discussing the possibility to use these APACHE telescopes for the 
purposes of a new asteroid photometric program aimed at greatly extending the work described
in the present paper. 
 
An immediate purpose of our observations was to obtain new asteroid lightcurves
to be used to increase the KAMPR database, namely the Kharkiv Asteroid
Magnitude-Phase
Relations\footnote{http://sbn.psi.edu/pds/resource/magphase.html}, a list
of asteroid phase - mag relations compiled at the Institute of Astronomy of
Kharkiv Kharazin University \citep{Shev_KAMPR}. At the same time, we wanted to 
derive for our targets the photometric parameters 
($H$, $G_1$, $G_2$) \citep{HG1G2} using the photometric system  
adopted at the 2012 IAU General Assembly. 
In principle, adding new phase - magnitude data is
useful to improve the $(H, G_1, G_2)$ system, in particular it helps to
better determine the values of its base functions.
Moreover, we wanted to compare the rms of the computed best-fit curves with those 
obtained using some other photometric systems still adopted by many authors. 

\section{The importance of phase - mag curves in the Gaia era}

Good-quality phase - mag curves are fundamental to determine asteroid 
absolute magnitudes\footnote{The absolute magnitude of a Solar system object is 
  defined as the (lightcurve-averaged) V magnitude reduced to unit distance from the 
  observer and the Sun, when the body is observed at ideal solar opposition (zero phase angle).}.
This can be an important information to complement the data produced by the Gaia
space mission of the ESA. Gaia is currently collecting a huge data-base of
sparse photometric measurements for tens of thousands Main Belt  asteroids.
Unfortunately, the spacecraft cannot observe these objects when they are
seen at phase angles smaller than about $10^\circ$ \citep{GDR2}. As a consequence, any
analysis of the opposition brightness surge is beyond the capability of Gaia,
and the photometric data collected by the spacecraft cannot be used in
principle to obtain accurate absolute magnitudes of the asteroids.

This is unfortunate, because a fundamental relation links
the absolute magnitude to the effective diameter
$D_e$, and the geometric albedo \citep{harrisharris97}: 
\begin{equation}
D_e = \frac{1329}{\sqrt p_V} 10^{-0.2 H_V} \label{DPH}
\end{equation}
where $H$ represents the absolute magnitude and $p_V$ the geometric albedo.
The latter is another key parameter, whose value is determined by composition
and texture of the surface regolith.

Gaia has not been designed to have the possibility to determine the albedo of
the tens of thousands asteroids that it observes in a variety of observing
circumstances, down to a nominal magnitude limit of $20.7$. On the other
hand, Gaia can derive phase - mag data taken in an interval of phase
angles where the relation between the two parameters is mostly linear. 

Interestingly, it has been proposed that the value of such linear slope
can be diagnostic of the geometric albedo \citep{Shev,Shevchenkoetal2016}. 
This proposed relation deserves confirmation and better calibration, 
because it opens the possibility to use the phase - mag curves observed by Gaia to infer the
corresponding geometric albedo values for tens of thousands asteroids.
For this reason, it is important to obtain new good-quality phase-mag
curves of asteroids for which the size is known with sufficiently good
accuracy, in order to determine for them the absolute magnitude and 
use Eqn.$\,$ \ref{DPH} to derive the corresponding albedo, to be used to
calibrate the \citet{Shev} relation.

%Moreover, it is important to derive the absolute
%magnitude of objects for which the geometric albedo has been determined
%by means of other techniques, including polarimetry and thermal radiometry,
%to check that the size derived using Eqn.$\,$ \ref{DPH} is compatible with
%the value known from observations.
%We should take into account that
The determination of the absolute magnitudes of the asteroids is not a
trivial affair. Due to the non-coplanarity of their orbits with that
of the Earth, the objects cannot be seen, as a rule,
at ideal Sun opposition, but at a minimum value of phase angle that changes
in different apparitions of the same object. As a consequence, the
determination of the
absolute magnitude is difficult, because the existence of the non-linear
brightness opposition effect makes it difficult to extrapolate
the observed magnitudes to zero phase angle. In other words, even small
differences in the treatment of the opposition effect may lead to important
differences in the determination of the absolute magnitude.

In addition, we should not forget that the absolute magnitude, based on its
definition, is not, strictly speaking, a really constant parameter, but
we can expect it to vary at different apparitions, due to the changing
cross section of an object having non-spherical shape when seen in different
geometric configurations.  
In this respect, obtaining the absolute magnitude in
different oppositions of the same object can help to improve the estimates
of its shape, and the accuracy of the albedo estimates based on thermal
radiometry data alone, in the absence of
any simultaneous measurements of the visible flux.

According to the above considerations, a systematic program of photometric
observations of asteroids to obtain sets of lightcurves obtained
at different phase angles is an important task that deserves an
investment of time and resources, and can represent
a fruitful use of telescopes of even modest aperture. The results of an
extensive investigation of the properties of the phase - mag curves of asteroids
belonging to different taxonomic classes have been presented by \citet{Shevchenkoetal2016}.
The observations that we present in this paper for a few objects are nothing but a 
first pilot program aimed at laying the foundations of a more ambitious long-term project 
to complement and extend the results obtained by the above authors. At the same
time, we consider in our analysis different possible sources of albedo values, 
from thermal radiometry and from polarimetry, and we consider possible relations 
between the albedo and a variety of parameters characterising different asteroid 
photometric systems. 

\section{Choice of the targets}
\label{2}

The observations presented in this paper have been the Master thesis subject
of one of us (SC), and were done using the 81-cm
reflector telescope of the OAVdA, located in north-western Italian Alps \citep{cal}.
We selected a limited sample of possible
targets, focusing on objects exhibiting apparent magnitudes
suitable for the OAVdA telescope, and having sizes well
constrained, based on accurate determinations by means of star
occultation measurements or reliable thermal infrared data \citep{Masiero2011},
as well as  reliable determinations of the geometric albedo, based whenever
possible upon polarimetric data \citep{MNRAS1, MNRAS2} or other data sources
\citep{she2}.

In choosing our targets, we had some obvious constraints related to the epoch of the 
solar opposition, to be between September and December 2012 on the basis
of the interval of time allocated to the project.
We made our target selection choosing among objects for which the
rotation period is known, and listed in the Asteroid Light Curve Data Base
by \citet{war09}. We preferentially chose objects having a
(moderately) fast rotation, preferred for an easier and faster determination
of the lightcurve. Another constraint was the minimum phase angle
expected during our observing window. We chose our targets among the
objects reaching phase angles around $2^\circ$. Unfortunately, weather
conditions did not allow us to reach such limit for all the targets of
our selected sample. On the other hand, 
minimum phase angles slightly larger than $5^\circ$ could still be
useful for the study of the linear part of the phase curve. 

We assigned higher priority to targets for which there is
a reasonably reliable determination of the geometric albedo,
determined using the $\Psi$ polarimetric parameter defined by \citet{MNRAS1}
or, when not available, taken from \citet{she2}. For all of them, an independent
albedo estimate based on thermal radiometry observations by the WISE satelllite,
was also available \citep{Masiero2011}.

The list of our targets is shown in Table \ref{ast}. The Table lists eleven
objects, but, as we will see below, could derived good-quality 
estimates of the absolute magnitude for only six of them.
%In particular, the number of observations
%for asteroid (135) was insufficient to derive a reliable phase - mag curve.
%For three other asteroids, (306), (308) and (925), the minimum phase angle
%turned out to be too large to put useful constraints on the behaviour of the
%phase - mag curve in the region of the opposition effect. Finally, in the
%case of (444), some unexpected anomalies also prevented us from deriving a
%reliable phase - mag curve.

\begin{table*}
  \caption{The observed asteroids with rotation periods, geometric albedos,
    opposition date, minimum phase angle and number of lightcurves.
    The geometric albedo corresponds to an updated and still unpublished
    determination of the polarimetric $\Psi$ parameter (as explained in
    \citet{MNRAS1}). In the case of asteroids (208), (306), (522) and (925),
    for which no reliable determination of $\Psi$ or other polarimetric
    parameters is available, we list albedo values taken from 
    \citep{she2}. The last column gives the number $N$ of different
    lightcurves obtained for each target. Note that, due to reasons explained
    in the text, not all the objects listed in this table were eventually used
  in our analysis of the phase - magnitude curve.}
\begin{center}
\begin{tabular}{lrcccc}
\hline
\\
Asteroid & \multicolumn{1}{c}{Rotation period (h)} & $p_v$ & Opp. Date
& $\alpha_{min}^{\circ}$ & $N$\\
\hline
\\
085 Io &  6.875 & $0.07 \pm 0.01$ & 2012-10-11 & 0.9 & 8 \\
135 Herta & 8.403 & $0.13 \pm 0.01$ & 2012-12-10 & 1.5 & 3 \\
208 Lacrimosa & 14.085 & $0.21 \pm 0.02$ & 2012-11-11 & 0.7 & 8 \\
236 Honoria & 12.333 & $0.19 \pm 0.01$ & 2012-09-21 & 0.9 & 9 \\
306 Unitas & 8.736 & $0.17 \pm 0.03$ & 2012-11-11 & 5.2 & 7 \\
308 Polyxo & 12.032 & $0.10 \pm 0.01$ & 2012-12-17 & 2.3 & 5 \\
313 Chaldaea & 8.392 & $0.07 \pm 0.01$ & 2012-09-22 & 0.3 & 6 \\
338 Budrosa & 4.608 & $0.12 \pm 0.01$ & 2012-12-11 & 1.5 & 6 \\
444 Gyptis & 6.214 & $0.09 \pm 0.01$ & 2013-01-03 & 5.3 & 7 \\
522 Helga & 8.129 & $0.06 \pm 0.01$ & 2012-10-02 & 1.8 & 7 \\
925 Alphonsina & 7.880 & $0.22 \pm 0.03$ & 2012-12-29 & 5.4 & 6 \\
\hline
\end{tabular}
\end{center}
\label{ast}
\end{table*}

\section{Instrument and reduction procedures}
\label{3}
{\color{red}
\begin{figure}
\includegraphics[width=16cm]{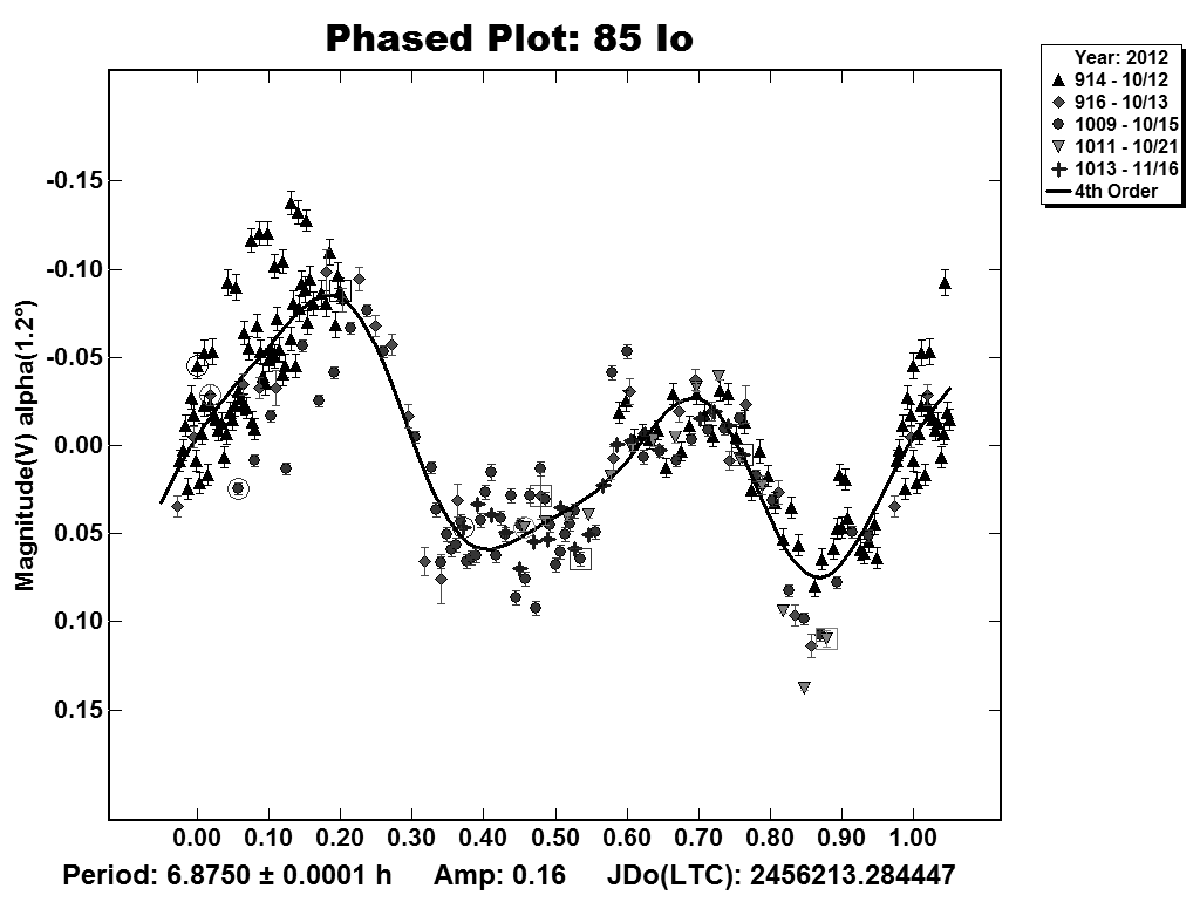}
\caption{First reference lightcurve for asteroid (85) Io ($JD_0$ = 2456214.50).}
\label{fig85}
\end{figure}
\begin{figure}
\includegraphics[width=16cm]{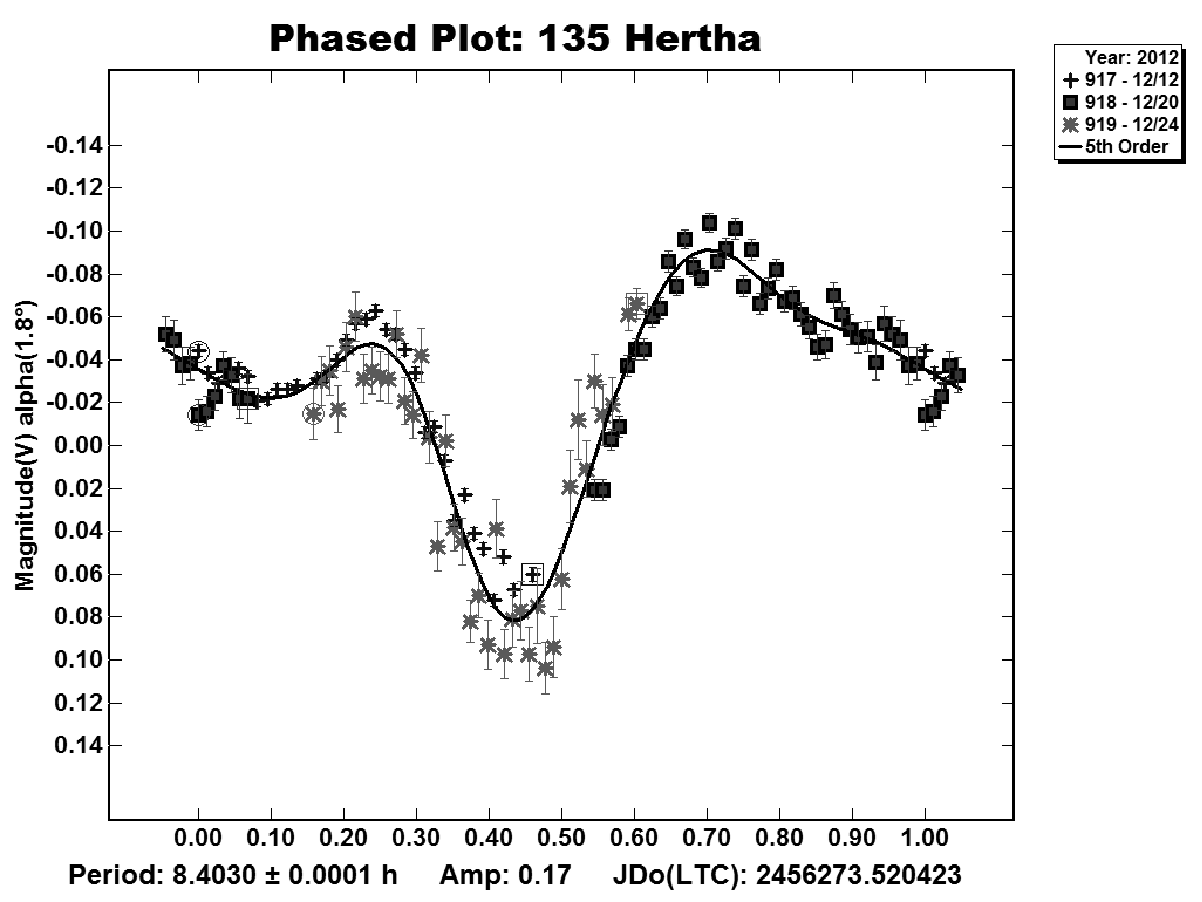}
\caption{Reference lightcurve for asteroid (135) Hertha ($JD_0$ = 2456280.50)).}
\label{fig135}
\end{figure}
\begin{figure}
\includegraphics[width=16cm]{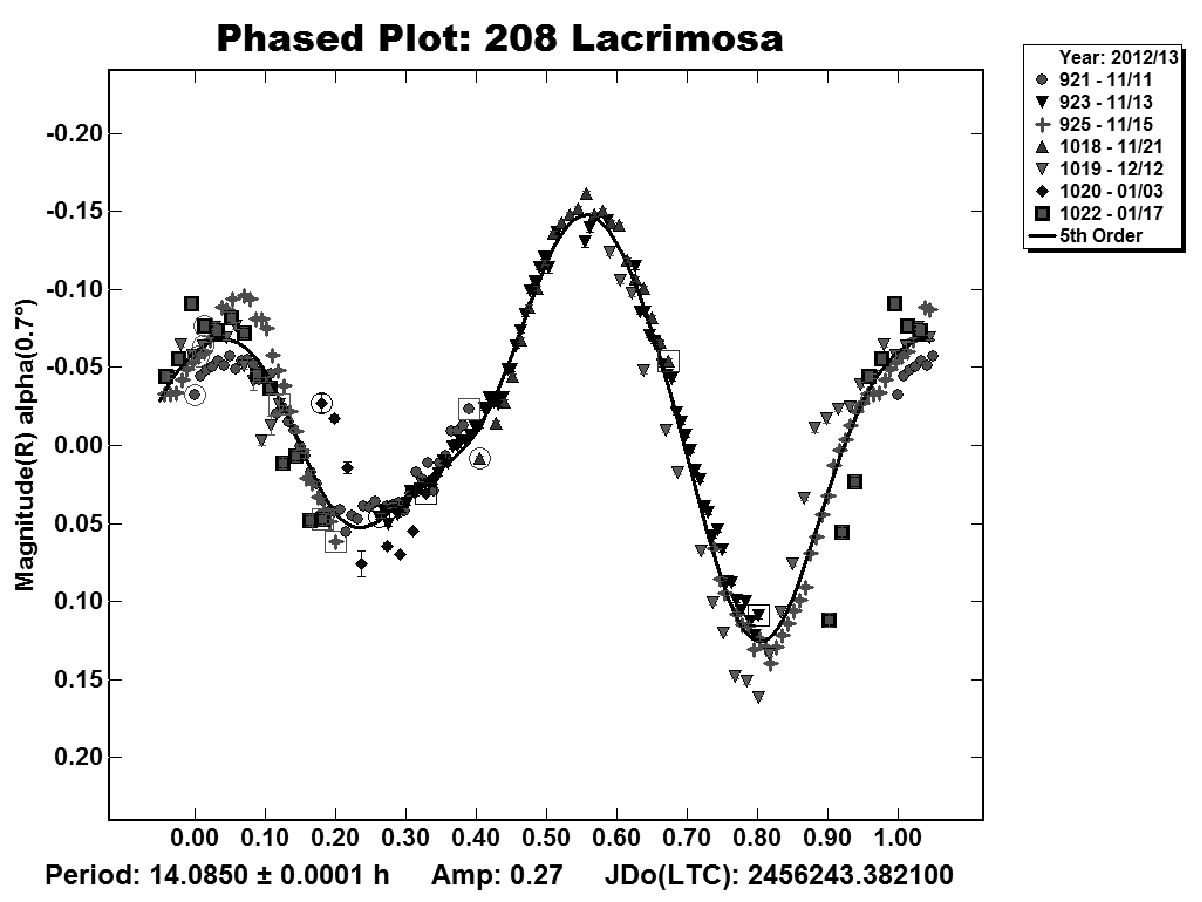}
\caption{Reference lightcurve for asteroid (208) Lacrimosa ($JD_0$ = 2456264.50).}
\label{fig208}
\end{figure}
\begin{figure}
\includegraphics[width=16cm]{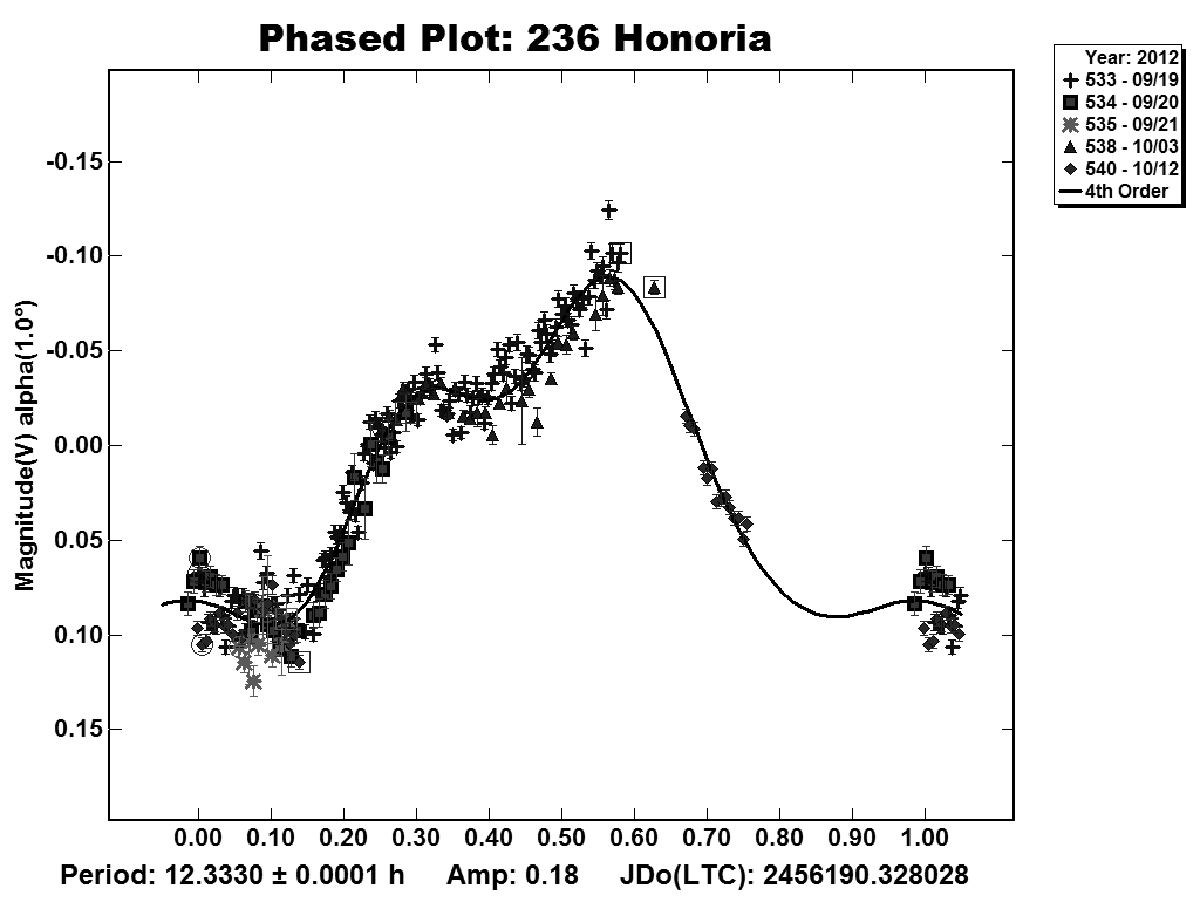}
\caption{First reference lightcurve for asteroid (236) Honoria ($JD_0$ = 2456197.50).}
\label{fig236}
\end{figure}
\begin{figure}
\includegraphics[width=16cm]{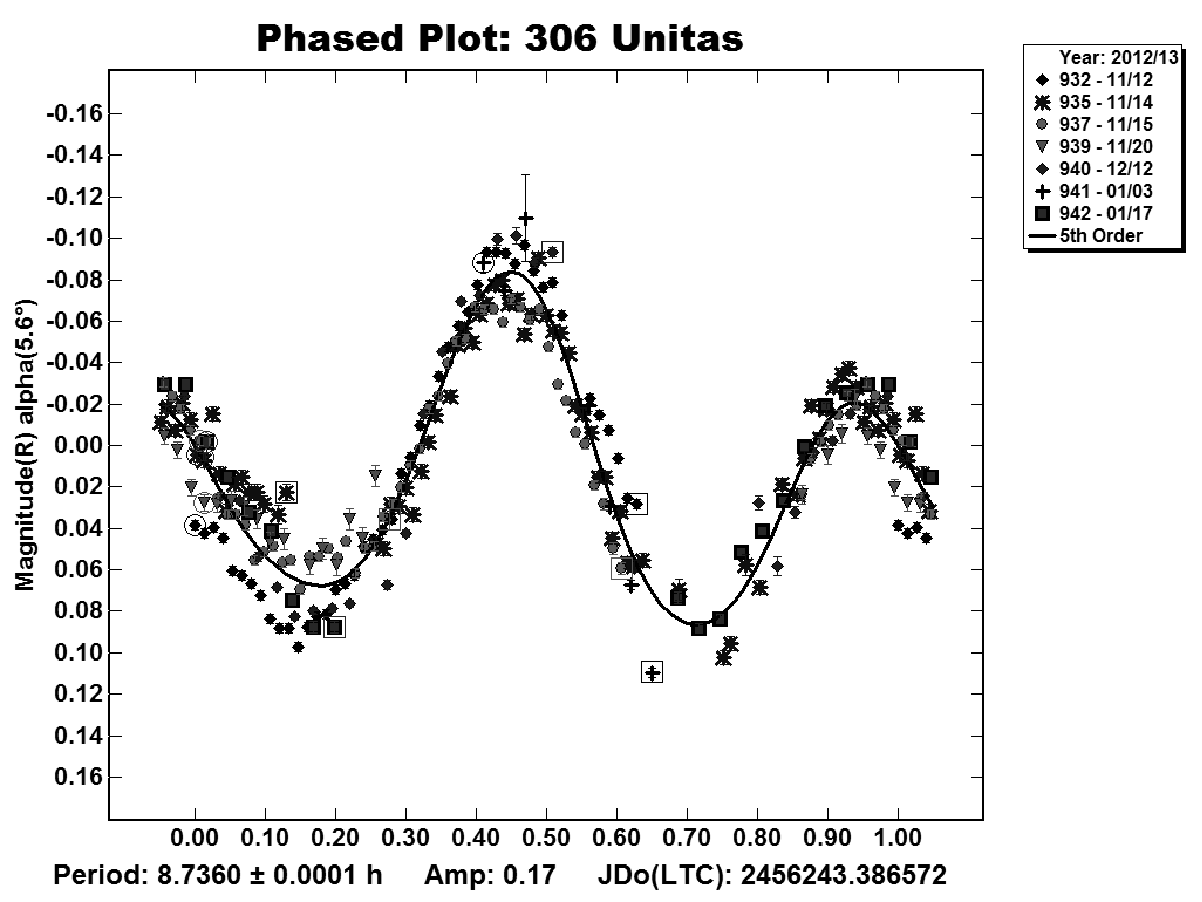}
\caption{Reference lightcurve for asteroid (306) Unitas ($JD_0$ = 2456255.50).}
\label{fig306}
\end{figure}
\begin{figure}
\includegraphics[width=16cm]{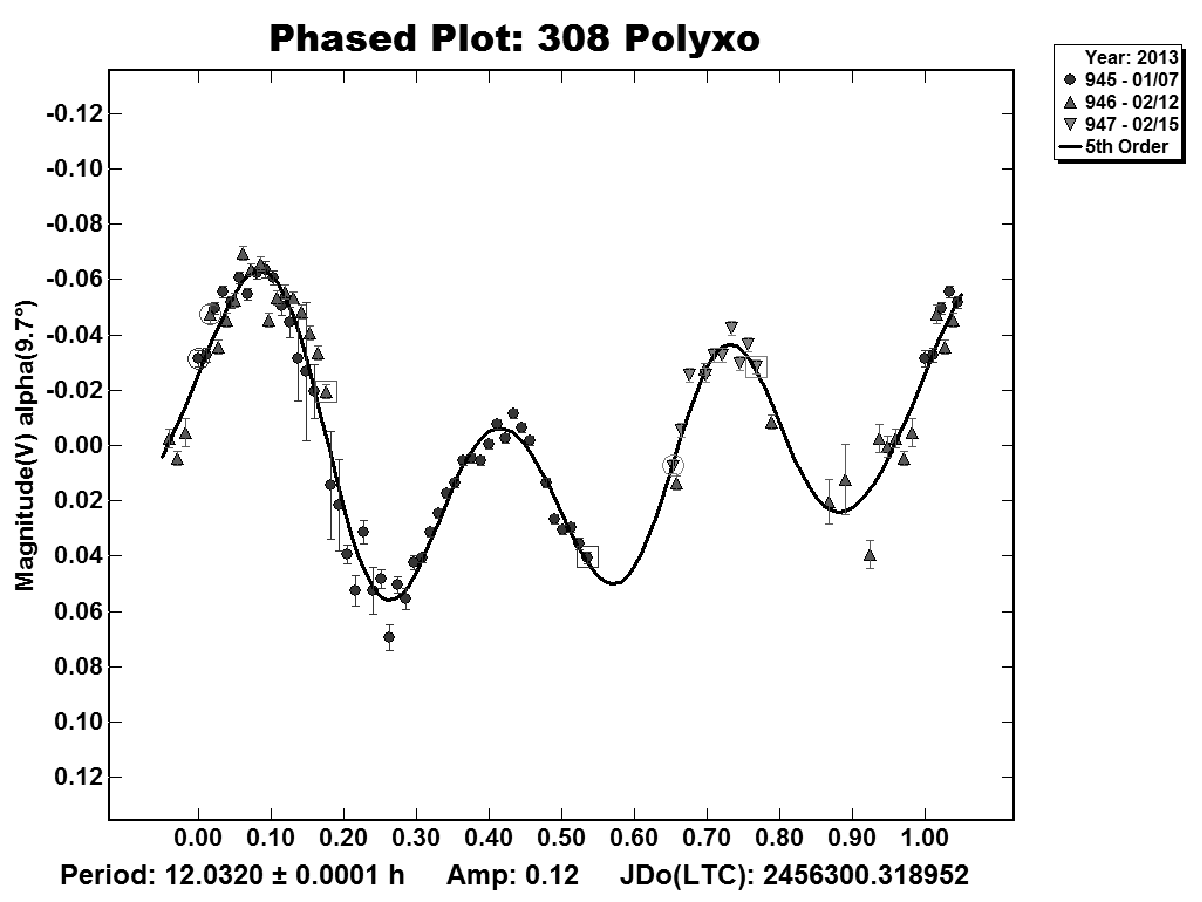}
\caption{Reference lightcurve for asteroid (308) Polyxo ($JD_0$ = 2456306.50).}
\label{fig308}
\end{figure}
\begin{figure}
\includegraphics[width=16cm]{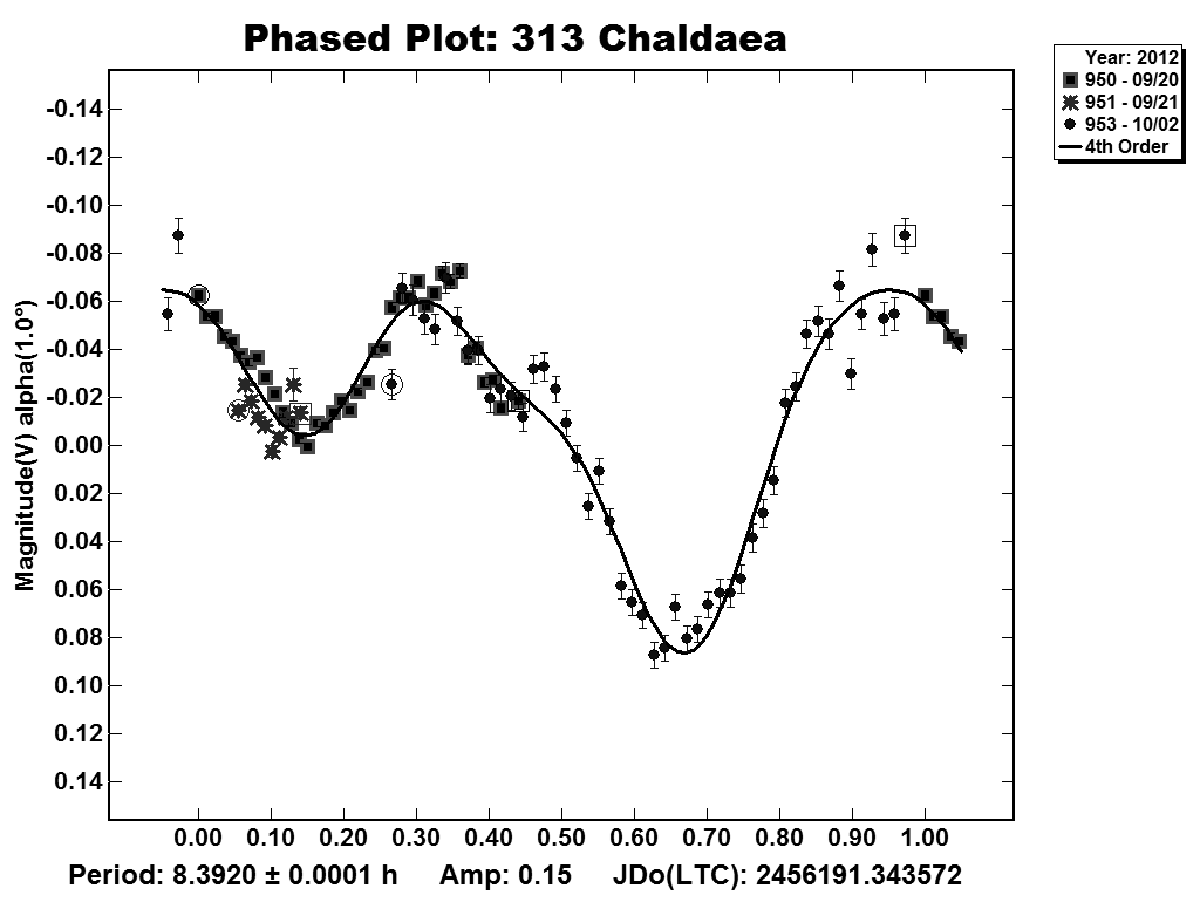}
\caption{First reference lightcurve for asteroid (313) Chaldaea ($JD_0$ = 2456197.50).}
\label{fig313}
\end{figure}
\begin{figure}
\includegraphics[width=16cm]{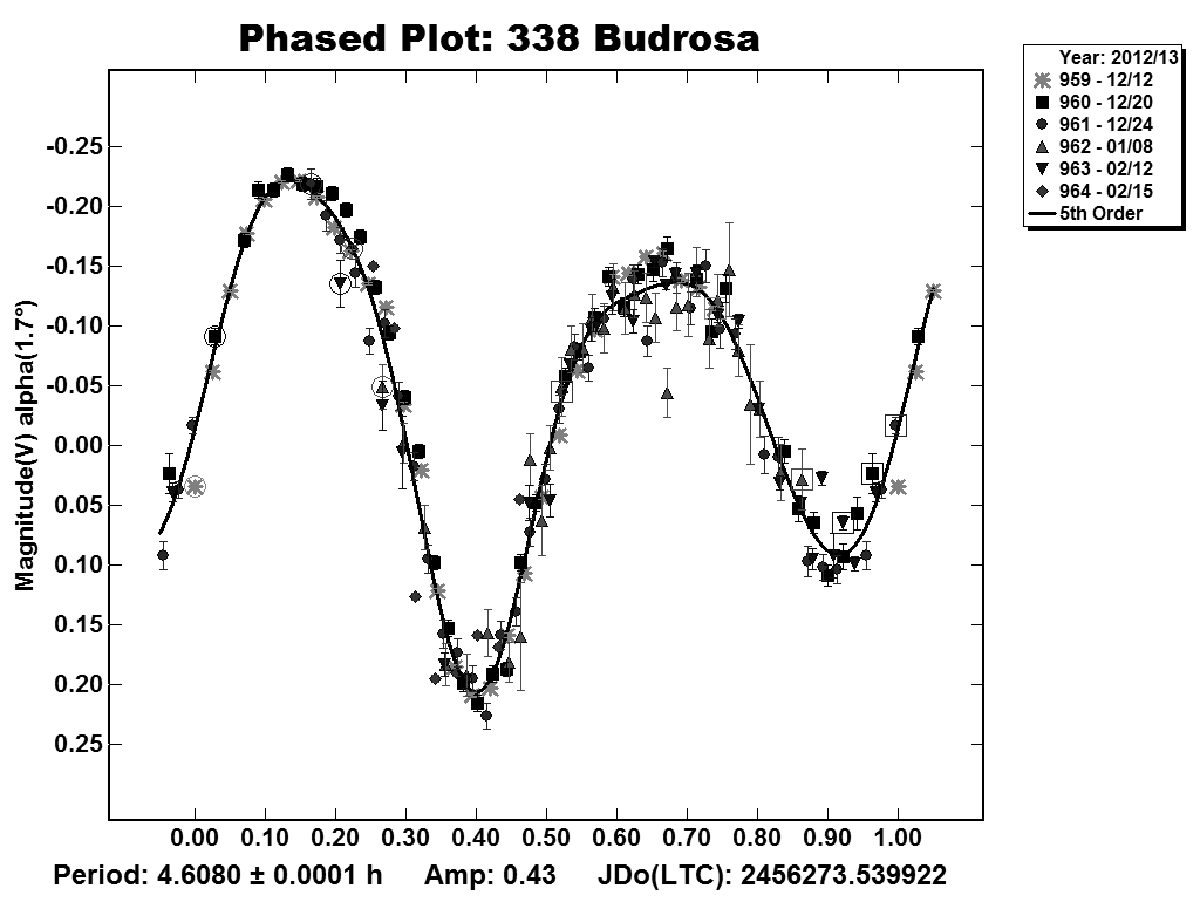}
\caption{Reference lightcurve for asteroid (338) Budrosa ($JD_0$ = 2456295.50).}
\label{fig338}
\end{figure}
\begin{figure}
\includegraphics[width=16cm]{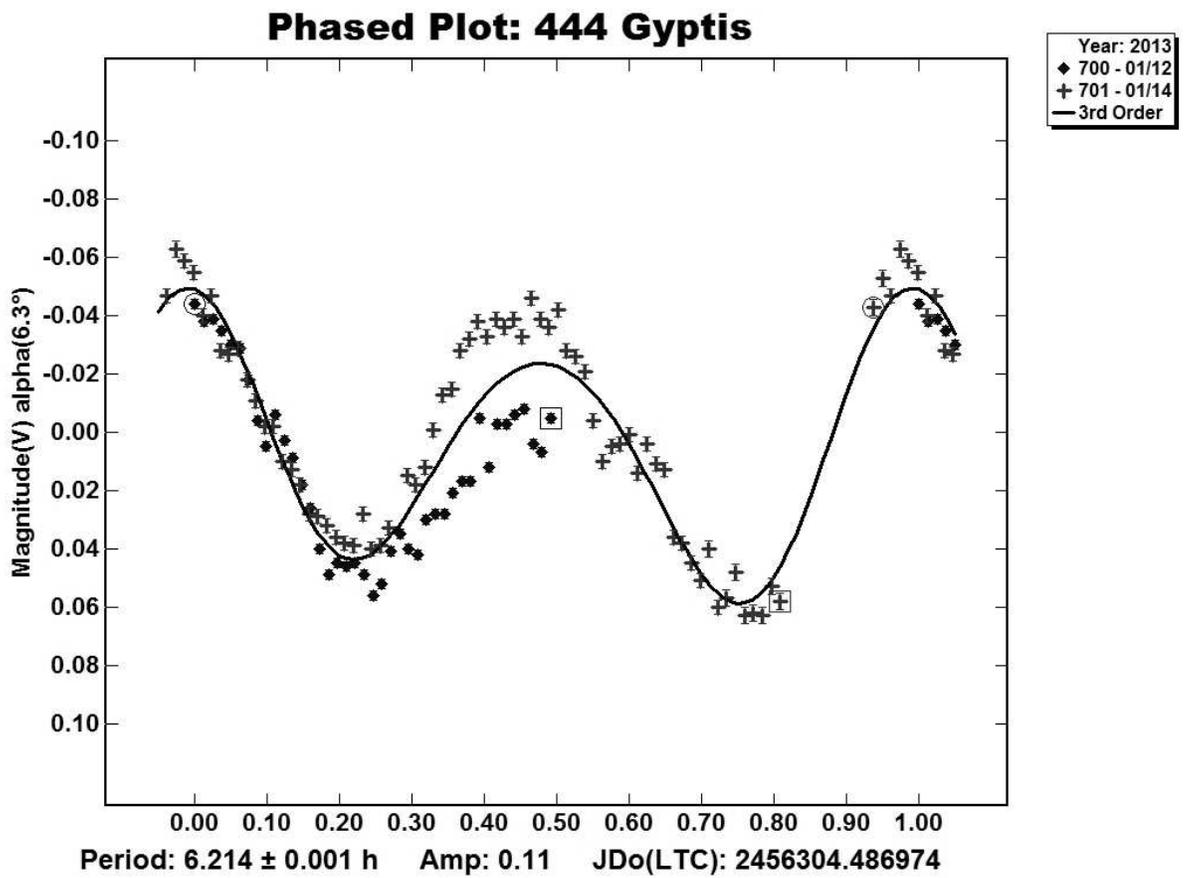}
\caption{The suspect binary event in lightcurve of (444) Gyptis.}
\label{fig444}
\end{figure}
\begin{figure}
\includegraphics[width=16cm]{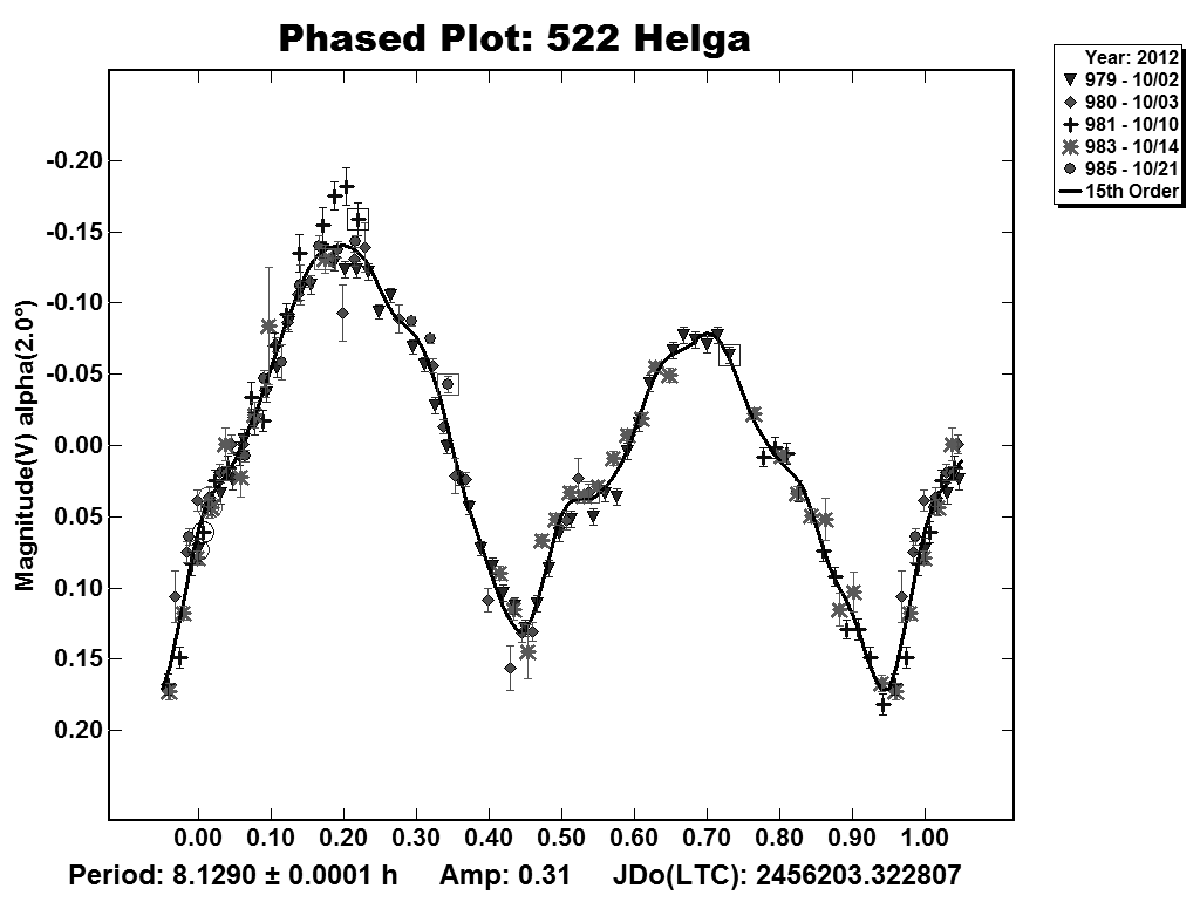}
\caption{First reference lightcurve for asteroid (522) Helga ($JD_0$ = 2456209.50).}
\label{fig522}
\end{figure}
\begin{figure}
\includegraphics[width=16cm]{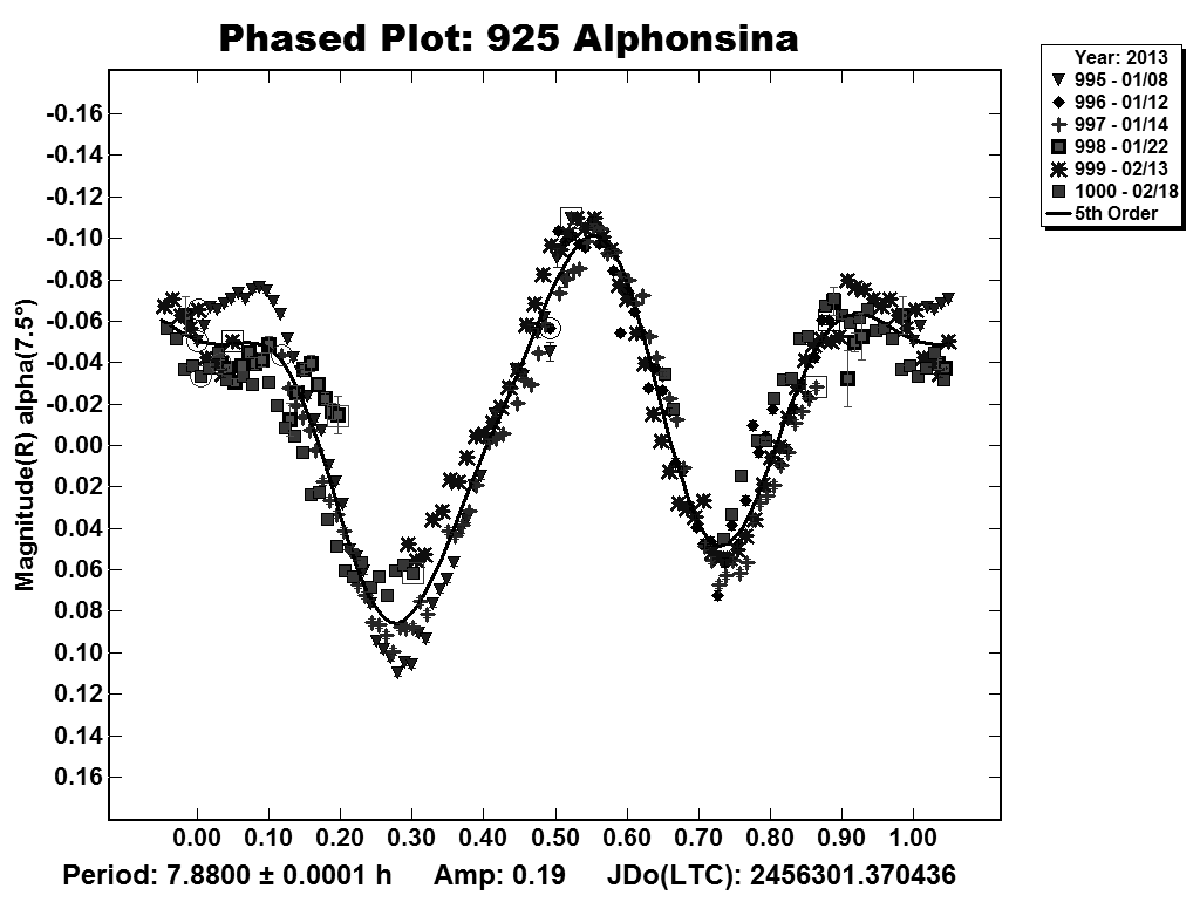}
\caption{Reference lightcurve for asteroid (925) Alphonsina ($JD_0$ = 2456318.50).}
\label{fig925}
\end{figure}

}
For each asteroid in our observing program, we obtained a maximum possible
number of lightcurves taken at different phase angles, as allowed by weather
conditions and available telescope time. All data were obtained, for each
object, at the same apparition, so they all correspond to a unique value of
the aspect angle (the angle between the line of 
sight and the direction of the rotation pole of the object).

Observations were performed between September 2012 and February 2013. We used an 810mm f/7.9 modified Ritchey-Chr\'etien reflector on a fork equatorial mount combined with a back illuminated CCD camera,
2048 $\times$ 2048 square pixel with size of 15 micron. The CCD was used in
binning mode 2 $\times$ 2 with a FOV (Field Of View) of 16.5 $\times$ 16.5 arcmin.
The telescope was equipped with a filter wheel with standard 
$B$, $V$, $R$, $I$ and $C$ (clear) filters. The work at the telescope was
divided in two steps:
\begin{itemize}
\item Asteroid photometric observations in $V$ and $R$ band.
\item Calibration of FOVs Landolt standard fields in $V$ and $R$ 
(all--sky photometry).
\end{itemize}
Reduction of the data and lightcurve analysis were done using 
MPO Canopus v10.7 \footnote{Warner, B.\ D. (2009). MPO Software, Canopus. Bdw Publishing. http://minorplanetobserver.com/}, a software package which carries out differential aperture photometry and Fourier period analysis using an algorithm developed by \citet{Harrisetal89}. 
In all cases, the 
computed rotation period of the targets was found to be in good agreement with the known value
available in the literature. Calibrated $V$ and $R$ magnitudes of the target
asteroids were obtained from calibrated photometry of the 
adopted comparison stars, based on measurements of Landolt calibration fields (see below). 
Note that we used $R$ measurements for the purposes of the determination of atmospheric
extinction (which depends upon the $V-R$ colour index), only, and we did not use them for
the purposes of building phase - mag curves. The resulting, calibrated $V$ magnitudes of 
the asteroids were converted to unit distance from both the Sun and the observer.\\ 

As all asteroids rotate, typically with periods from a few hours to a few days, they should 
all show some rotational modulation superposed on top of the phase curve. Without correction, these rotational
modulations will cause deviations from a smooth phase curve.

For each object, a full lightcurve was obtained by merging together, whenever possible,
phase-calibrated magnitudes taken in consecutive nights
at approximately the same phase angle. A Fourier best-fit was computed to derive the 
lightcurve morphology, in particular the magnitude values at the lightcurve maximum and 
minimum $m_V(max)$ and $m_V(min)$. %\citep{proc}.
These reference lightcurves, shown in Figs.~\ref{fig85} - \ref{fig925}, were then used 
to compute the rotational phases and phase-dependent magnitude shifts of a number of 
partial lightcurves of the same object taken at different phase angles, whenever 
a full lightcurve could not be obtained in the same night due to time or wheather 
constraints. In so doing, each night of observation could be eventually used to derive values
for $m_V(max)$ and $m_V(min)$, and use them to build the phase - mag 
curves. 

In this work we were mainly interested in comparing different photometric systems between 
them. To do this we used in our analysis the magnitudes corresponding to lightcurve maxima, although we are aware that 
we could have used instead the magnitudes at lightcurve minima or, as done in papers of other authors, 
the mean magnitudes, that can be in principle the best option. 
On the other hand, taking into account that we did not obtain at each epoch and for each object complete lightcurves,
as explained above, something that certainly introduces some uncertainty, we do not think
that the main results of this analysis are strongly dependent upon our choice of working
in terms of lightcurve maxima. When looking at our results, however, one should take into account that
we use the symbol $H$ to refer to maximum brightness, and not to lightcurve-averaged values.\\

The implicit requirement of this approach is that the morphology of the lightcurve 
does not change substantially at different phase angles. For sake of safety, in most cases 
two (or three, as in the case of (313) Budrosa) distinct Fourier reference lightcurves were obtained, one at low and another at high 
phase angle. For more details see \ref{7}. We note that not all our reference lightcurves are of the same quality. In some 
cases, the data obtained in consecutive nights do not fit perfectly with each other, and
this effects seems to be larger than one would expect by considering the pure effect
of little differences in phase angle. These effects, quite usual in asteroid phtometry,
may be due to different reasons, mostly related to changes in atmospheric conditions. As
a general rule, however, the reference lightcurves are reasonably stable and well defined.
%
%In this way, we could estimate for each partial lightcurve the
%corresponding value of $m_V(max)$ and $m_V(min)$, and we obtained the resulting phase - mag
%curve of each asteroid of our sample.

\begin{figure}[t]
%\resizebox{\hsize}{!}{\includegraphics[clip=true]{O-K.eps}}
\includegraphics[clip=true]{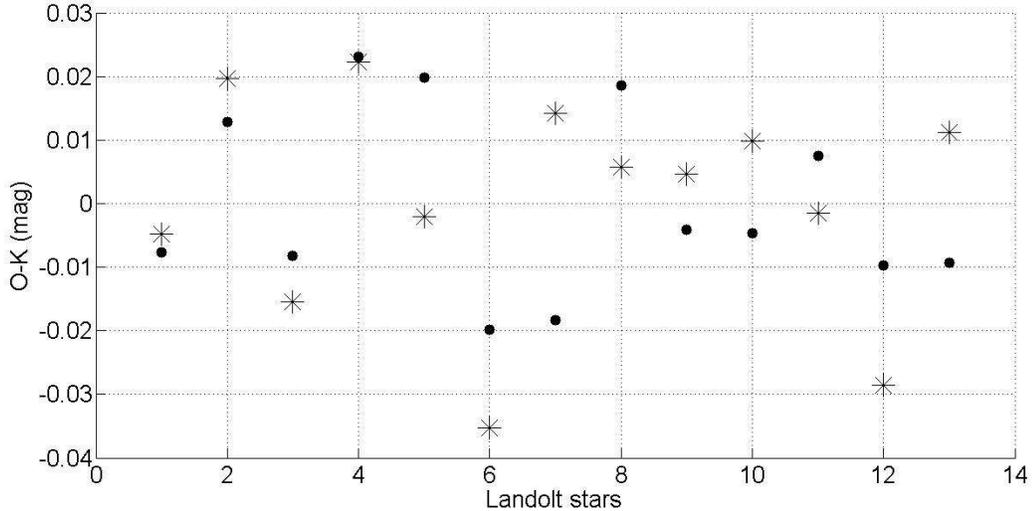}
\caption{
\footnotesize
Comparison between the observed (O) and known (K) magnitudes of the Landolt
stars used on 2012, Oct 22 to compute the atmospheric model used to reduce
the observations of the asteroid 236 Honoria (asterisk = $R$ filter,
dots = $V$ filter).}
\label{O-K}
\end{figure}

\subsection{The Landolt field calibration}
\label{4}
Every time the night was photometric, we carried out photometric
measurements of Landolt field stars in $V$ and $R$ to
derive a correct calibration of the instrumental magnitudes.
This was done to derive reliably calibrated magnitudes
of the comparison stars used to derive the correspondingly calibrated magnitudes 
of the asteroids for each night of observation. 
Of course, the comparison stars are in the same FOV as the asteroid.
Images of various Landolt fields were taken at various air masses in order to calibrate 
the atmospheric extinction
\citep{modAtm}.\\
We then measured the $V$ and $R$ instrumental magnitude for each Landolt star
by computing the average of $5$ images per filter. From the instrumental
magnitude, a four parameters atmospheric model was fit using the following
linear equations:
\begin{equation}
V-v=-[k_{1v}+k_{2v}(V-R)]X+c_{v}(V-R)+V_{0} \label{atmV}
\end{equation}
\begin{equation}
R-r=-[k_{1r}+k_{2r}(V-R)]X+c_{r}(V-R)+R_{0} \label{atmR}
\end{equation}
where:
$k_{1}$ is the first-order atmospheric extinction for the V or R filter;
$k_{2}$ is the second-order atmospheric extinction for the given filter;
$c$ is the instrumental color correction coefficient for the given colour index;
$R_0$ and $V_0$ are the unknown zero point magnitudes;
$V$, $R$ and $V-R$ are the known, tabulated magnitudes and color index from Landolt fields;
$v$ and $r$ are the measured instrumental magnitudes;
$X$ is the air-mass, i.e. the optical path length through the Earth's
  atmosphere, expressed as a ratio relative to the path length at the zenith.

To get the four unknown coefficients it is sufficient to have at least four
different Landolt stars, so that we can write Eqn. (\ref{atmV}) and
Eqn. (\ref{atmR}) for each star and get a system of linear equations to
be solved. In general, to increase the accuracy, we processed about ten stars
in such a way to have an overdetermined linear system. The method of ordinary
least squares was then used to find an approximate solution.

After finding the various constant parameters of the atmospheric model, we
derived the calibrated $V$ and $R$ magnitudes of the comparison stars used 
to compute asteroids' magnitudes. 
%by taking some images, in the same night, of the FOV in which asteroids were located. 
We show in Fig.~ \ref{O-K} a comparison between the observed  
and known magnitude of stars located within some of the considered Landolt
fields analyzed to process the observations of asteroid (236) Honoria. The
resulting $O-C$ differences are at most few hundredths of a magnitude both
in $V$ and in $R$ filter.

In applying this data reduction procedure we assumed that the $V-R$ of the
comparison stars are about equal (within few hundredths of mag)
to the $V-R$ of the asteroid. For this reason, we used as comparison star
only solar-type stars. Under this 
assumption, from Eqn. (\ref{atmV}) and Eqn. (\ref{atmR})) we obtain 
($a$=asteroid, $c$=comparison star):
\begin{equation}
V_c-V_a=v_c-v_a \label{Vast}
\end{equation}
\begin{equation}
R_c-R_a=r_c-r_a \label{Rast}
\end{equation}
%From which we derived the apparent magnitudes of our asteroids.

\begin{table*}
  \caption{Summary of the lightcurve analysis for our sample of asteroids.
    $V_{max}$ is the reduced magnitude of the lightcurve
    maximum, $V_{min}$ is the reduced magnitude of the lightcurve minimum, 
    $V_{mean}$ is the mean magnitude of the lightcurve
    while $\sigma_m$ is the standard deviation of the mean between the Fourier
    fit and the observed magnitudes.}
\begin{center}
\begin{tabular}{lcccc}
\hline
 Phase angle ($^{\circ}$) & $V_{max}$ (mag) & $V_{min}$ (mag) & $V_{mean}$ (mag) & $\sigma_m$ (mag)\\
\hline
236 Honoria & & & & \\
\hline
1.27 	&      8.17 	&      8.35 & 8.27 & 0.06\\
1.95 	&      8.22 	&      8.40 & 8.32 & 0.06\\
0.86 	&      8.10 	&      8.28 & 8.20 & 0.09\\
6.39 	&      8.63 	&      8.82 & 8.74 & 0.05\\
9.77 	&      8.73 	&      8.91 & 8.83 & 0.06\\
10.69 	&      8.78 	&      8.97 & 8.89 & 0.07\\
22.14 	&      9.12 	&      9.30 & 9.22 & 0.04\\
24.72 	&      9.16 	&      9.34 & 9.27 & 0.05\\
\hline
313 Chaldea & & & & \\
\hline
0.99 	&      9.06 	&      9.21 & 9.11 & 0.02\\
0.57 	&      8.94 	&      9.09 & 8.99 & 0.01\\
4.74 	&      9.21 	&      9.36 & 9.26 & 0.05\\
8.37 	&      9.41 	&      9.53 & 9.46 & 0.04\\
10.50 	&      9.54 	&      9.66 & 9.60 & 0.05\\
20.33 	&      9.86 	&      9.99 & 9.94 & 0.06\\
\hline                                
522 Helga & & & & \\                  
\hline
2.00 	  &    9.23 	&      9.54 &  9.37 & 0.08\\
2.19 	  &    9.28 	&      9.59 &  9.42 & 0.09\\
4.07 	  &    9.36 	&      9.67 &  9.50 & 0.11\\
4.99 	  &    9.41 	&      9.72 &  9.55 & 0.08\\
7.40 	  &    9.55 	&      9.86 &  9.69 & 0.09\\
13.67 	&      9.85 	&     10.14 &  9.99 & 0.12\\
16.24 	&      9.88 	&     10.17 & 10.11 & 0.13\\
\hline                                
85 Io & & & & \\
\hline
0.89 	  &    7.62 	&      7.80 & 7.73 & 0.06\\
1.18 	  &    7.67 	&      7.85 & 7.77 & 0.07\\
2.07 	  &    7.82 	&      8.00 & 7.92 & 0.04\\
5.11 	  &    8.01 	&      8.19 & 8.11 & 0.05\\
16.24 	&      8.48 	&      8.66 & 8.58 & 0.04\\ 
17.49 	&      8.53 	&      8.76 & 8.65 & 0.08\\
21.24 	&      8.66 	&      8.89 & 8.78 & 0.08\\
\end{tabular}                         
\end{center}
\label{phase_data}
\end{table*}

\setcounter{table}{1}
\begin{table*}
  \caption{(continued). Summary of the lightcurve analysis for our sample of
    asteroids.}
\begin{center}
\begin{tabular}{lcccc}
\hline
Phase angle ($^{\circ}$) & $V_{max}$ (mag)& $V_{min}$ (mag)& $V_{mean}$ (mag) & $\sigma_m$ (mag)\\
\hline
208 Lacrimosa & & & & \\
\hline
0.64 	  &    9.20 	&      9.48 &   9.33 & 0.04\\
1.01 	  &    9.24 	&      9.52 &   9.37 & 0.07\\
1.74 	  &    9.39 	&      9.66 &   9.52 & 0.08\\
3.75 	  &    9.69 	&      9.96 &   9.82 & 0.09\\
11.85 	&      9.73 	&     10.01 &   9.87 & 0.08\\ 
17.28 	&      9.97 	&     10.25 &  10.10 & 0.08\\
19.18 	&      9.99 	&     10.27 &  10.12 & 0.08\\
\hline
306 Unitas & & & & \\
\hline
5.45 	  &    9.18 	 &     9.37 & 9.28 & 0.07\\
5.83 	  &    9.17 	 &     9.36 & 9.27 & 0.05\\
6.34 	  &    9.21 	 &     9.39 & 9.31 & 0.05\\
8.00 	  &    9.28 	 &     9.46 & 9.38 & 0.02\\
16.03 	&      9.61 	&      9.79 & 9.71 & 0.06\\ 
21.23 	&      9.60 	&      9.78 & 9.70 & 0.10\\
22.85 	&      9.69 	&      9.87 & 9.79 & 0.05\\
\hline
338 Budrosa & & & & \\
\hline
1.54 	  &    8.74 	 &     9.16 &  8.93 & 0.16\\
3.86 	  &    8.93 	 &     9.36 &  9.12 & 0.14\\
%5.40 	  &    9.53 	 &     9.95 &  9.72 & 0.13\\
10.89 	&      9.49 	&      9.91 &  9.68 & 0.11\\ 
18.93 	&      9.85 	&     10.28 & 10.04 & 0.11\\
19.25 	&      9.77 	&     10.20 &  9.97 & 0.16\\
\hline                                 
925 Alphonsina & & & & \\
\hline
7.16 	  &    8.84 	&      9.01 & 8.92 & 0.07\\
8.15 	  &    8.79 	&      8.97 & 8.87 & 0.05\\
9.22 	  &    8.87 	&      9.05 & 8.96 & 0.06\\
12.16 	&      9.03 	&      9.21 & 9.11 & 0.03\\ 
18.93 	&      9.17 	&      9.35 & 9.25 & 0.05\\
20.04 	&      9.11 	&      9.29 & 9.19 & 0.05\\
\end{tabular}                          
\end{center}
\end{table*}

%\begin{figure}[p]
%\resizebox{\hsize}{!}{\includegraphics[clip=true]
%{img/444_Gyptis_suspect_binary_event.eps}}
%\caption{
%\footnotesize
%The suspect binary event for (444) Gyptis.}
%\label{Gyptis}
%\end{figure}

\section{Asteroid photometric systems}
\label{1a}
Here we briefly summarize some of the most commonly adopted
photometric systems to describe the observed phase - mag curves of asteroids.
By knowing the orbit, and hence the distance of an asteroid observed
at a given epoch, if we call $\alpha$ the phase angle and $m_V$ the 
measured apparent magnitude, we can immediately compute  
$V(\alpha, 1)$, namely the conversion of $m_V$ into the magnitude corresponding 
to unit distance from both the Sun and the observer:
\begin{equation}
V(\alpha,1) = m_V-5\, log_{10}\left(r \Delta \right) \label{reduced}
\end{equation}
where $r$ is the asteroid heliocentric distance, and $\Delta$ is
the distance from the observer, both distances being expressed in AU.
Since in this paper we always use magnitudes reduced to unit distance, let us
simplify the notation by writing henceforth $V(\alpha)$ instead of
$V(\alpha,1)$.

In analyzing our data, we have considered three main photometric systems that
are or have been used in asteroid science to describe the variation
of $V(\alpha)$ as a function of $\alpha$. The first one is the so-called
$(H, G)$ system, that was officially adopted by the International
Astronomical Union between 1985 and 2012 \citep{Bowelletal89}.
By indicating as $H$ the value of $V(0^\circ)$, namely, by definition, the
absolute magnitude of the asteroid, the $(H, G)$ 
system is defined by the following Equation:

\begin{equation}
  V(\alpha) = H +2.5\, log_{10} \left[ \left( 1-G \right)\Phi_1(\alpha)+
    G\Phi_2(\alpha) \right] \label{HGsystem}
\end{equation}
where the so-called slope parameter $G$ is a function describing the
variation of $V(\alpha)$ measured at different phase angles. 
$\Phi_1(\alpha)$ and $\Phi_2(\alpha)$ are two ancillary functions of the
phase angle $\alpha$, with $\Phi_1(0^\circ)=\Phi_2(0^\circ)\equiv 1$. A
comprehensive description of the $(H, G)$ system, including an
explicit mathematical formulation of the $\Phi_1$ and $\Phi_2$ functions
of $\alpha$, can be found in \citet{HG1G2}.

The second photometric system that we consider in this paper is
the  $(H, G_1, G_2)$ system, that has been more recently proposed by
\citet{HG1G2} as an improvement of the older $(H, G)$ system, and
has been later officially adopted by the IAU during the
General Assembly in 2012. It is defined as:
\begin{equation}
 V(\alpha) = H + 2.5\, log_{10} \left[ G_1\overline\Phi_1(\alpha)+G_2\overline\Phi_2(\alpha)+\left(1-G_1-G_2\right)\overline\Phi_3(\alpha) \right] \label{HG1G2system}
\end{equation}
where $\overline{\Phi}_1(\alpha)$, $\overline{\Phi}_2(\alpha)$ and
$\overline{\Phi}_3(\alpha)$ are three base functions of the phase angle
$\alpha$, with $\overline{\Phi}_1(0^\circ)=\overline{\Phi}_2(0^\circ)=\overline{\Phi}_3(0^\circ)\equiv 1$ A comprehensive description of this photometric
system, that has been proposed to improve the accuracy of the derived values
of H, is given by \citet{HG1G2}. A constrained non linear least-squares algorithm to be used in estimating the parameters in the ($H$, $G_1$, $G_2$) phase function has been
published more recently by \citet{Penttilaetal16}. Note that we decided to consider
only the full ($H$, $G_1$, $G_2$) phase function in this paper, and we did
not use the simplified ($H$, $G_{12}$) phase function that can replace
($H$, $G_1$, $G_2$) when the coverage of the phase - mag is not optimal. We took this decision
because since the beginning our goal was to obtain reasonably well-sampled phase - mag
curves, suitable to derive the full set of $H$, $G_1$ and $G_2$ parameters. Although 
in few cases this was not really possible, we decided that adding the simplified 
system would not have been so advantageous.

Moreover, we also consider the empirical photometric system proposed by
\cite{she}, expressed through the following Equation:
\begin{equation}
V(\alpha) = V_{lin}(0)-\frac{a}{1+\alpha}+b\alpha \label{Vabsystem}
\end{equation}
In this formulation, the opposition effect corresponds to the difference
between a simple extrapolation to zero phase of the linear part of the
mag - phase curve, described by the $b$ parameter, and the value that is
actually observed and is determined by the presence of a brightness surge
described by the term including the parameter $a$.
$V_{lin}(0)$ represents therefore the extrapolation to zero
phase angle of a purely linear mag - phase relation having angular
coefficient $b$. The absolute magnitude, taking into account the presence
of an opposition effect, is then given by $H = V_{lin}(0)-a$, 
$a$ and $b$ being parameters to be determined for each object. Note also that
the form of the term with $a$ in Eqn.$\,$\ref{Vabsystem} is such that
the profile of the opposition effect has a fixed shape, determined by
the adopted form for the denominator: for instance, at the phase of $1^\circ$
the magnitude opposition effect becomes about one half the value assumed
at zero phase angle. In principle, this might be changed, although we do
not make any attempt to explore alternative formulations.

Finally, we note that we also computed simple linear least-squares fits
for the objects of our sample, but using only observations obtained
at phase angle $\alpha \ge 10^\circ$. The reason was to explore the
relation between such kind of slope, that will be routinely determined
by Gaia observations, and the albedo, to improve the calibration of the
proposed \citet{Shev} relation.

\section{The results}
\label{res}

%In so doing, we used both the older $(H,G)$ (Eq. \ref{HGsystem}) and
%the newly adopted $H,G_{1},G_{2}$ (Eq. \ref{HG1G2system}) photometric systems,
%as well as the Shevchenko model (Eq. \ref{Vabsystem}) as reported in
%Section \ref{1a}.
We built our Phase - mag curves by using as magnitude values those corresponding
to the maxima of the obtained lightcurves. In particular, we estimated that in all
the different cases the uncertainty in the determination of this parameter was 
not exceeding $\pm 0.015$ mags. Therefore, we adopted this value in the computations 
of the best-fit representations of our phase - mag curves using
the different photometric systems described above.

Apart from an easy computation to 
determine the slope $\beta$ of a linear relation between magnitude and phase for 
phase angles $> 10^\circ$, in the most difficult case of the three main photometric 
systems described in the previous Section, nonlinear regression methods were needed. 
In our case, we used different independent approaches, and we
checked that the results were coincident within the nominal error bars.

A first approach was the use of a genetic algorithm developed in the past by some of
us for other purposes, and optimized for the present problem. Genetic algorithms
are relatively simple and are well suited to determine sets of
best-fit parameters using even more complicated relations than those considered
in this application. These algorithms have the
advantage of making an extensive exploration of the parameters' space, and
to be scarcely prone to fall into local best-fit minima. The draw-back is
some difficulty in estimating the resulting uncertainties of
the solution parameters. Essentially
the same approach, originally developed for the purpose of inversion of
sparse asteroid photometric data \citep{Cellino2009} has also been more
recently used in the computation of best-fit approximation of asteroid
phase - linear polarization data \citep{MNRAS1, MNRAS2}.

In order to obtain independent results and to obtain a better estimate
of the uncertainties on the derived parameters of the different photometric systems, 
we adopted also more classical least-squares approaches by using standard
numerical routines available in the literature and implementing them in algorithms 
either written by ourselves,
or using, as a check, routines included in the MATLAB$^{\textregistered}$
package by MathWorks\footnote{http://www.mathworks.com}. In particular, we
used the MATLAB statistical toolbox ``nlinfit'', that uses a minimisation tool 
based on the Levenberg-Marquardt algorithm. We also note that, in fitting 
our data using the ($H$, $G_1$, $G_2$) system, we followed the procedures
recommended by \citet{Penttilaetal16}, apart from the fact that in a few cases
we allowed the value of either $G_1$ or $G_2$ to reach negative values, provided that,
in any case, $1 - G_1 -G_2 \le 1$

%Like other numerical minimization algorithms, the Levenberg–Marquardt
%algorithm is an iterative procedure that converges to the global minimum
%only if the initial guess is already somewhat close to the final solution.
%In our case as starting parameters we have chosen ($min = minimum$):
%\begin{itemize}
%\item $H_{guess}=\left[min(V_{max})+min(V_{min})\right]/2$, $G_{guess}=1$ for
%  the $H,G$ model 
%\item $H_{guess}=H_{HG}$ and $G_{1guess}=G_{2guess}=G_{HG}$ for the $H,G_{1},
%G_%{2}$ model ($H_{HG}$ and $G_{HG}$ are the best parameters for $H, G$ model) 
%\item $V(0, 1)_{guess}=H_{HG}$ and $a_{guess}=b_{guess}=1$ for the Shevchenko
%model
%\end{itemize}
%The choice of the initial values for the parameters was not so critical as
%regards the final results, that will be presented in Section \ref{5}.

The results obtained with the different approaches were found to be
in good agreement, giving coincident results within the derived error bars.
We also note that for some objects no solution could be found using one
or more of the adopted photometric systems. This happened in the cases 
of asteroids (306) Unitas and (925) Alphonsina, for which
the low-phase angle region was not adequately covered by our observations
making it impossible to obtain a best-fit solution using the ($H$, $G_1$, $G_2$)
photometric system. In this respect, we remind that the ($H$, $G_1$, $G_2$)
system has been developed precisely to improve the accuracy in the determination of
the absolute magnitude $H$, taking into account the behaviour exhibited by the objects
at small phase angles.
%In such cases, however, we could in any case compute a simple linear
%best-fit of the data obtained at phase angles larger than $10^\circ$.

Conversely, in the case of (338) Budrosa, we could not compute a simple
linear fit, due to a lack of data at large phase angles.
Among the other asteroids of our original target list shown in Table \ref{ast},
we were forced to exclude {\em a priori} from our analysis (135) Hertha, due 
to an insufficient number of data points. We also discarded (308) Polyxo
and (444) Gyptis due to problems of data quality. In particular, in the case of Gyptis
we obtained some anomalous lightcurve morphology on January 13-14, 2013, when
we observed a magnitude variation looking like a mutual event in a binary
system (Fig. \ref{fig444}). 
This is certainly to be confirmed by future observations,
but it is clear that in any case the possible presence of mutual events would alter the
lightcurve morphology, therefore we did not take into account this asteroid for any 
further analysis in this paper.

\begin{figure}[p]
\begin{center}
\includegraphics[width=6.5cm]{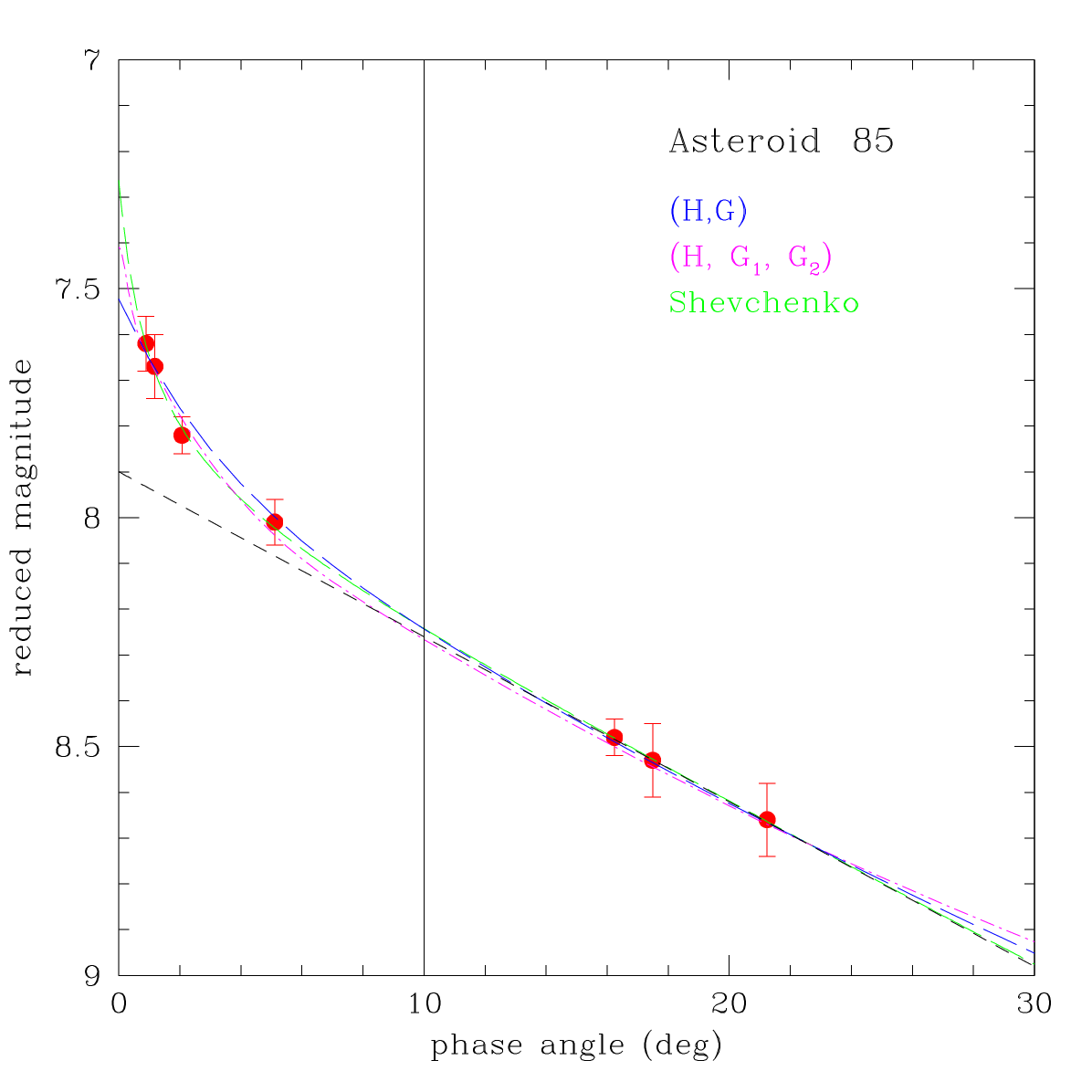}
\includegraphics[width=6.5cm]{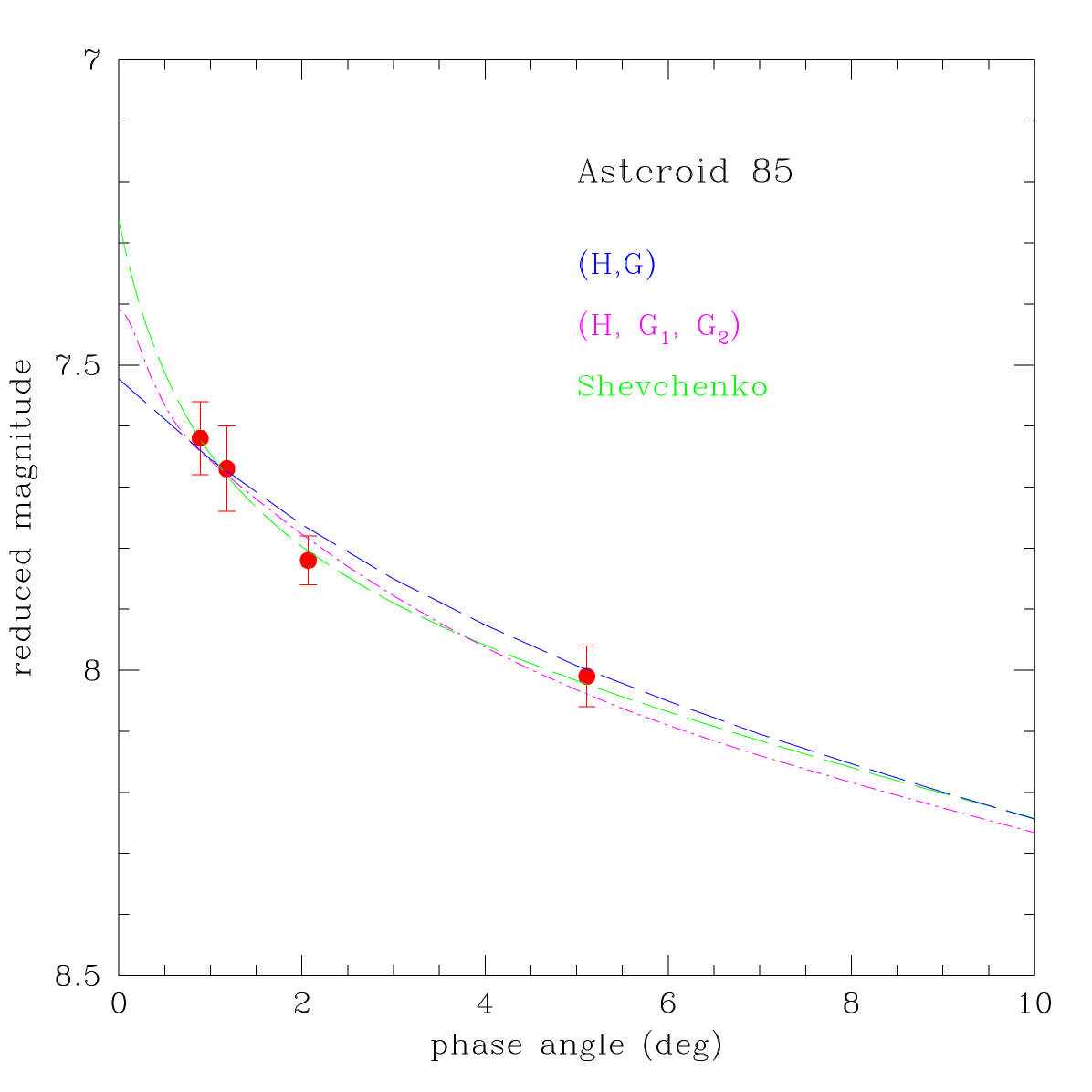}
\end{center}
\caption{Magnitude maximum brightness - phase curve in $V$ band for (85) Io, and resulting
  best-fit curves corresponding to the three photometric systems considered
  in this paper. Left Panel: whole curve. The vertical line separates
  the range of phase angles smaller than $10^\circ$, the lower limit in phase angle 
  attainable by Gaia. The linear fit of the measurements obtained above 
  $10^\circ$ are also shown. Right Panel: zooming on the solar
  opposition region, at phase angle $\le$ 10 degrees}
\label{phaseV85}
\end{figure}

\begin{figure}[p]
\begin{center}
\includegraphics[width=6.5cm]{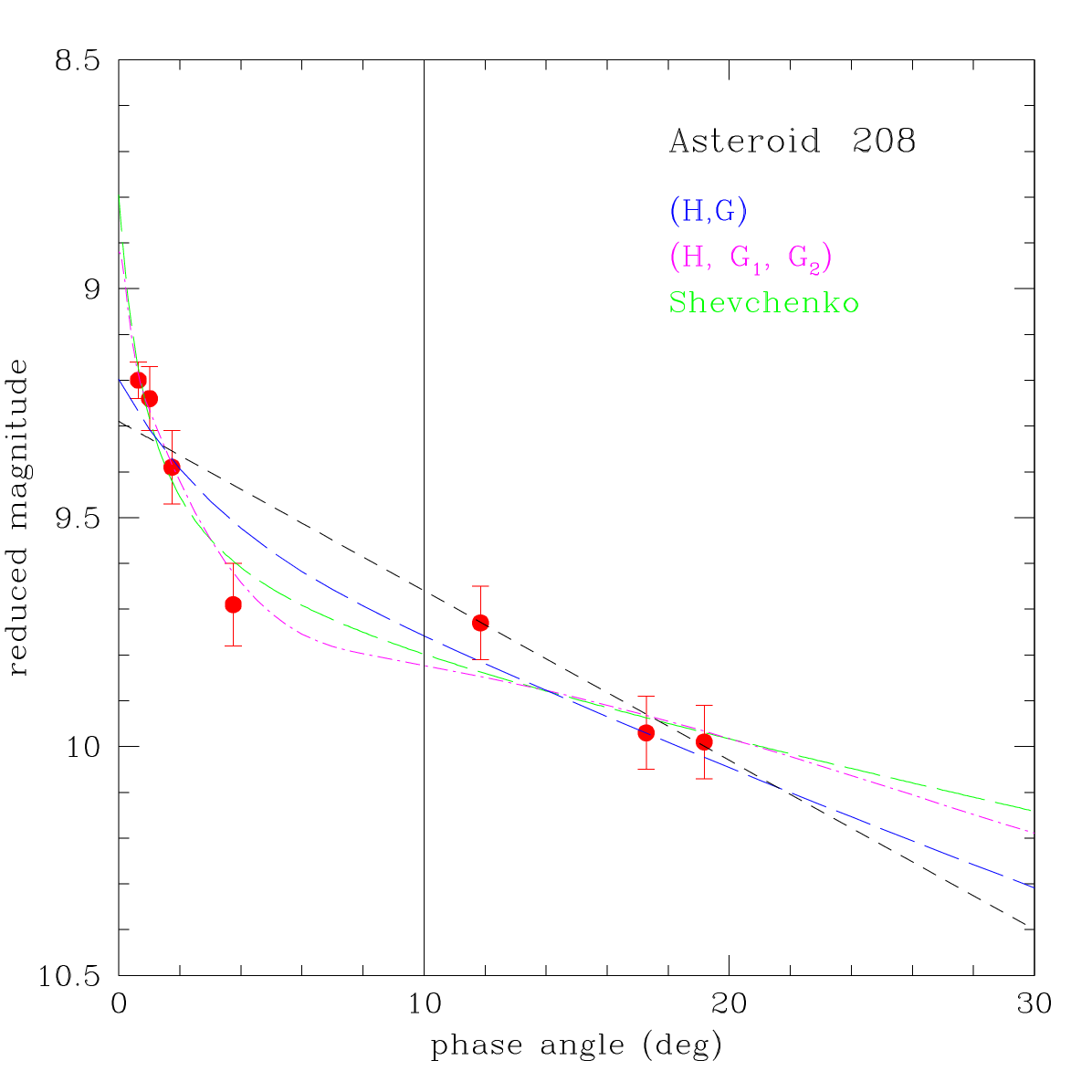}
\includegraphics[width=6.5cm]{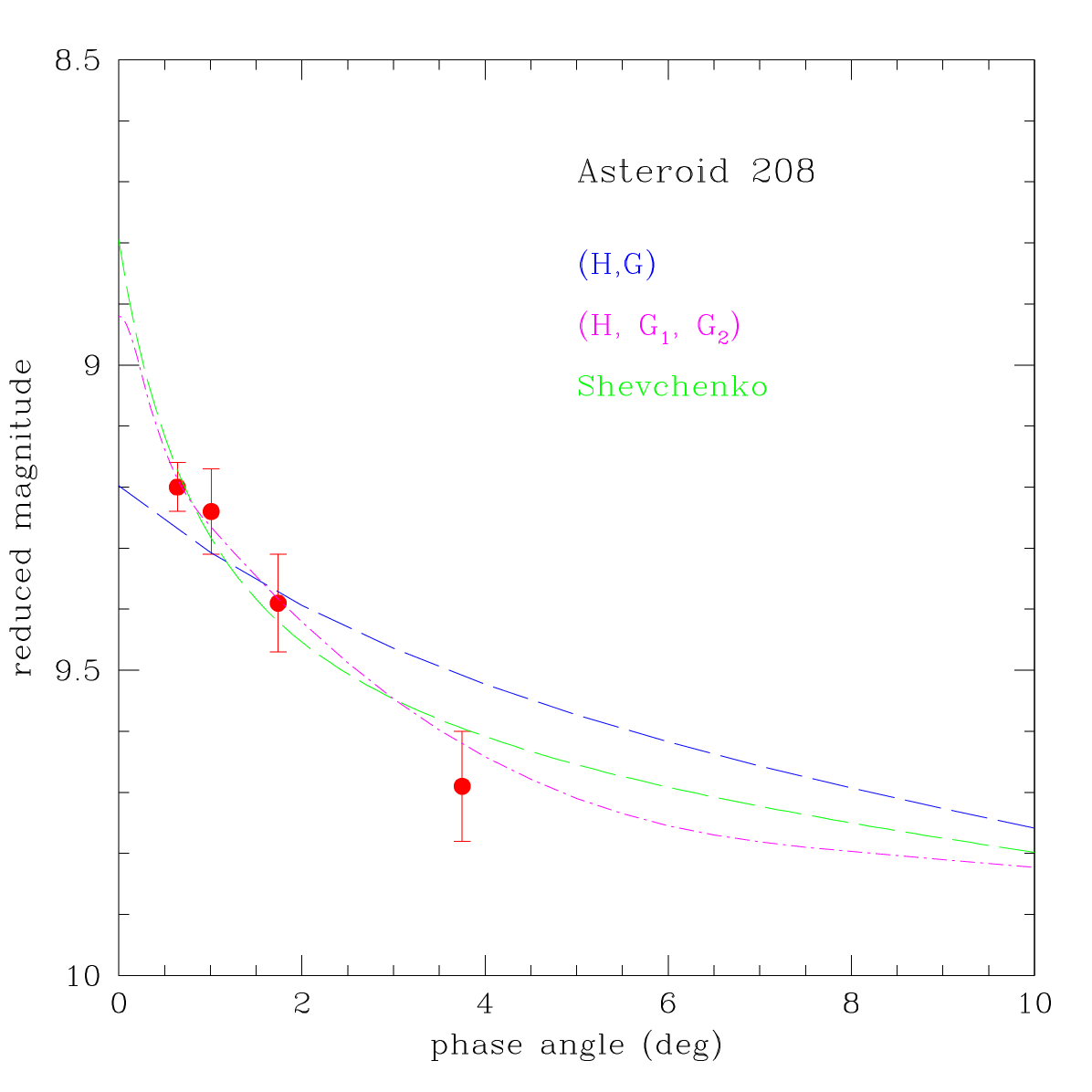}
\end{center}
\caption{The same as Fig. \ref{phaseV85}, but for asteroid (208) Lacrimosa.
Note that the point at phase angle of 3.75 degrees is responsible of the 
negative value of the $G_1$ parameter found in the best-fit solution.}
\label{phaseV208}
\end{figure}

\begin{figure}[p]
\begin{center}
\includegraphics[width=6.5cm]{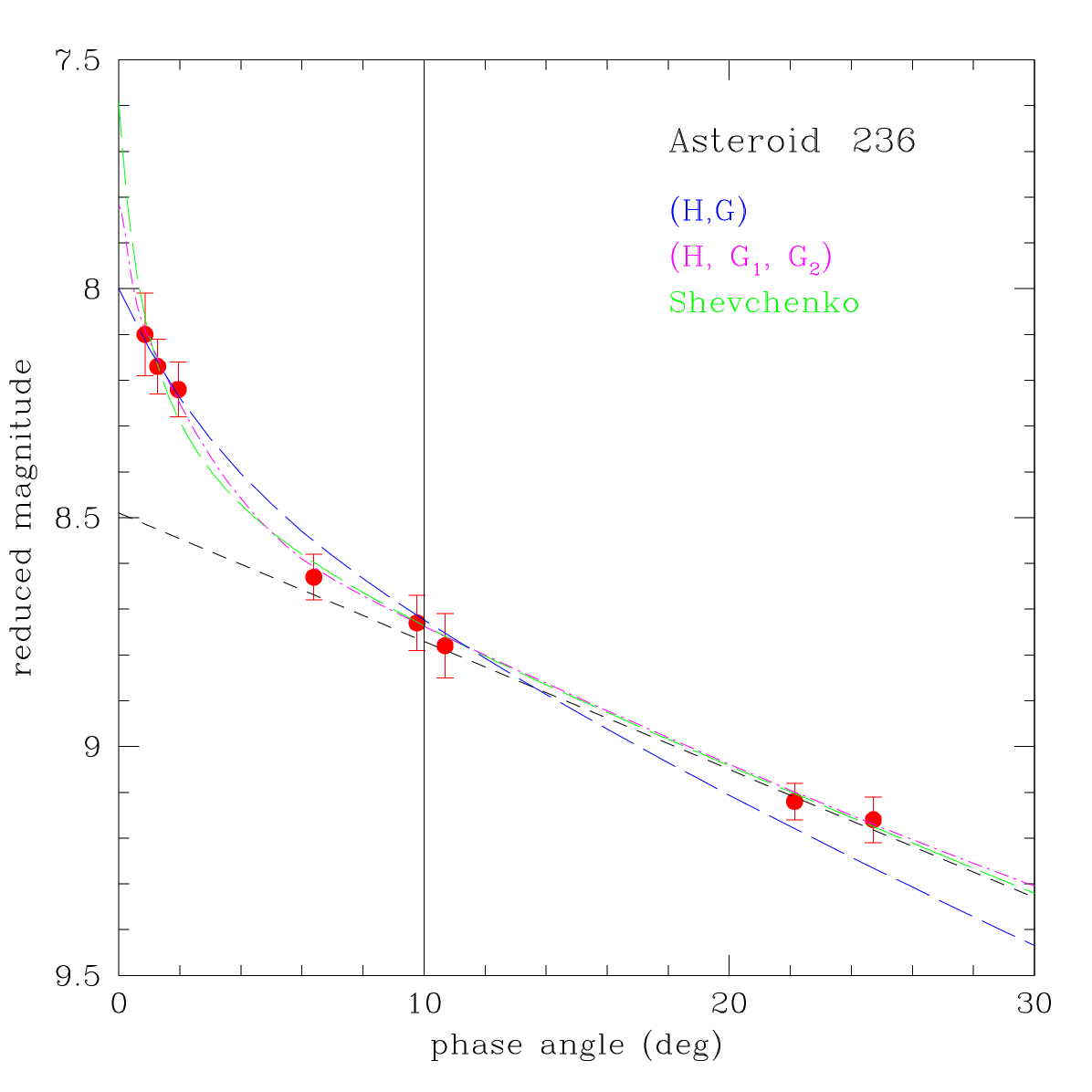}
\includegraphics[width=6.5cm]{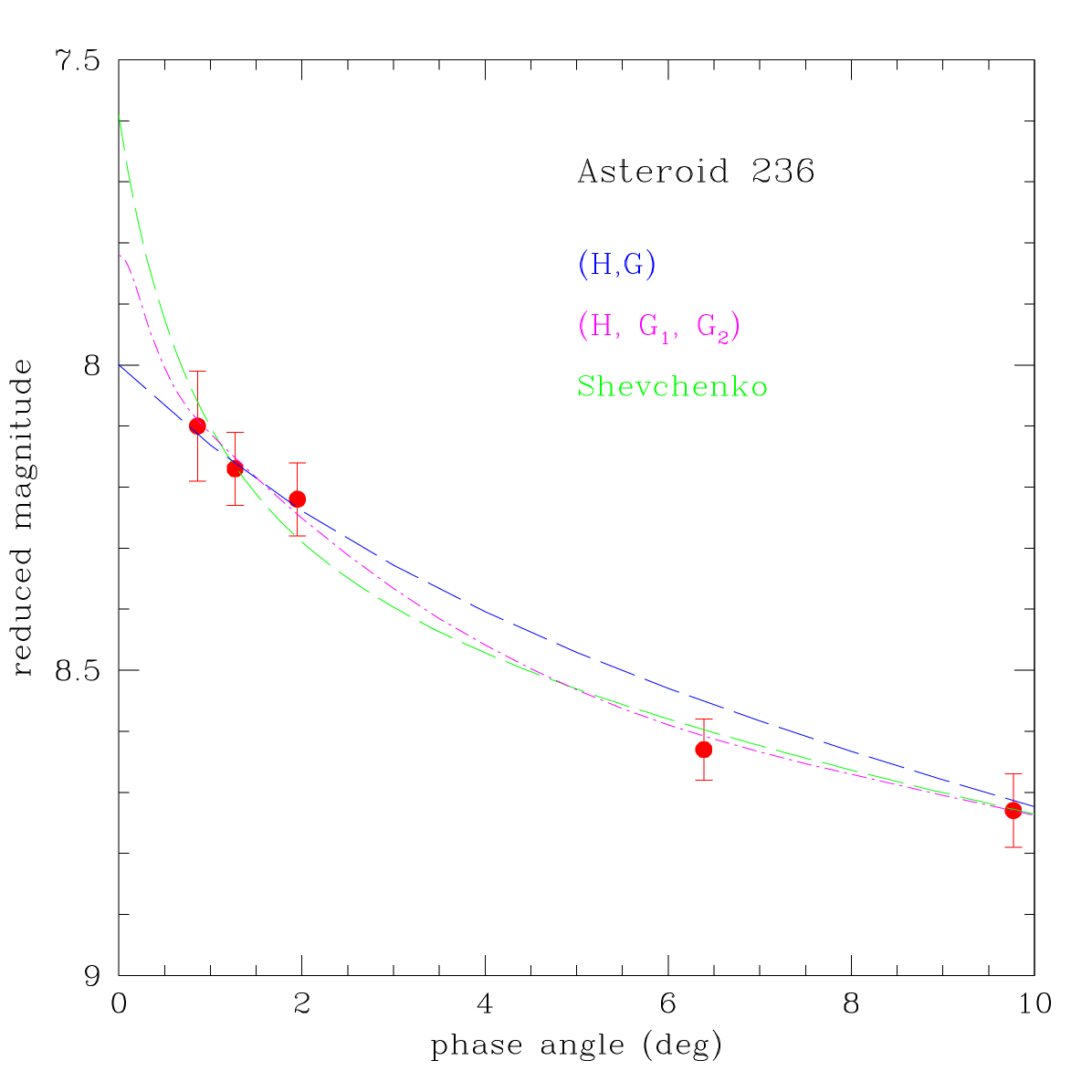}
\end{center}
\caption{The same as Fig. \ref{phaseV85}, but for asteroid (236) Honoria.}
\label{phaseV236}
\end{figure}

\begin{figure}[p]
\begin{center}
\includegraphics[width=6.5cm]{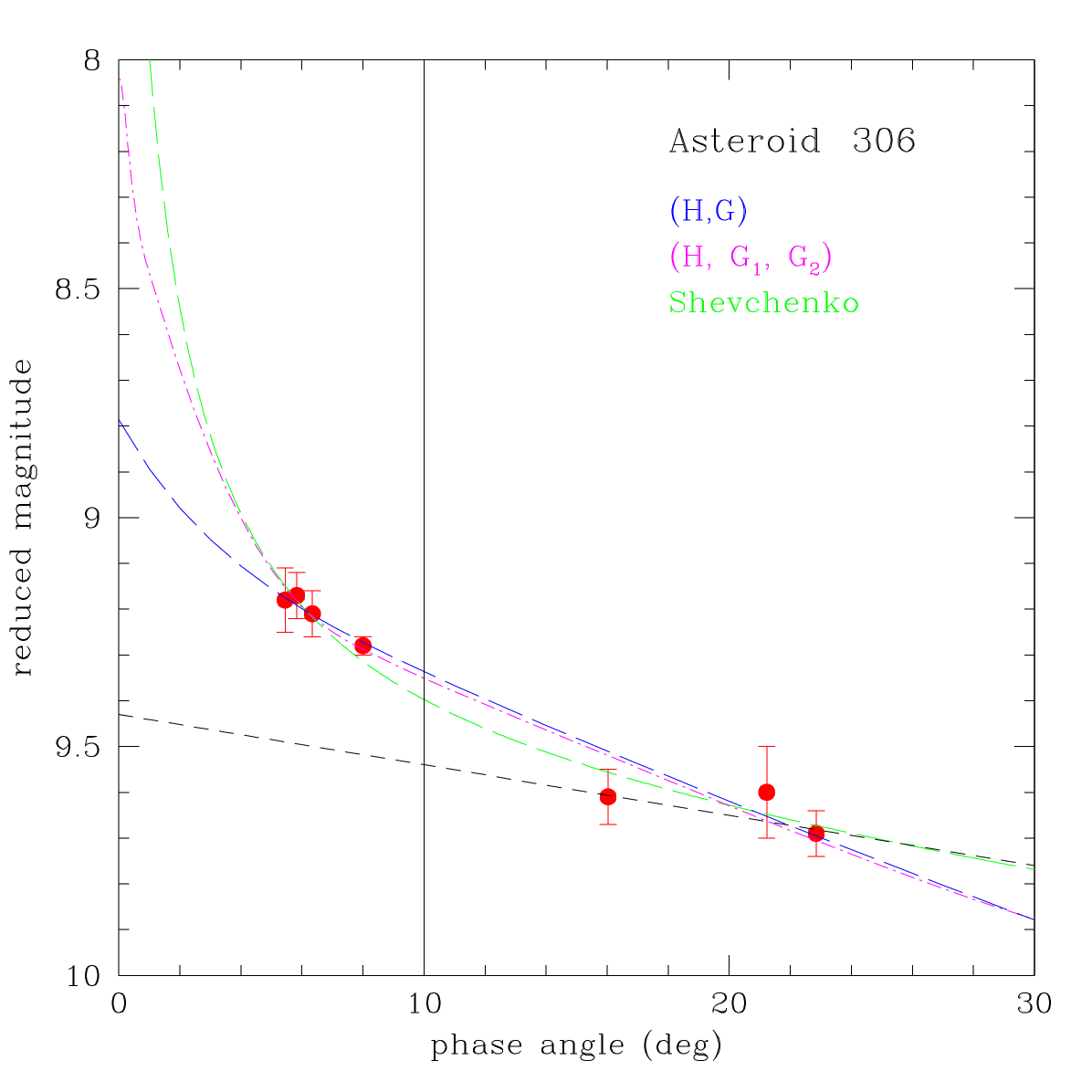}
\includegraphics[width=6.5cm]{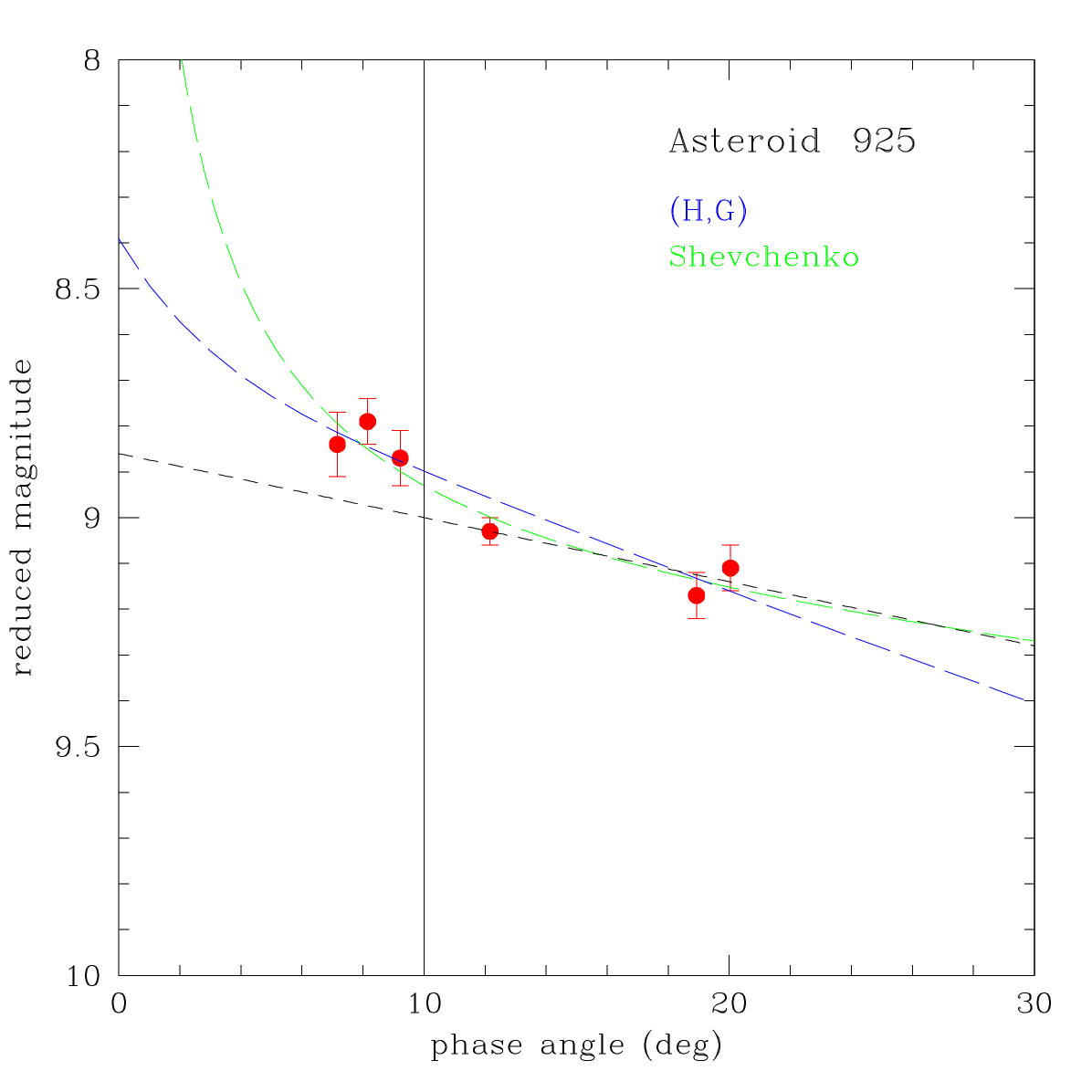}
\end{center}
\caption{Left Panel: Magnitude maximum brightness - phase curve in $V$ band for (306) Unitas, and resulting
  best-fit curves corresponding to the three photometric systems considered in this paper. 
  The region
  at small phase angles is poorly sampled, leading to big differences in the derived values  
  of $H$, and is shown separately. Right Panel: the same, but for asteroid (925) Alphonsina.
  For this object, no solution using the ($H$, $G_1$, $G_2$) photometric system could be 
  found. Not that for both objects, the linear fit of data at phase angles $> 10^\circ$ 
  are also shown.}
\label{phaseV306925}
\end{figure}

\begin{figure}[p]
\begin{center}
\includegraphics[width=6.5cm]{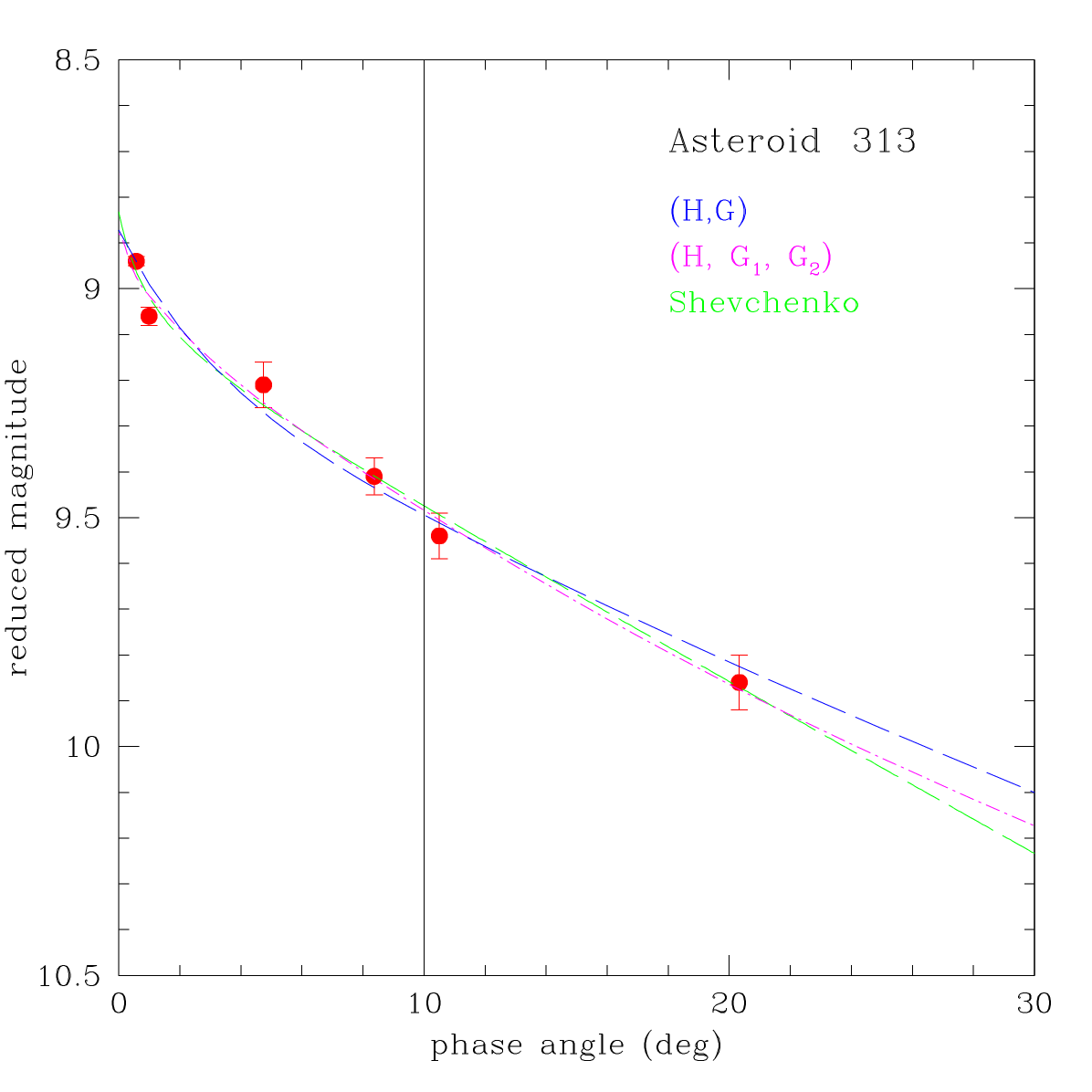}
\includegraphics[width=6.5cm]{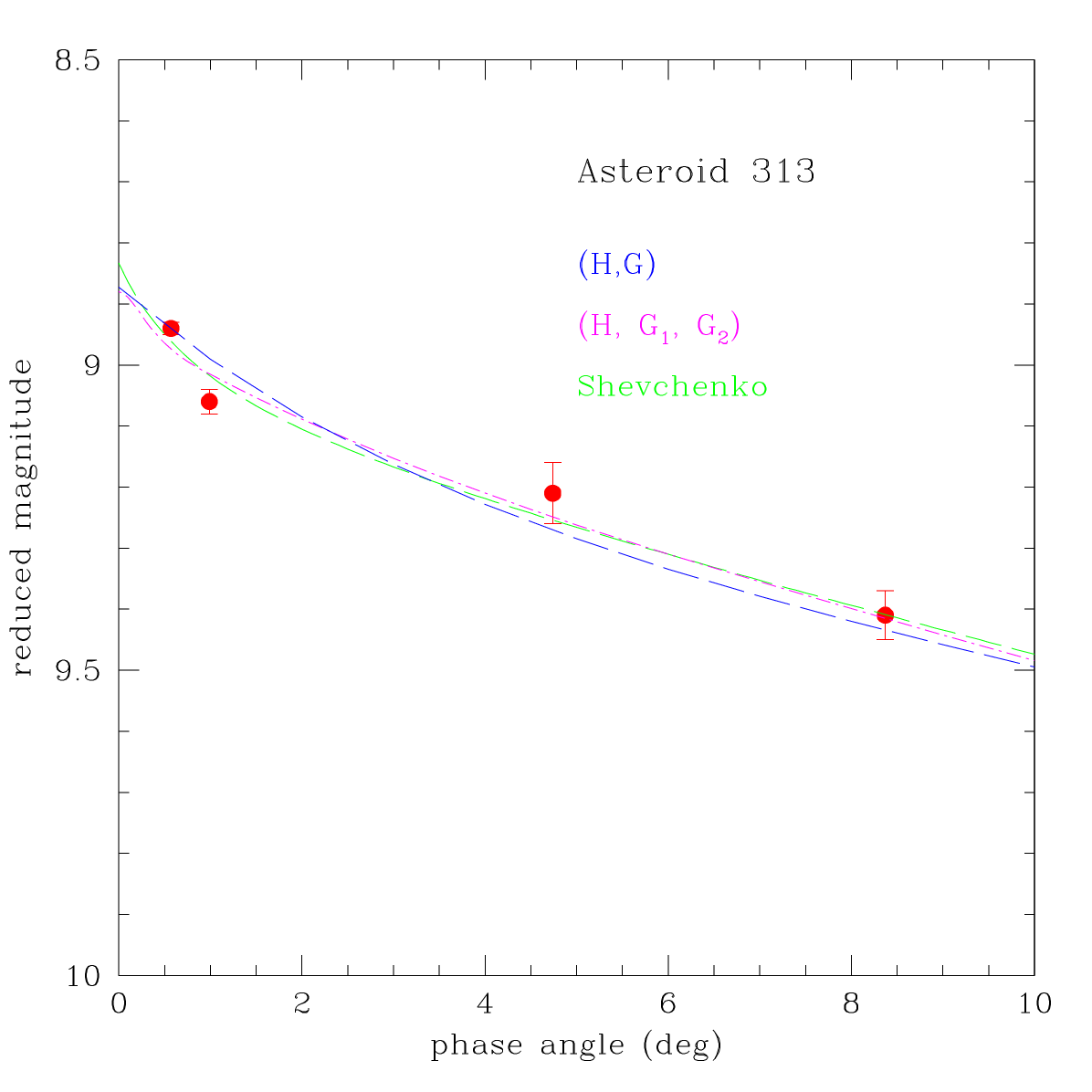}
\end{center}
\caption{The same as Fig. \ref{phaseV85}, but for asteroid (313) Chaldaea. Note that for this 
asteroid, only two measurements have been obtained at phase angles $> 10^\circ$, therefore
no linear-fit of these measurements has been computed.} 
\label{phaseV313}
\end{figure}

\begin{figure}[p]
\begin{center}
\includegraphics[width=6.5cm]{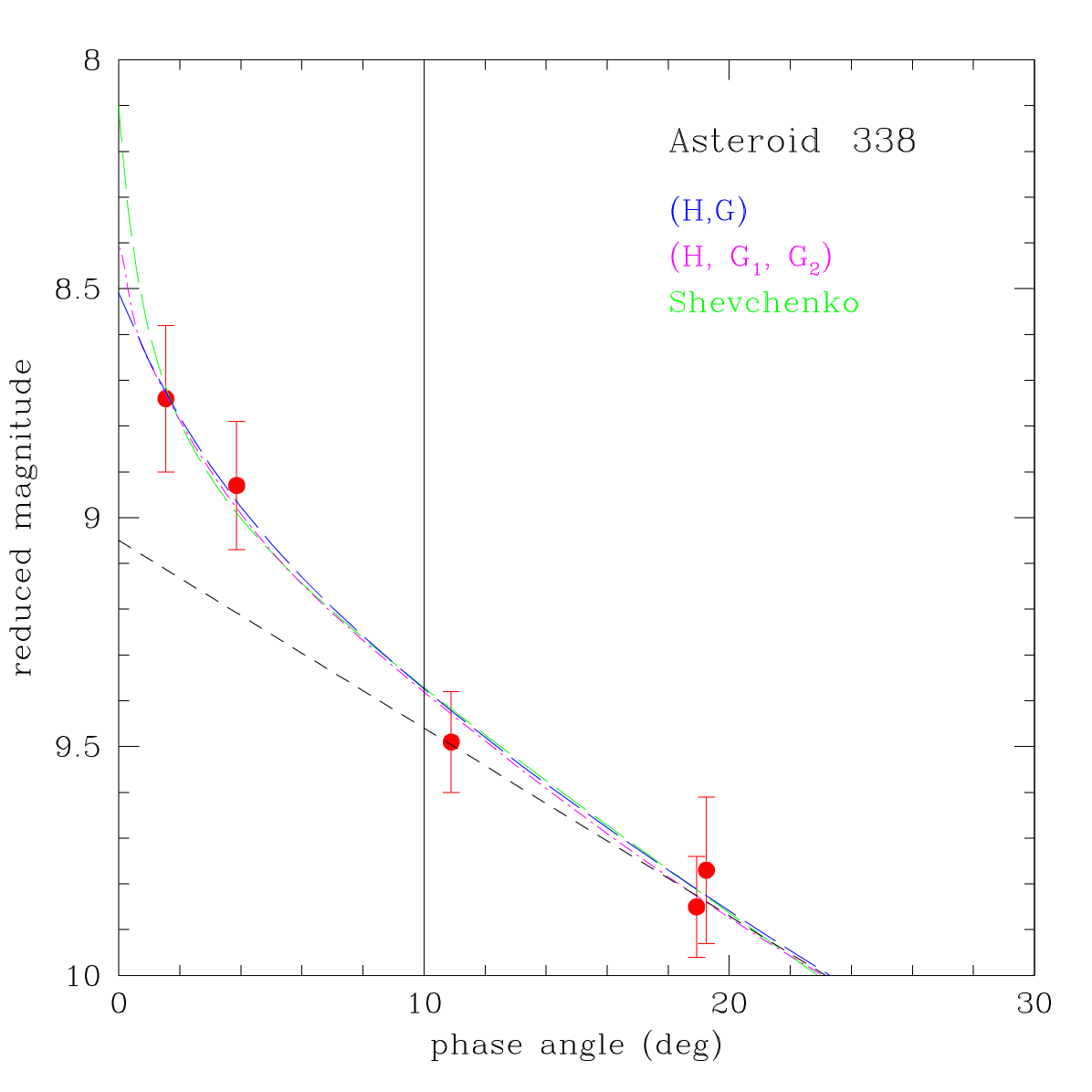}
\includegraphics[width=6.5cm]{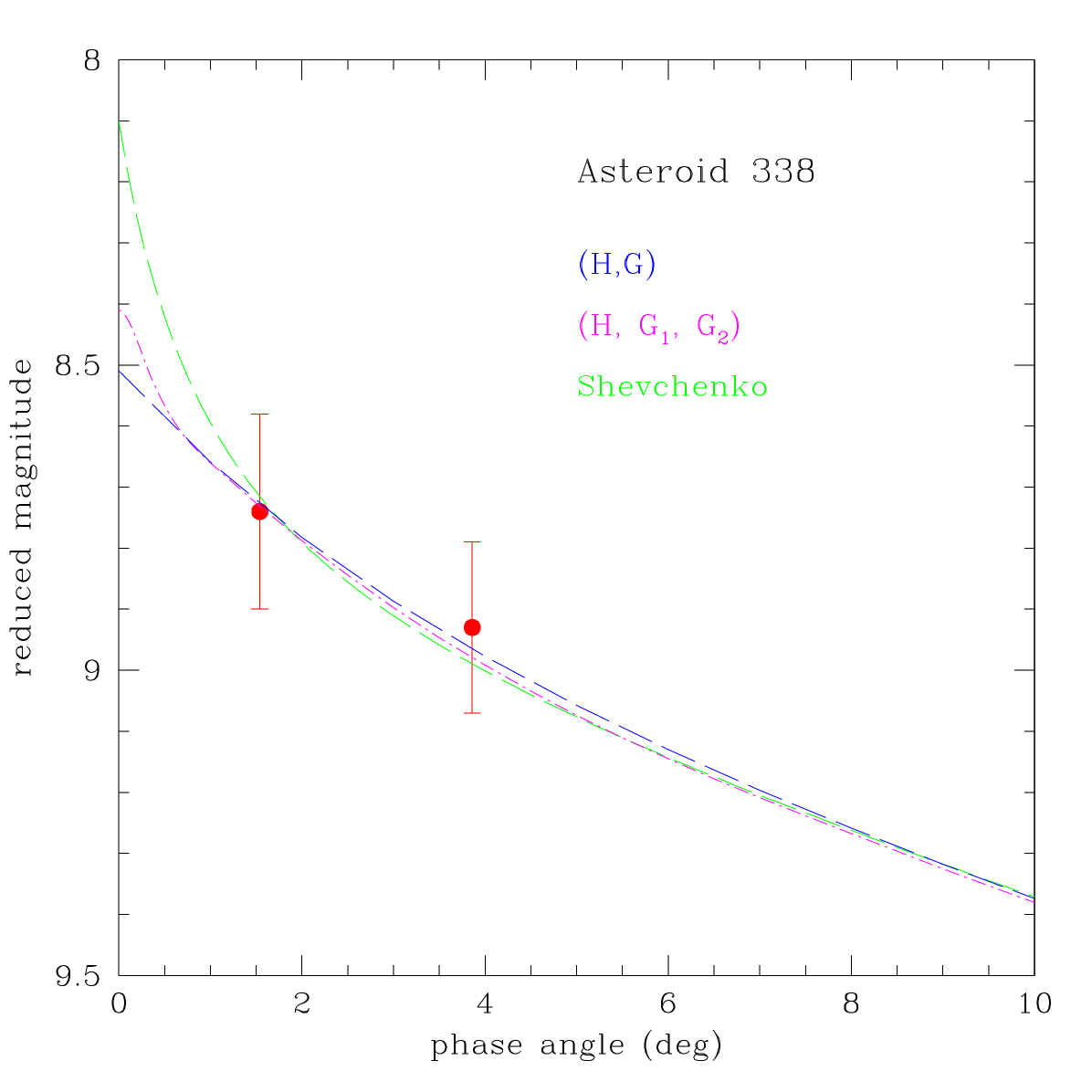}
\end{center}
\caption{The same as Fig. \ref{phaseV85}, but for asteroid (338) Budrosa.
  Only two observations cover the phase angle interval between $0^\circ$ and $10^\circ$.}
\label{phaseV338}
\end{figure}

\begin{figure}[p]
\begin{center}
\includegraphics[width=6.5cm]{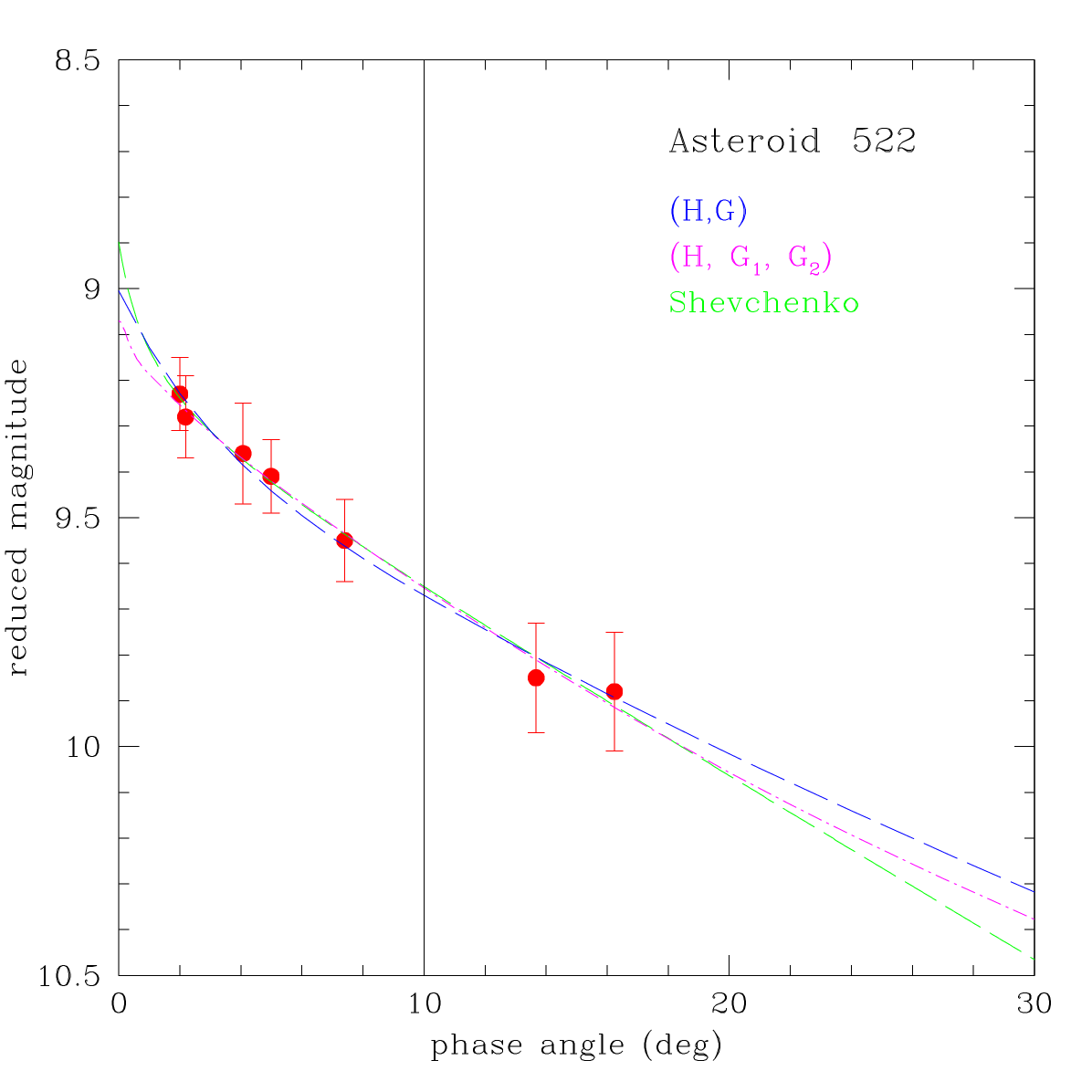}
\includegraphics[width=6.5cm]{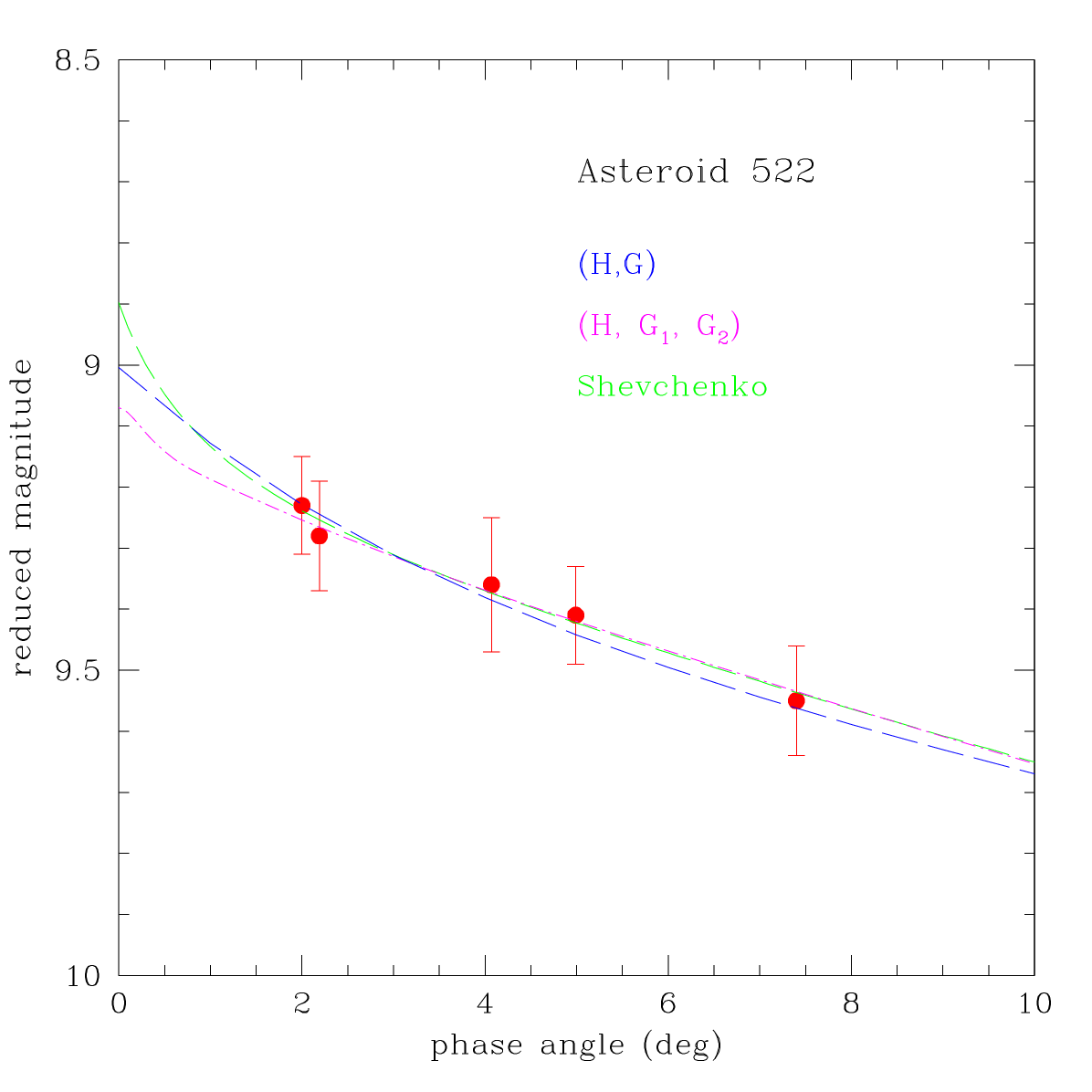}
\end{center}
\caption{The same as Fig. \ref{phaseV85}, but for asteroid (522) Helga. As in the case of 
(313) Chaldaea, only two measurements have been obtained at phase angles $> 10^\circ$, and 
no linear-fit of these measurements has been computed.}
\label{phaseV522}
\end{figure}

The results of our best-fit computations using the ($H, G$), ($H, G_1, G_2$)
and the Shevchenko systems are shown in Figs. \ref{phaseV85}--\ref{phaseV522}.
In all the figures, the best-fit obtained using the ($H, G$), 
($H$, $G_1$, $G_2$), and Shevchenko photometric systems are shown using different
colours. We show also the linear best-fits of the measurements obtained
at phase angles larger than $10^\circ$, for the objects having at least three measurements
in this interval. We remind that $10^\circ$ is approximately the lower limit in
phase angle attainable by Gaia observations of main belt asteroids. We are 
therefore interested in studying how a simple linear fit of data obtained at phase angles
$> 10^\circ$ can be used to derive some information about the main properties of the
phase - mag curves, and in particular how the slopes of these linear fits can be 
used to derive reasonable values for the geometric albedo.
%, as suggested by \citet{Shev, Shevchenkoetal2016}.
 
The obtained best-fit parameters corresponding to the different photometric
systems considered in this paper are also listed in Tables
\ref{resultsHG} - \ref{resultsbeta}. Note that asteroid (925) Alphonsina
is not included in Table \ref{resultsHG1G2}, because no ($H$, $G_1$, $G_2$)
fit could be found for this object.
The average rms residuals of the phase - mag data of each object in our sample, 
obtained using the three different photometric systems, are listed in 
Table \ref{resultsbeta}.

%----------------------------------------------------------------

\begin{table*}
  \caption{Results of the best-fit solutions of the phase-mag maximum brightness curves
    using the $(H, G)$ photometric system.}
\begin{center}
\begin{tabular}{rrccc}
\hline
object & N(obs) & $H$ & $G$ & rms (mag)\\
\hline
 85 & 7 & $7.52 \pm 0.10$ & $0.0753 \pm 0.0210$ & 0.022 \\
208 & 7 & $9.20 \pm 0.45$ & $0.2761 \pm 0.0853$ & 0.085 \\
236 & 8 & $8.00 \pm 0.14$ & $0.0715 \pm 0.0311$ & 0.053 \\
306 & 7 & $8.79 \pm 0.27$ & $0.2891 \pm 0.0452$ & 0.043 \\
313 & 6 & $8.87 \pm 0.28$ & $0.1954 \pm 0.0512$ & 0.043 \\
338 & 5 & $8.51 \pm 0.18$ & $-0.0812 \pm 0.0437$ & 0.046\\
522 & 7 & $9.00 \pm 0.16$ & $0.1411 \pm 8.0317$ & 0.027 \\
925 & 6 & $8.39 \pm 0.48$ & $0.3498 \pm 0.0830$ & 0.047 \\
average & &               &                     & 0.046\\
\hline
\end{tabular}
\end{center}
\label{resultsHG}
\end{table*}

\begin{table*}
  \caption{Results of the best-fit solutions of the phase-mag maximum brightness curves
    using the $(H, G_1, G_2)$ photometric system.}
\begin{center}
\begin{tabular}{rrcccc}
\hline
object & N(obs) & $H$ & $G_1$ & $G_2$ & rms (mag)\\
\hline
 85 & 7 & $7.41 \pm 0.08$ & $0.3358 \pm 0.0401$ & $0.2147 \pm 0.0298$ & 0.022 \\
208 & 7 & $8.92 \pm 0.07$ & $-0.3116 \pm 0.0675$ & $0.6598 \pm 0.0418$ & 0.053\\
236 & 8 & $7.82 \pm 0.06$ & $0.1155 \pm 0.0331$ & $0.3436 \pm 0.0223$ & 0.019\\
306 & 7 & $8.04 \pm 0.43$ & $-0.1309 \pm 0.4369$  & $0.3624 \pm 0.6269$ & 0.043\\
313 & 6 & $8.88 \pm 0.08$ & $0.6240 \pm 0.0386$ & $0.1510 \pm 0.0336$ & 0.026\\
338 & 5 & $8.41 \pm 0.10$ & $0.5607 \pm 0.0457$ & $-0.0037 \pm 0.0365$ & 0.047\\
522 & 7 & $9.07 \pm 0.12$ & $0.7302 \pm 0.0623$ & $0.0879 \pm 0.0450$ & 0.020 \\
average & &               &                     &                     & 0.033 \\
\hline
\end{tabular}
\end{center}
\label{resultsHG1G2}
\end{table*}

\begin{table*}
  \caption{Results of the best-fit solutions of the phase-mag maximum brightness curves
    using the Shevchenko photometric system.}
\begin{center}
\begin{tabular}{rrccccc}
\hline
object & N(obs) & $V(1,0)$ & $a$ & $b$ & $H = V(1,0)-a$ & rms (mag)\\
\hline
85 & 7 & $7.96 \pm 0.03$ & $0.70 \pm 0.07$ & $0.035 \pm 0.002$ &
$7.26 \pm 0.07$ & 0.009 \\
208 & 7 & $9.74 \pm 0.03$ & $0.95 \pm 0.06$ & $0.014 \pm 0.002$ &
$8.79 \pm 0.07 $ & 0.060\\
236 & 8 & $8.56 \pm 0.02$ & $0.97 \pm 0.05$ & $0.026 \pm 0.001$ &
$7.59 \pm 0.05$ & 0.030\\
306 & 7 & $9.60 \pm 0.09$ & $3.23 \pm 0.53$ & $0.009 \pm 0.003$ &
$6.37 \pm 0.54$ & 0.033 \\
313 & 6 & $9.13 \pm 0.02$ & $0.30 \pm 0.05$ & $0.037 \pm 0.002$ &
$8.83 \pm 0.05$ & 0.033\\
338 & 5 & $9.00 \pm 0.04$ & $0.90 \pm 0.11$ & $0.045 \pm 0.002$ &
$8.10 \pm 0.12$ & 0.053 \\
522 & 7 & $9.29 \pm 0.04$ & $0.39 \pm 0.12$ & $0.040 \pm 0.002$ &
$8.90 \pm 0.13$ & 0.025 \\
925 & 6 & $9.21 \pm 0.15$ & $3.74 \pm 1.00$ & $0.006 \pm 0.006$ &
$5.47 \pm 1.01$ & 0.042 \\
average &                 &                 &                    & & & 0.036 \\
\hline
\end{tabular}
\end{center}
\label{resultsShev}
\end{table*}

\begin{table*}
  \caption{Results of the best-fit solutions of the phase-mag maximum brightness curves
    using a simple linear relation for observations at phase angle $> 10^\circ$.
    Objects having a number N(obs) $< 3$ of measurements at phase angle $> 10^\circ$
    were not included in the analysis.}
\begin{center}
\begin{tabular}{rrcc}
\hline
object & N(obs) & $\beta$ & rms (mag)\\
\hline
 85 & 3 & $0.036 \pm 0.001$ & 0.003 \\
208 & 3 & $0.037 \pm 0.007$ & 0.022 \\
236 & 3 & $0.028 \pm 0.002$ & 0.014 \\
306 & 3 & $0.011 \pm 0.012$ & 0.036 \\
338 & 3 & $0.041 \pm 0.010$ & 0.040 \\
925 & 3 & $0.014 \pm 0.008$ & 0.031 \\
average &  &                & 0.024 \\
\hline
\end{tabular}
\end{center}
\label{resultsbeta}
\end{table*}

\section{Discussion}

The results shown in the previous Section indicate that the
old (H, G) system gives the worst residuals when it is adopted to fit
our limited sample of phase - mag data. This is not unexpected, because
both the ($H$, $G_1$, $G_2$) and the Shevchenko photometric systems
have been developed to replace (H, G), and provide better fits
of existing data, and in particular more accurate estimates of the
absolute magnitudes. 

In terms of average residuals, listed in Table \ref{resultsbeta},
at face values the best-fits obtained using ($H$, $G_1$, $G_2$)
are slightly better than in the case of using the Shevchenko photometric system,
but the differences are really small. By looking at Figs. \ref{phaseV85}--\ref{phaseV522},
however, one can see that in terms of the resulting value of the absolute magnitude $H$,
the differences can be non-negligible, attaining typical 
values of 0.2 mag in most cases. In a couple of cases, however, we obtained
an ($H$, $G_1$, $G_2$) best-fit solution with a negative value
for either $G_1$ or $G_2$. We cannot rule out the possibility that in these cases 
the obtained phase - mag curves could include some abnormal magnitude value,
obliging the flexible ($H$, $G_1$, $G_2$) system to obtain a best-fit
representation including anomalous values for some of its parameters.
Even in the other cases, however, the resulting
differences in $H$ are not negligible. By looking at the plots shown in Figs. \ref{phaseV85}--\ref{phaseV522} it turns out 
that the $H$ value found by using the Shevchenko photometric system is
systematically brighter than the value corresponding to an ($H$, $G_1$, $G_2$)
(and, even more, the (H, G)) 
solution. At least in some cases, a purely visual inspection of the data would
suggest that the ($H$, $G_1$, $G_2$) best-fit solution looks slightly more
credible, but this is not, of course, an acceptable criterion. At face value,
the average rms values indicate that the Shevchenko-based fits are sligthly
worse, but the differences do not appear to be sufficient to conclude
that the $H$ values given by the ($H$, $G_1$, $G_2$) system are more realistic.
 
We also note that, if we make a comparison between the values of the linear coefficient 
$\beta$ of a purely linear best-fit of data, taking into account only lightcurves obtained 
at phase angles $> 10^\circ$ (shown in Table \ref{resultsbeta}) and the corresponding
value of the $b$ parameter in the Shevchenko system (listed in Table \ref{resultsShev}),
we can see that in most cases, including asteroids (85), (236), (338), and (925), 
there is a good agreement. In only one case, that of (208), 
the differences are significant. In this case, however, it seems that the difference
between $b$ and $\beta$ would tend to disappear if the point of the phase - mag
curve at $3.75^\circ$, which is also responsible of the negative value derived for the
$G_1$ parameter, would be removed.

These results, though very preliminary, tend to 
suggest that using a simple linear fit of data obtained at large phase angles, as in 
the case of Gaia asteroid data, could lead
in many cases to linear coefficient values that would be in reasonable agreement 
with the linear part of the phase - mag curve derived by the more
refined Shevchenko photometric system. The latter takes into account the
existence of a non-linear brightness surge at small phase angles, and it would not
be obvious {\em a priori} that the corresponding values of the linear part of the 
Shevchenko system must be necessarily found to be in good agreement with a simple 
linear fit of data not including magnitude values affected by the 
opposition effect. This can be
an important result from the point of view of the future treatment of Gaia data,
in the case a relation between geometric albedo and the slope of a linear 
variation of the phase - mag curves will be used to determine estimates of the albedo 
from Gaia data. The point is that the relation suggested by \citet{Shev}
was based on the use of the $b$ parameter of the Shevchenko photometric system, and 
not on a simple $\beta$ slope obtained from an analysis based on large-phase data, only. 

\begin{table*}
  \caption{Results of the linear best-fit solutions for the relations 
  $y = m\, \log_{10}(p_V) +q$
  where $p_V$ is the geometric albedo and $y$ represents the photometric parameters
  indicated in column 1. $N_{curves}$ is the number of phase - mag curves used
  in computing the linear best-fit.}
\begin{center}
\begin{tabular}{rrccr}
\hline
\multicolumn{1}{c}{y} & \multicolumn{1}{c}{$N_{curves}$} & \multicolumn{1}{c}{$m$} & \multicolumn{1}{c}{$q$} 
& Correlation \\
\hline
$G_1$   & 36  & $-0.6855 \pm 0.1019$ &  $-0.1383 \pm 0.0907$ &  0.756\\
$G_2$   & 36  & $ 0.5445 \pm 0.0571$ &  $ 0.7299 \pm 0.0508$ &  0.853\\
$b$     & 23  & $-0.0241 \pm 0.0041$ &  $ 0.0135 \pm 0.0035$ &  0.790\\
$\beta$ &  7  & $-0.0152 \pm 0.0060$ &  $ 0.0196 \pm 0.0057$ &  0.748\\
\hline
\end{tabular}
\end{center}
\label{computedfits}
\end{table*}

\begin{figure}[p]
\includegraphics[width=12cm]{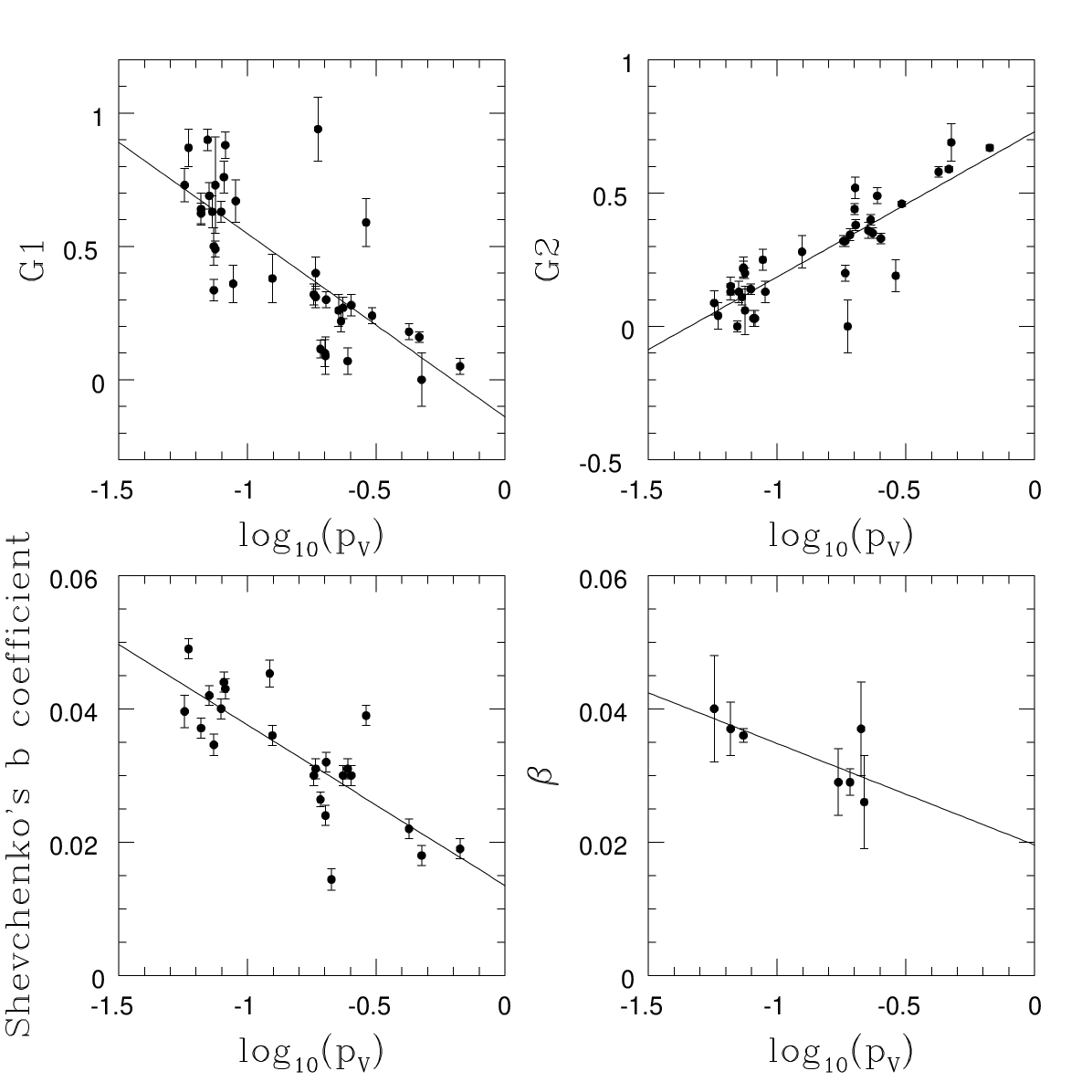}
\caption{results of the best-fit computations for the relations of the kind
$y = m\, \log_{10}(p_V) +q$
  where $p_V$ is the geometric albedo and $y$ represents the photometric parameters
  $G_1$ (top.left), $G_2$ (top-right), $b$ (bottom-left) and $\beta$ (bottom-right),
  respectively. See the text for the meaning of the above parameters.}
\label{Mosaicfits}
\end{figure}

In this respect, following \citet{Shevchenkoetal2016}, we explored some possible relations 
between photometric parameters and geometric albedo, taking profit of the data published 
by the above authors. In particular, we considered relation of the type:
\begin{equation}
y = m\, \log_{10}(p_V) +q
\end{equation}
where $p_V$ is the geometric albedo, and $y$ is one among $G_1$, $G_2$, 
$b$, and $\beta$. With respect to the analysis performed by \citet{Shevchenkoetal2016},
a difference is that, in our analysis we do not use albedo values coming from 
WISE thermal radiometry data, because we think that these data can be affected by
too big uncertainties to be used for calibration purposes, for the reasons explained by 
\citet{MNRAS1}. Instead, in the present analysis we use albedo values computed either
by exploiting the proposed $\Psi$ - albedo relation, where $\Psi$ is the polarimetric
parameter proposed by \citet{MNRAS1}, and computed for a sample of asteroids for which
we have estimates of the $G_1$, $G_2$ and $b$ photometric parameters, either given
by \citet{Shevchenkoetal2016}, or found in the present paper. We adopt values of
$\Psi$ based on the results of \citet{MNRAS2}, whenever possible updated using 
still unpublished polarimetric values obtained by the Calern Asteroid Polarimetric
Survey (CAPS) \citep{CAPS}. 
In some cases in which 
the polarimetric $\Psi$ parameter is not known, we use the albedo values published
by \citet{she2}.

\begin{figure}[p]
\includegraphics[width=12cm]{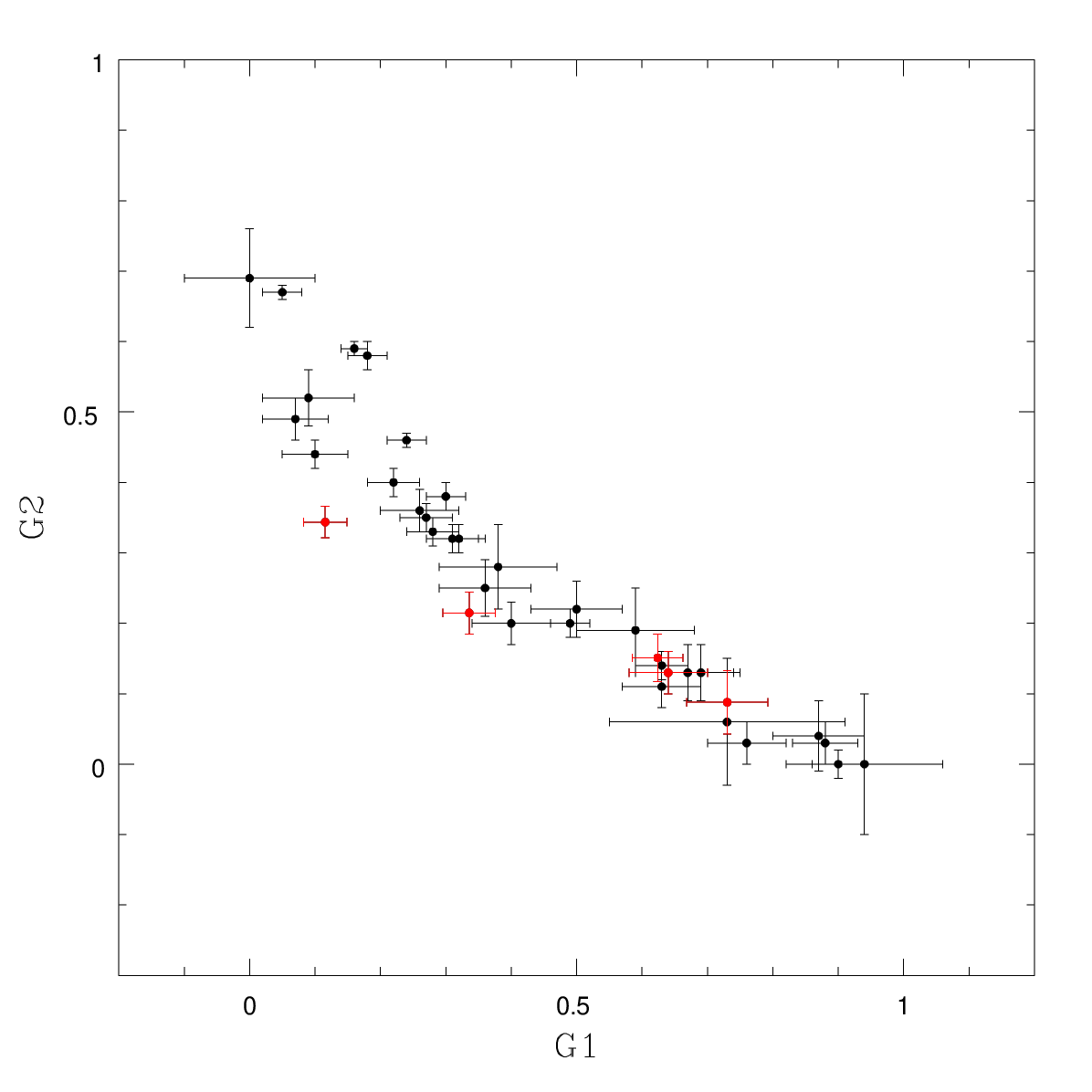}
\caption{Plot of the relation between the 
  $G_1$ and $G_2$ best-fit parameters. The points in red indicate four 
  objects for which the values of the parameters could be derived from the
  observations presented in this paper (asteroids (85), (236),(313) and (522)).
  Note that, in the case of (313), we have two independent estimates, one having
  been computed by \citet{Shevchenkoetal2016}. The two points, having both $G_1$ equal to
  0.62 and 0.64, and $G_2$ equal to 0.15 and 0.13, respectively, in the two cases, are 
  in good agreement.}
\label{Fig:G1G2}
\end{figure}
In Table \ref{computedfits} the results of this exercise are shown. Note that, in the 
case of the $\beta$ parameter, that has not been used in previous investigations, 
we only use the few data presented in the present analysis.
Figure \ref{Mosaicfits} shows the same results in graphical form.
The results show that we find the highest correlation when considering the relation
between albedo and $G_2$. A reasonable correlation between albedo and $G_1$ is also found and 
this is not unexpected, because \citet{Shevchenkoetal2016} found a strict correlation between
the $G_1$ and $G_2$ photometric parameters, a correlation that we confirm, and show in 
Fig.~ \ref{Fig:G1G2}.

A correlation between albedo and $G_2$ is interesting {\em per se}, but it is 
something that cannot be of practical application in future analyzes of Gaia
photometric data, because Gaia, alone, cannot provide values for $G_2$. On the other
hand, it is interesting to note that, in terms of correlation, the second best 
relation we find is the one between the albedo and the $b$ parameter of the
Shevchenko photometric system. Very interesting is then the fact that, although using 
a much smaller and certainly still insufficient data-base, we find a very similar
correlation also for the relation between albedo and $\beta$, the latter being a
photometric parameter that will be derived by Gaia data. The problem here seems to
be that the correlation between albedo and $b$, or $\beta$, is not very high, little less
than $0.8$. Taken at face value, this means that the albedo values derived by 
the adopted photometric parameters have 
large uncertainties, of the order of about 30\%. We note, however, that in the case of the
results obtained using $\beta$, the situation could be better than
it appears. In particular, the obtained correlation value listed in Table \ref{computedfits}
in the case of $\beta$ is influenced by two facts that might be largely improved in principle: 
(1) the high uncertainties affecting some of the $\beta$ determinations, and (2) the fact that the
range of albedos covered by the few asteroids in our sample is significantly 
narrower than in the case of the other photometric parameters listed in Table \ref{computedfits}
and shown in Figure \ref{Mosaicfits}. In the case of $G_1$, $G_2$ and $b$ a large fraction
of data points are not well represented by the best-fit line, according to nominal
errors, whereas this does not happen in the case of the still extremely limited
number of cases in which we use $\beta$. We do not rule out, therefore, the 
possibility that the determination of the $\beta$ parameter obtained by Gaia data could be easier 
to obtain and more accurate than in the few cases considered in our preliminary analysis. 
If this will be confirmed by
future investigations, it might be possible that the value of albedo 
obtained from knowledge of $\beta$ will be affected by much lesser uncertainty than what we
have found here. Even if the accuracy in albedo determination
will not be found to improve, the current error bars can be sufficient 
to allow us at least to distinguish between
quite bright and quite dark objects, and this, coupled with spectroscopic data that
will be also obtained by Gaia, will be in any case useful for taxonomic classification, in particular
to distinguish between objects (the old $E$, $M$, $P$ classes defined in the 80's)
which cannot be separated on the basis of visible reflectance spectra alone, but 
require some information about the geometric albedo.

\section{Conclusions}
\label{6}
In this paper we add just a few phase - mag data to the data-base already available in
the literature.
Though not quantitavely important, our analysis has some elements of interest, because 
it explores the relation between geometric albedo and parameters characterizing
phase - mag curves, using albedo values derived from polarimetry, instead of 
(likely more uncertain, see \citet{MNRAS1}) values derived by WISE thermal radiometry
measurements.

We also focused on the possibility to determine values for the $\beta$ coefficient
of the linear part of
the phase - mag relation using only data obtained at phase angles corresponding 
to those that characterize the Gaia observations of main belt asteroids. We find
some preliminary evidence that the obtained $\beta$ values can be in many cases in
good agreement with the values for the analogous $b$ parameter in the Shevchenko 
photometric system, that is derived using also data obtained at 
small values of the phase angle. This means that, hopedly, Gaia photometric data 
can be used
to determine at least rough estimates of the geometric albedo of the asteroids, through 
the relation between the slope of the linear part of the phase - mag curves and the albedo 
itself.

On the average, each asteroid will be observed a number of times of of the order of 70 
by Gaia during the nominal lifetime of the mission (that will be probably extended). Each transit
on the Gaia focal plane will correspond to non-repeatable observation circumstances.
Since the aspect angle of the objects will be very different at different epochs of
transit on the Gaia FOV, this means that we cannot simply use all the data together to
derive a unique phase - mag curve. On the other hand, it will be possible to select 
Gaia data obtained at similar aspect angles, in particular around equatorial view,
because the analysis of the Gaia data will allow us to compute a solution for the
spin properties of each object, including rotation period and pole coordinates
\citep{Santana}. In this
way, we hope that it will be possible to select subsets of Gaia photometric 
observations of asteroids to be used to compute phase - mag curves from which
some useful estimates of the albedo will be derived. The present investigation
is a first, preliminary step forward to investigate this possibility. \\
 
\textbf{Acknowledgements}\\
The Astronomical Observatory of the Autonomous Region of the Aosta Valley (OAVdA) is managed by
the Fondazione Cl\'ement Fillietroz-ONLUS, which is supported by the Regional Government of the
Aosta Valley, the Town Municipality of Nus and the ``Unit\'e des Communes vald\^otaines 
Mont-\'Emilius''. The research was partially funded by a 2017 ``Research and Education'' grant from Fondazione CRT.\\

\appendix

\section{Fourier models}
\label{7}
In this appendix we shown some of the Fourier reference lightcurves, superimposed on individual observed sessions, used to compute the maximum brightness magnitude of the asteroids in case it was not directly observed \citep{Harrisetal89b}.\\

\begin{figure}[p]
\begin{center}
\includegraphics[width=6.5cm]{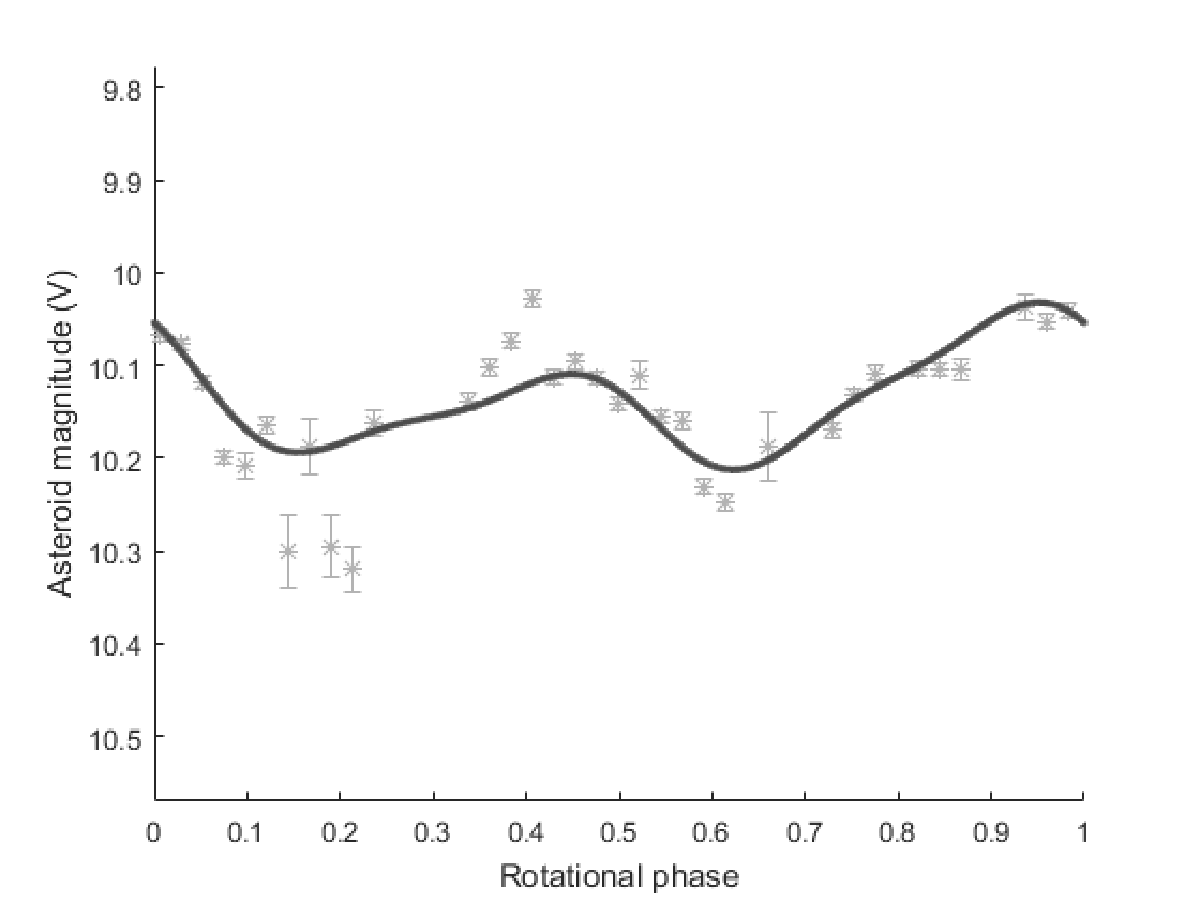}
\includegraphics[width=6.5cm]{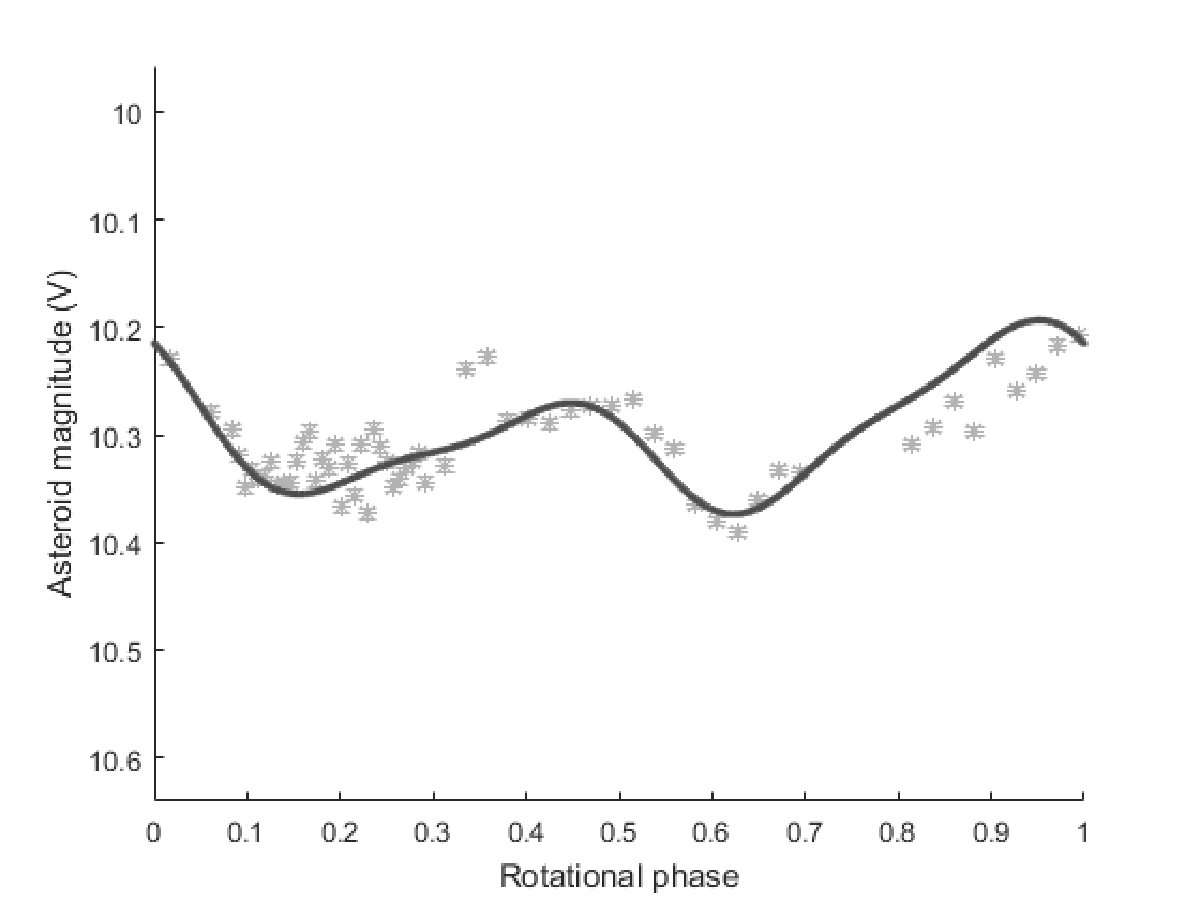}
\includegraphics[width=6.5cm]{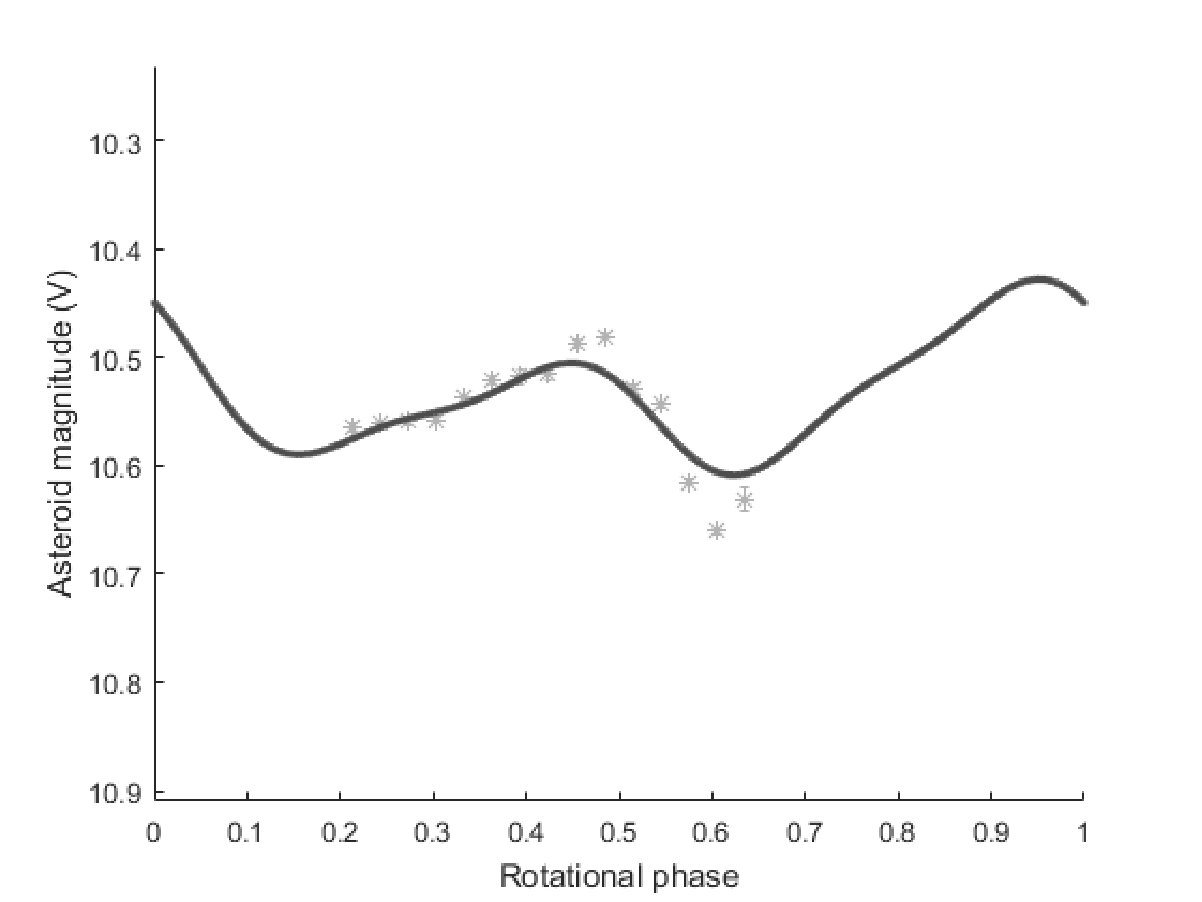}
\includegraphics[width=6.5cm]{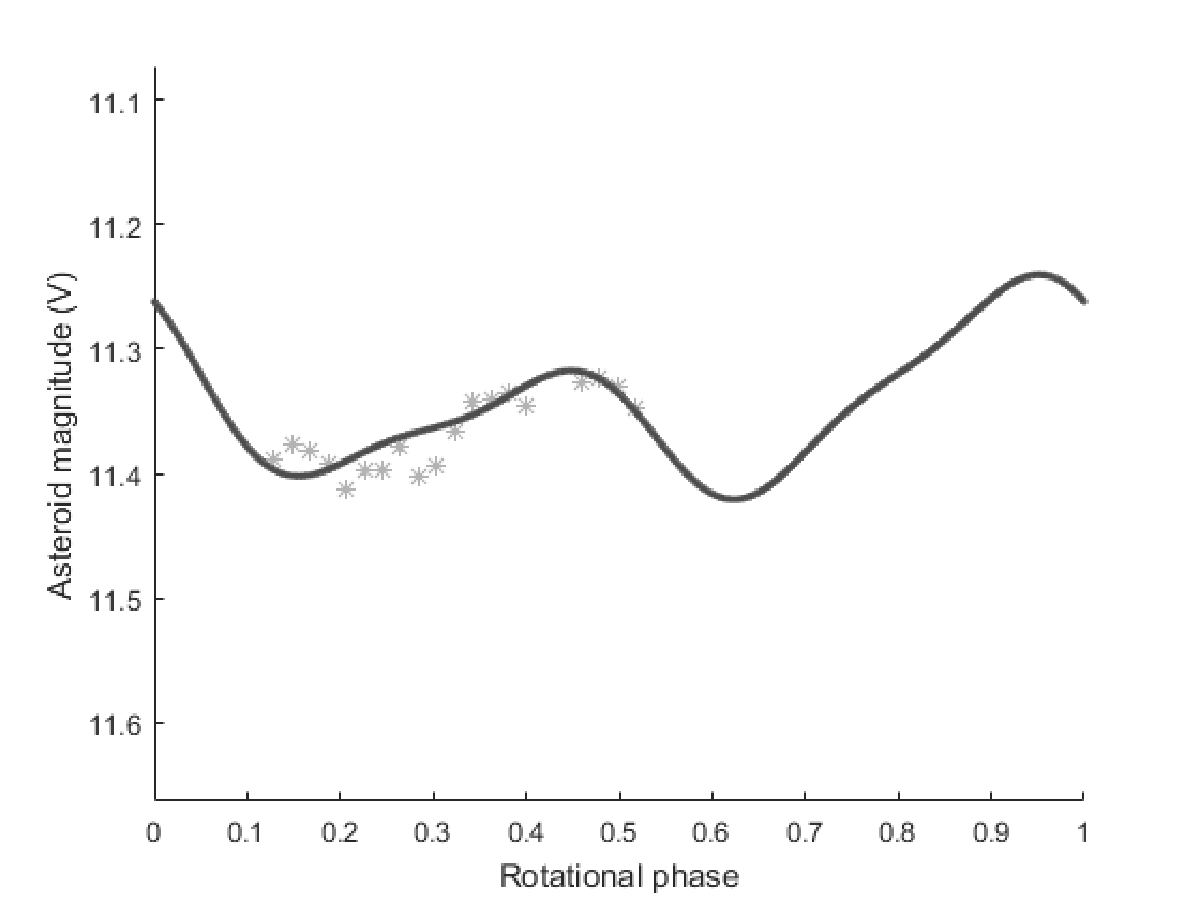}
\includegraphics[width=6.5cm]{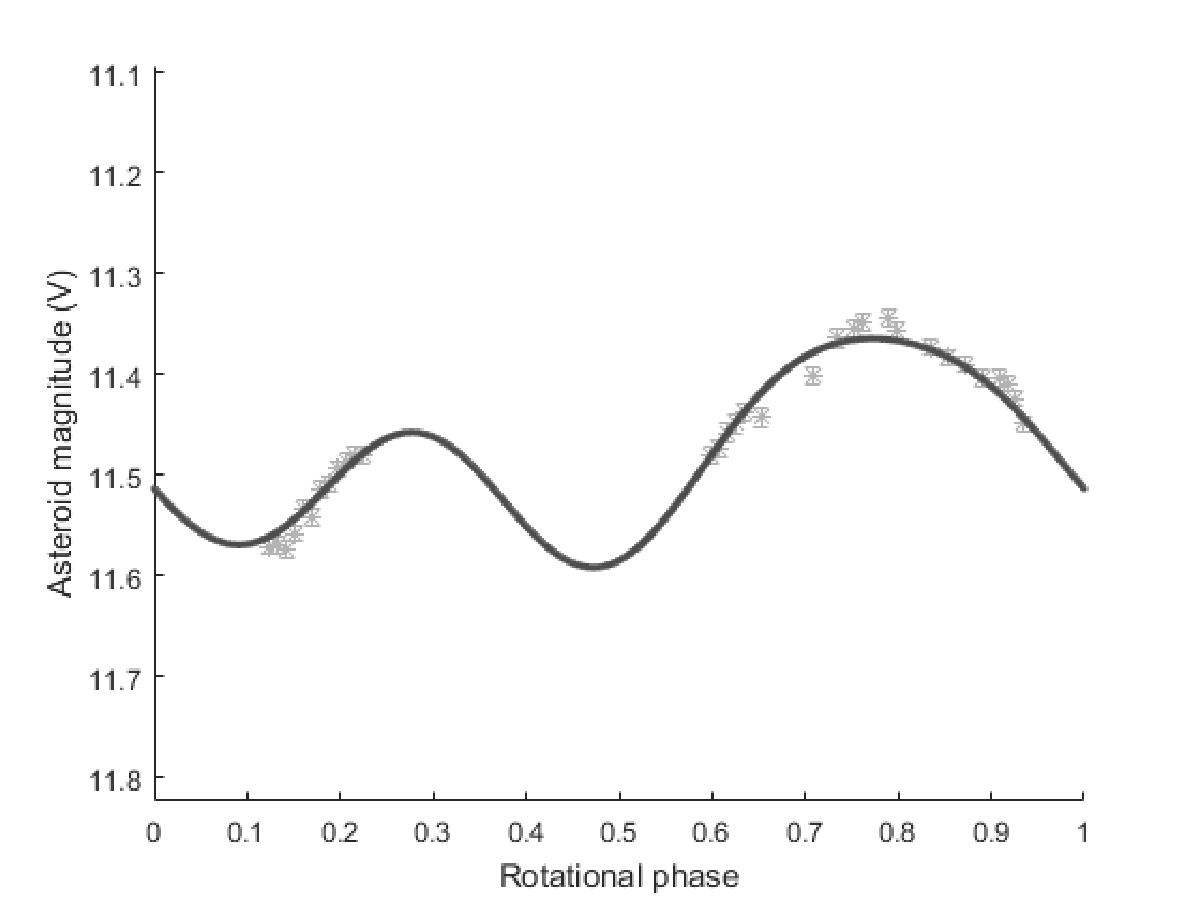}
\includegraphics[width=6.5cm]{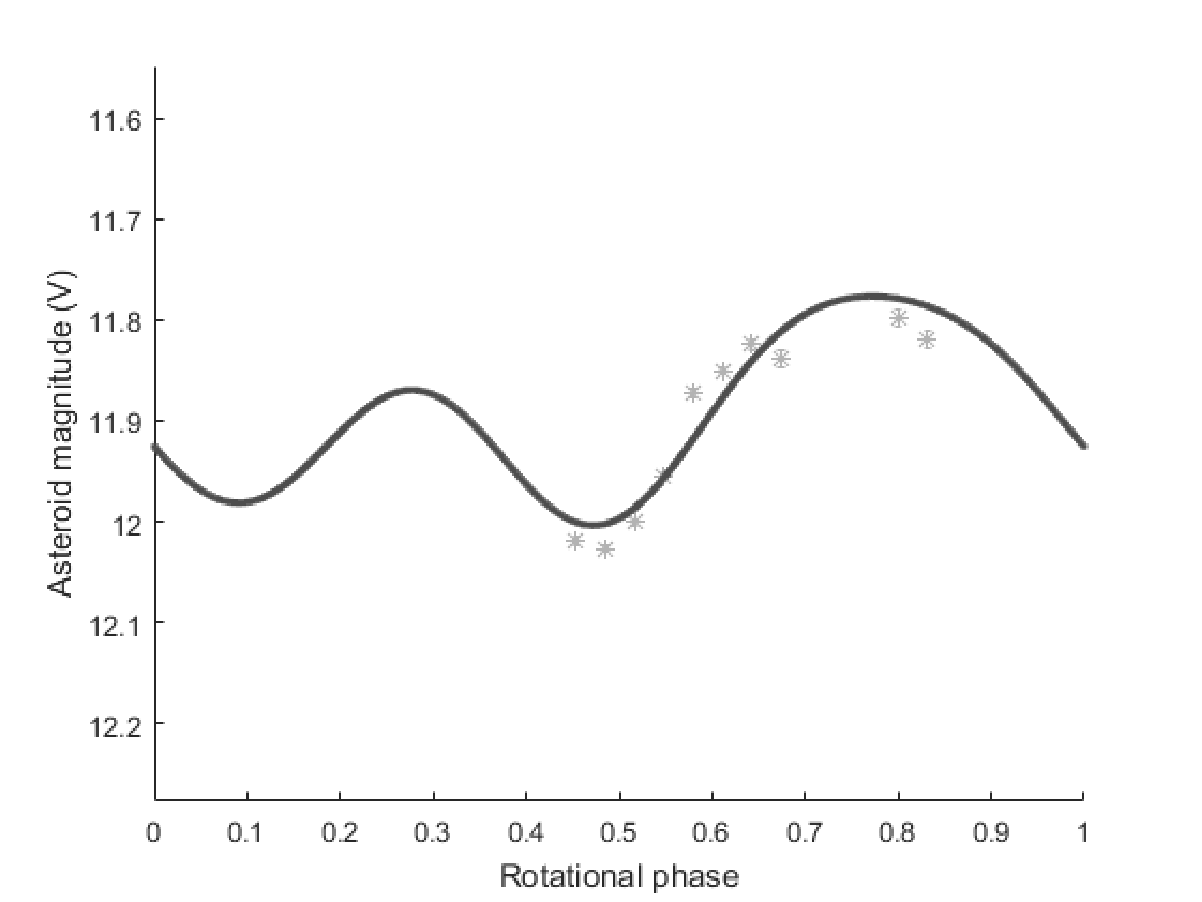}
\end{center}
\caption{The two Fourier fit models for 85 Io superimposed on the observed data sessions (yyymmdd): 20121013, 20121015, 20121021, 20121116, 20121120, 20121206, from left to right, top to bottom.} 
\label{FourierFit85}
\end{figure}

\begin{figure}[p]
\begin{center}
\includegraphics[width=6.5cm]{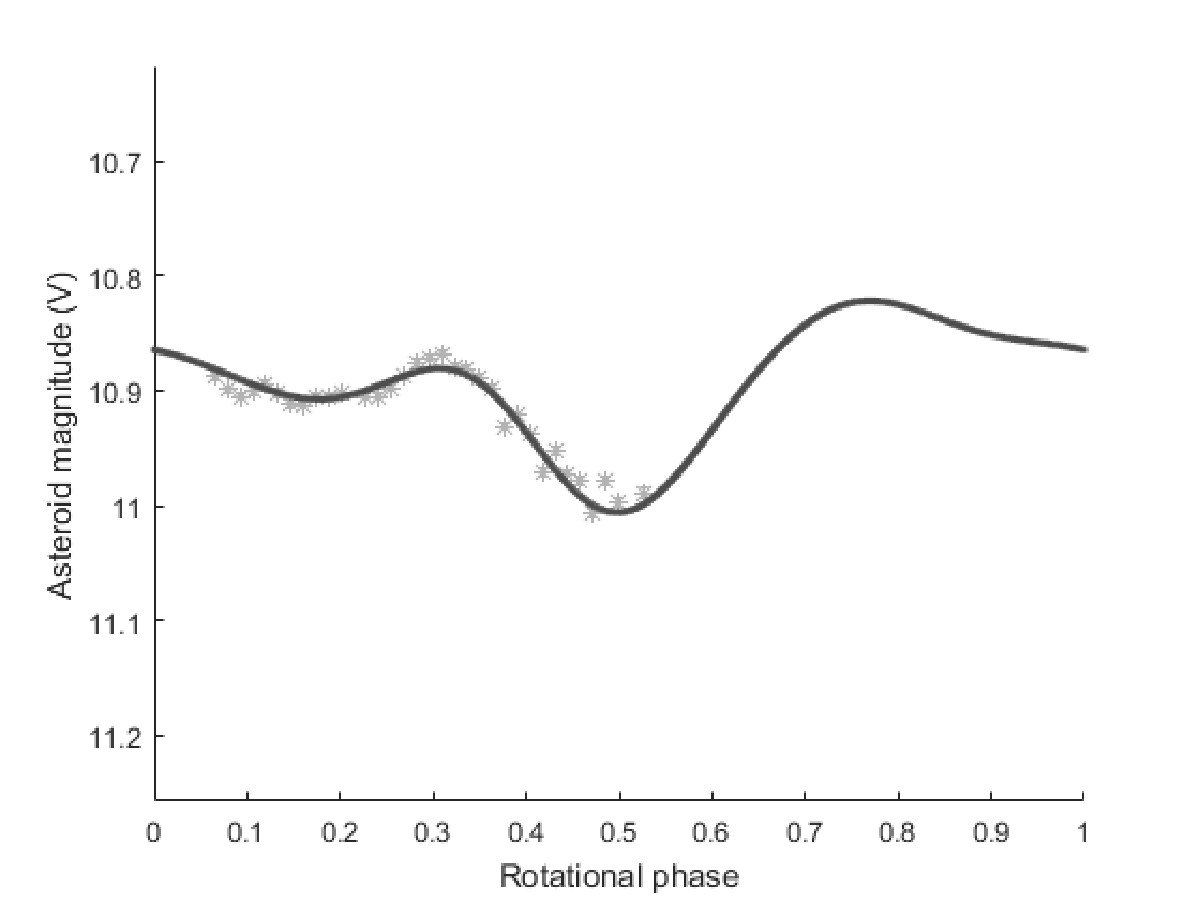}
\includegraphics[width=6.5cm]{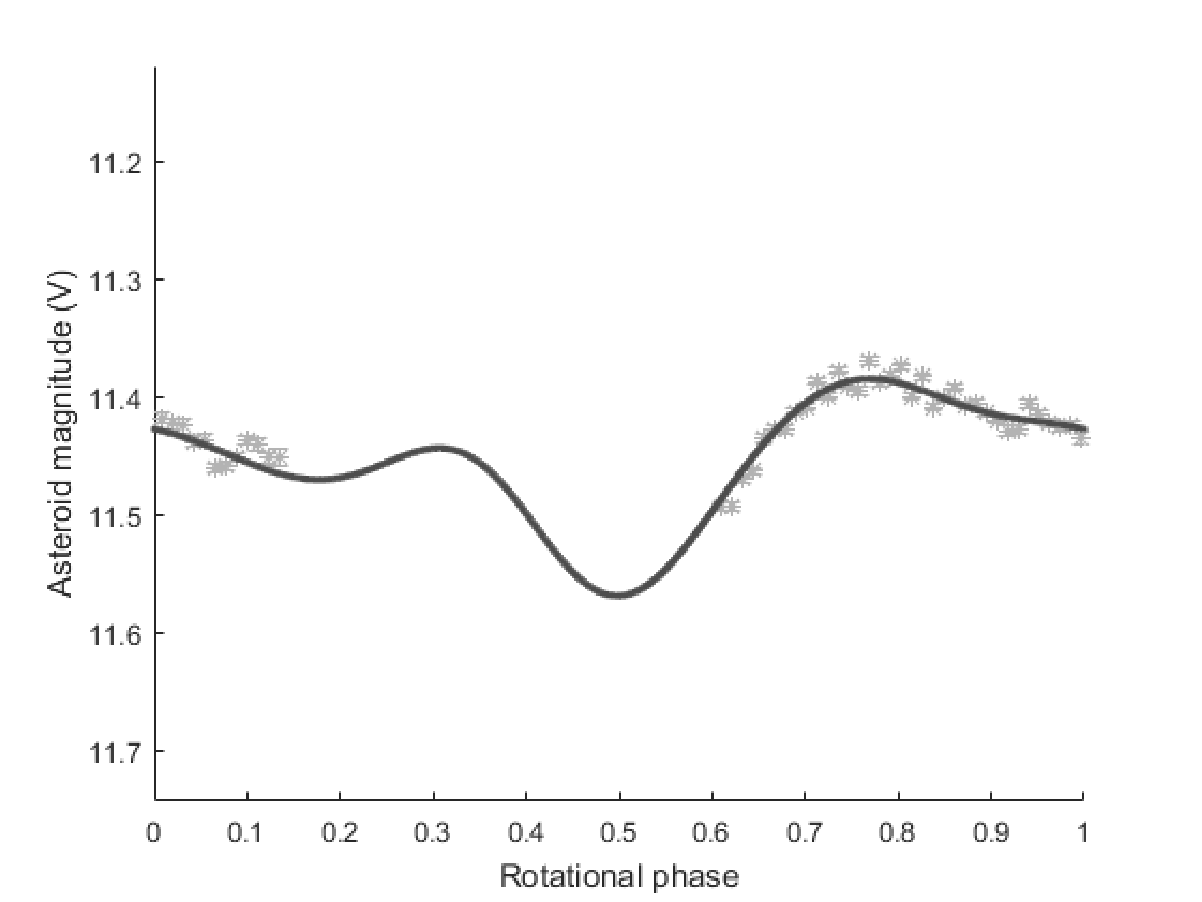}
\includegraphics[width=6.5cm]{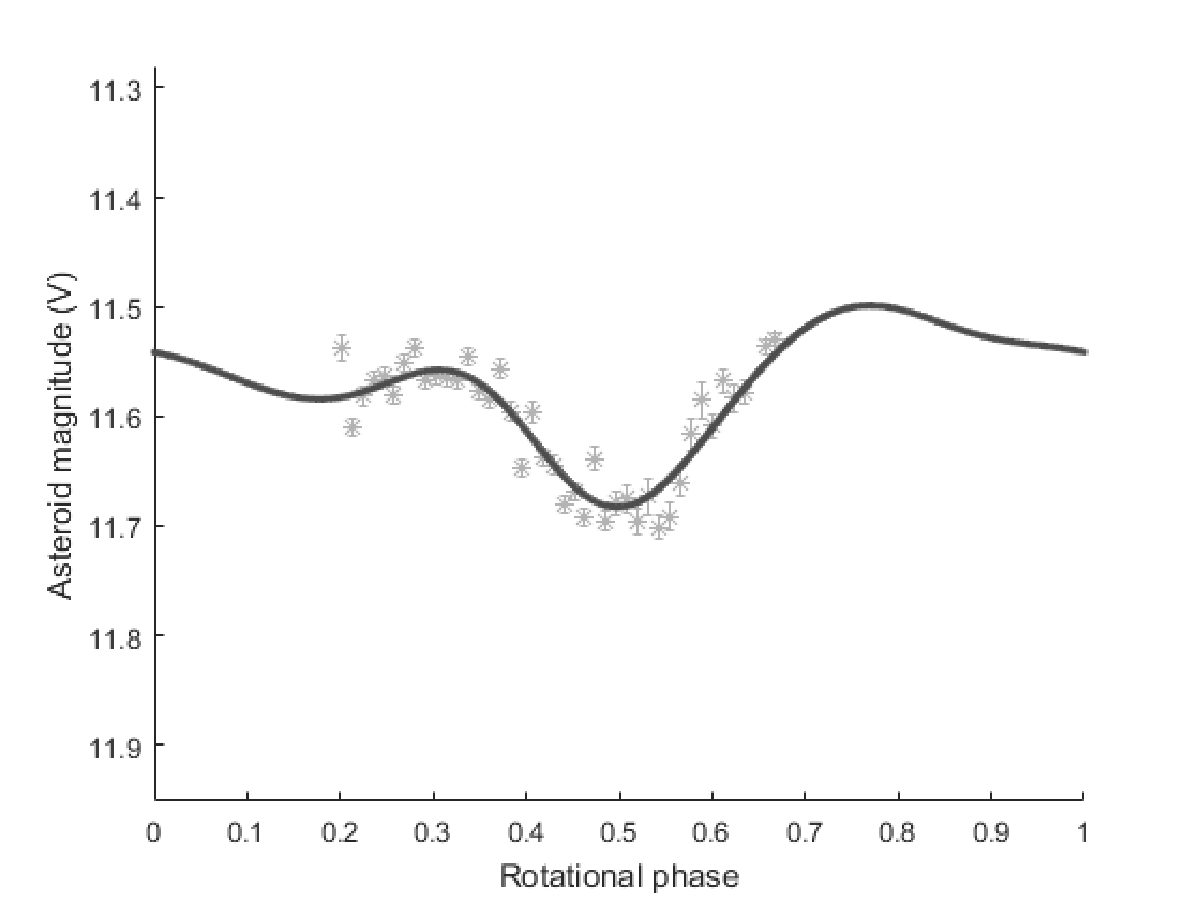}
\end{center}
\caption{The Fourier fit model for 135 Herta superimposed on the observed data sessions (yyymmdd): 20121211, 20121219, 20121223, from left to right, top to bottom.} 
\label{FourierFit135}
\end{figure}

\begin{figure}[p]
\begin{center}
\includegraphics[width=6.5cm]{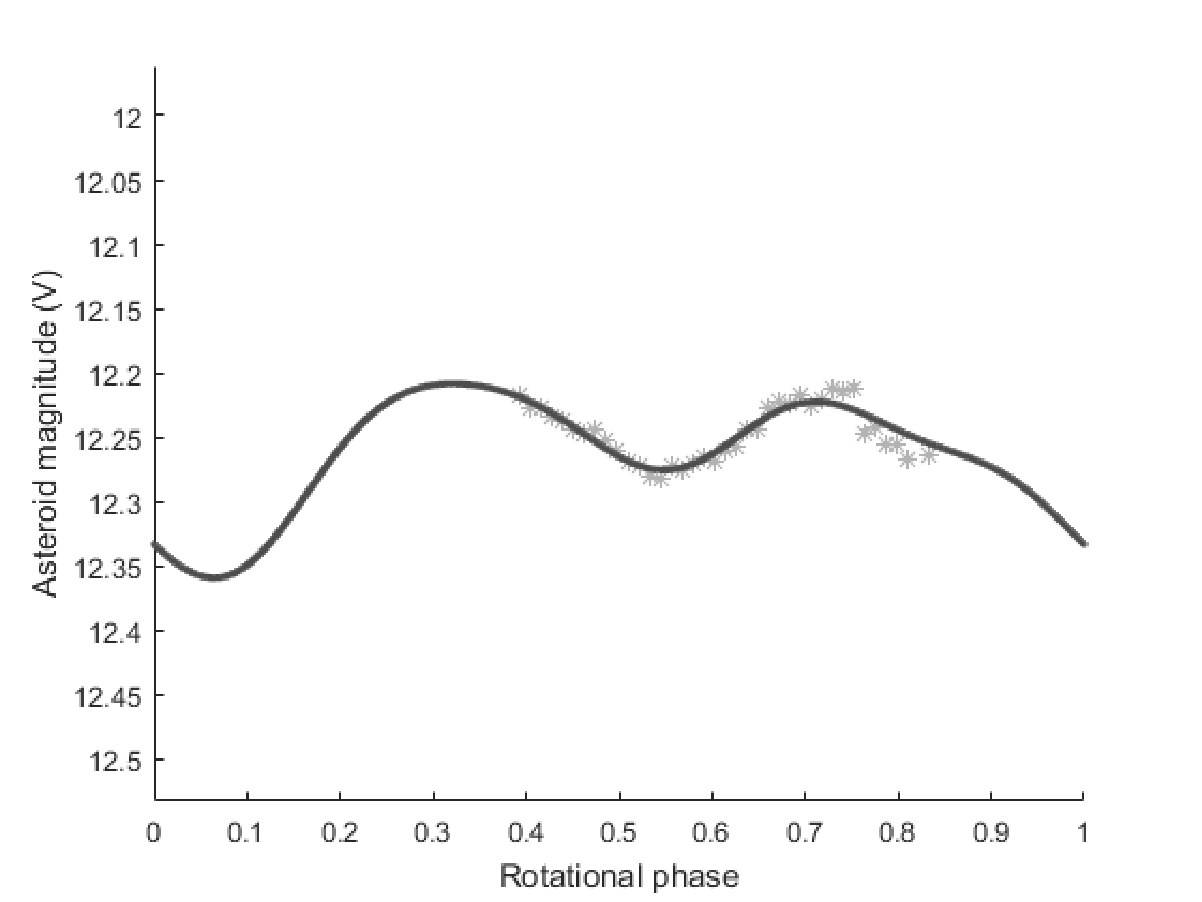}
\includegraphics[width=6.5cm]{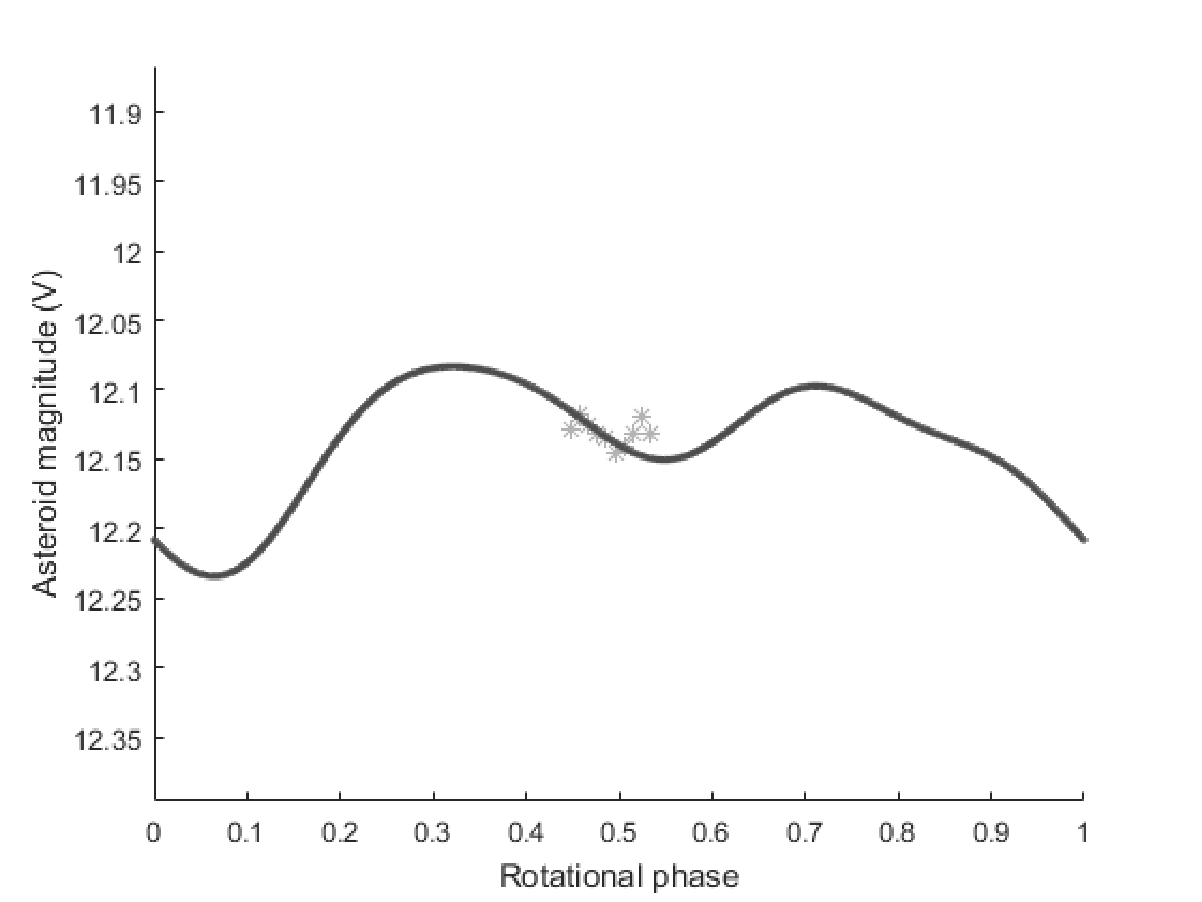}
\includegraphics[width=6.5cm]{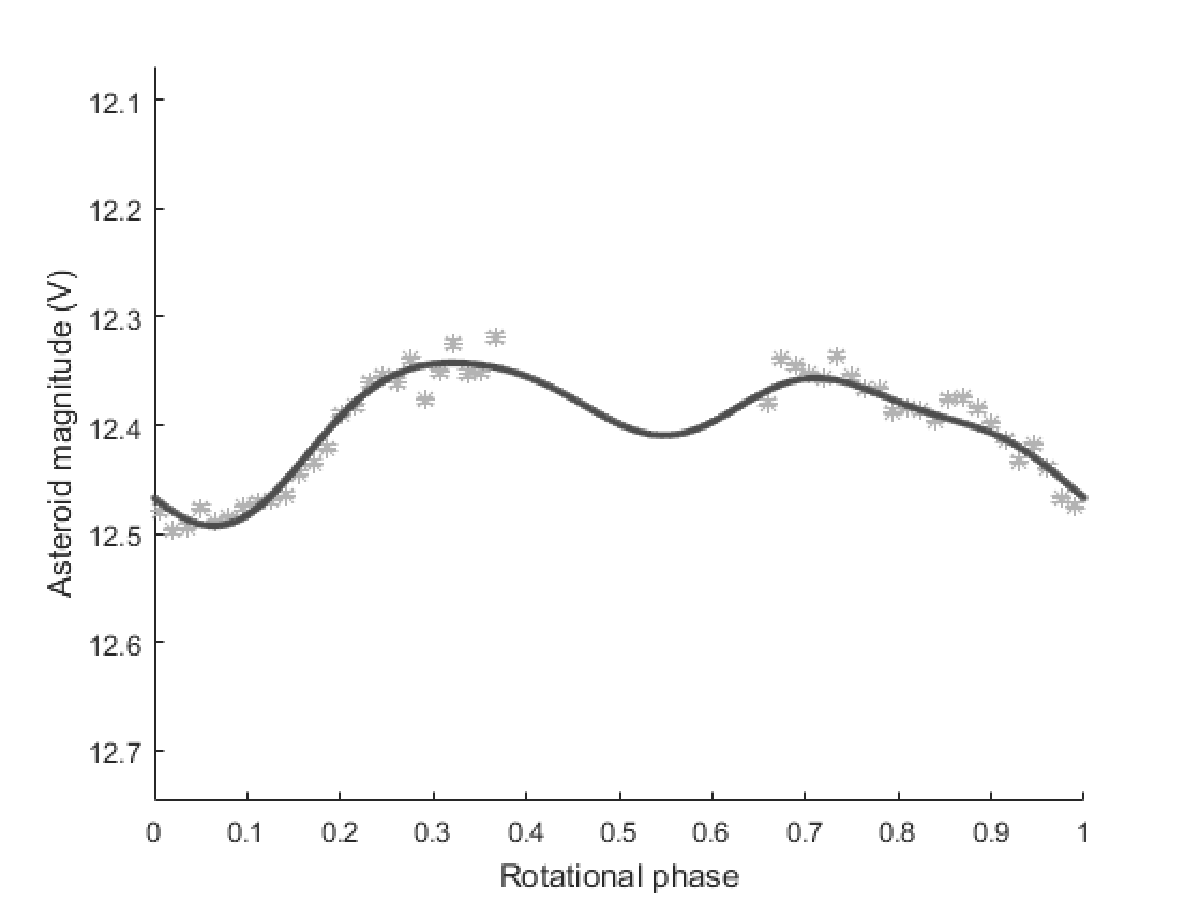}
\includegraphics[width=6.5cm]{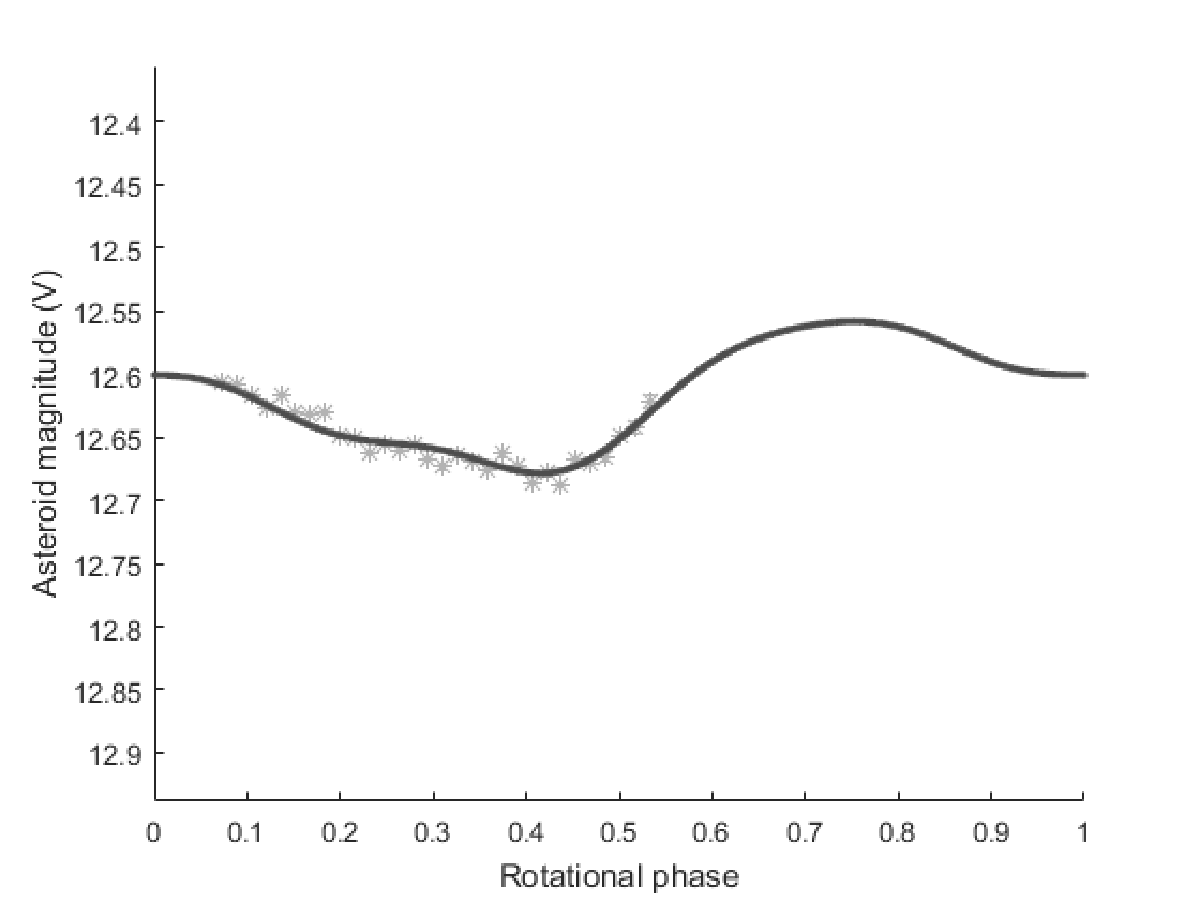}
\includegraphics[width=6.5cm]{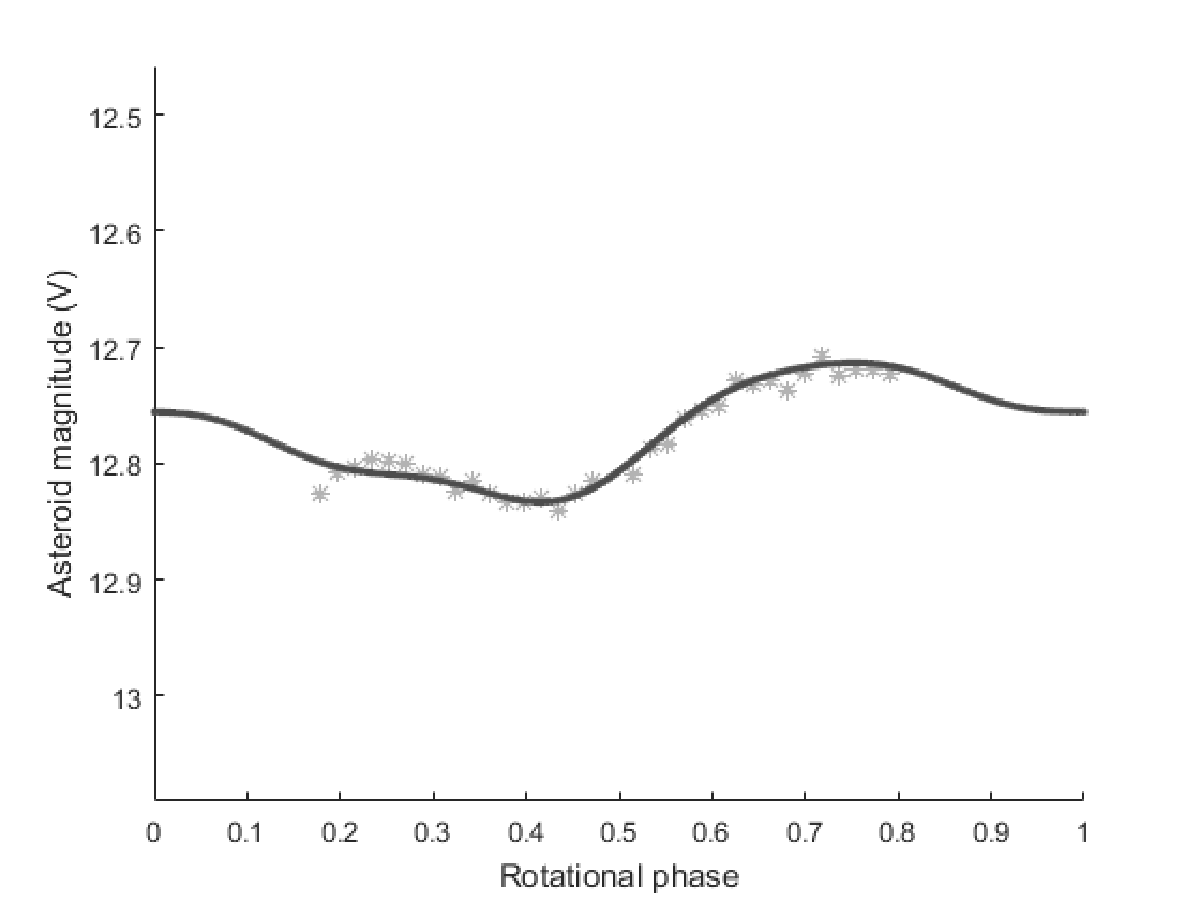}
\includegraphics[width=6.5cm]{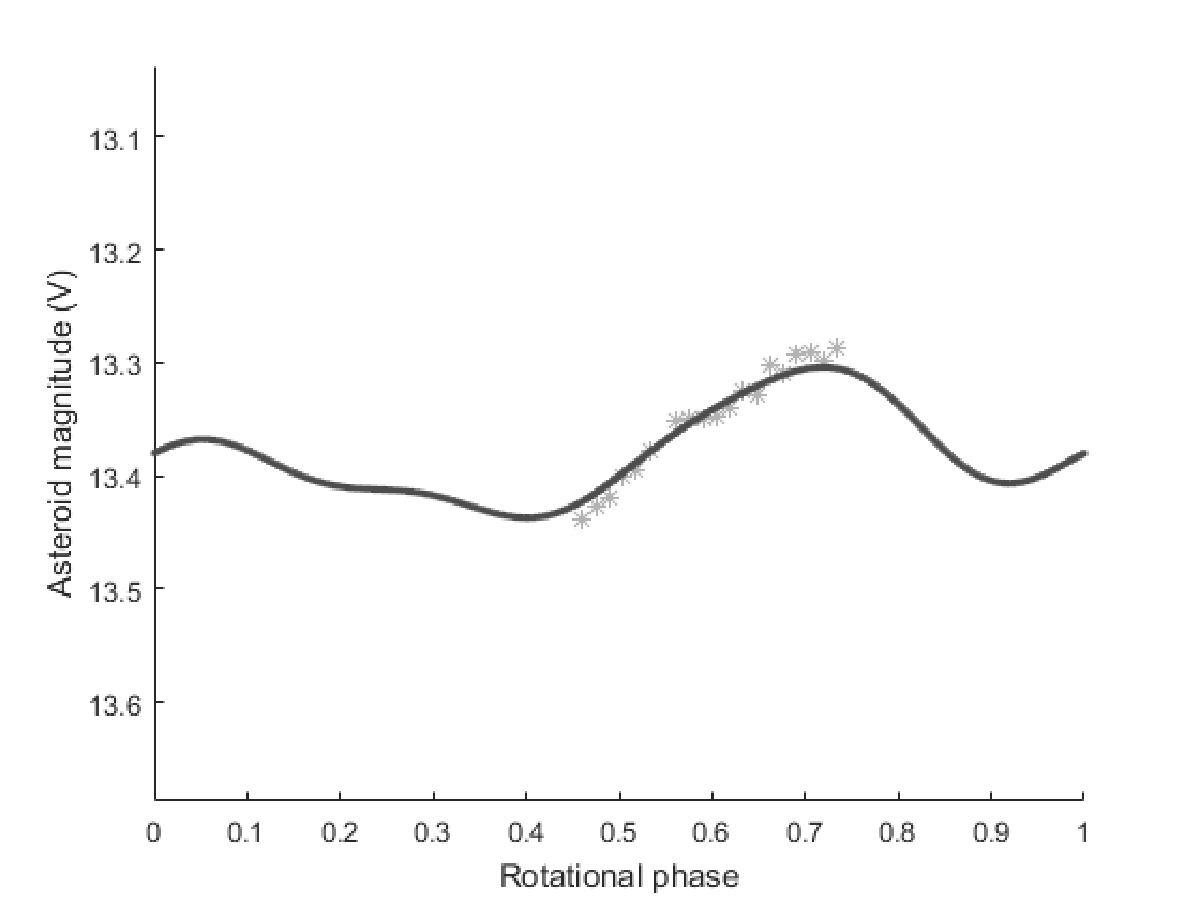}
\end{center}
\caption{The three Fourier fit models for 313 Chaldea superimposed on the observed data sessions (yyymmdd): 20120920, 20120921, 20121002, 20121010, 20121015, 20121116, from left to right, top to bottom.} 
\label{FourierFit313}
\end{figure}

\begin{figure}[p]
\begin{center}
\includegraphics[width=6.5cm]{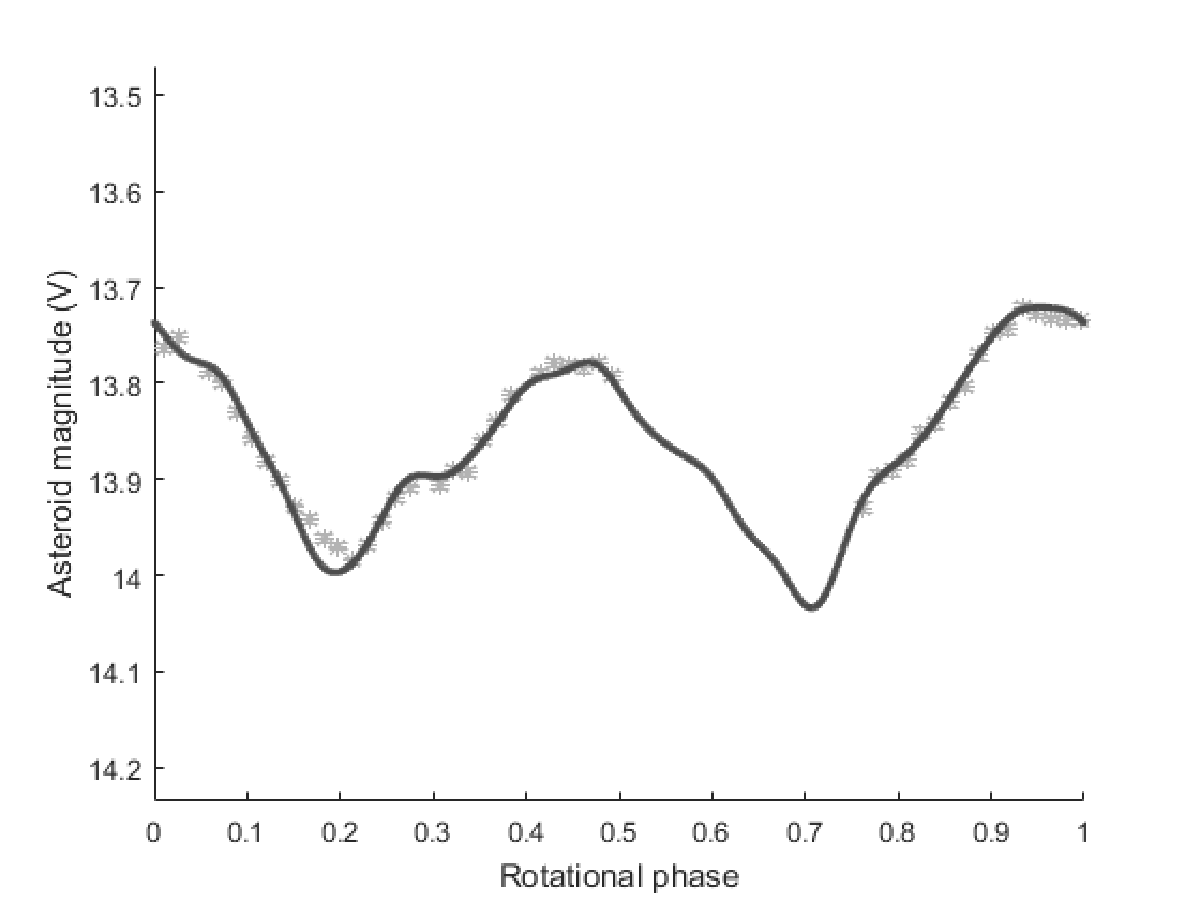}
\includegraphics[width=6.5cm]{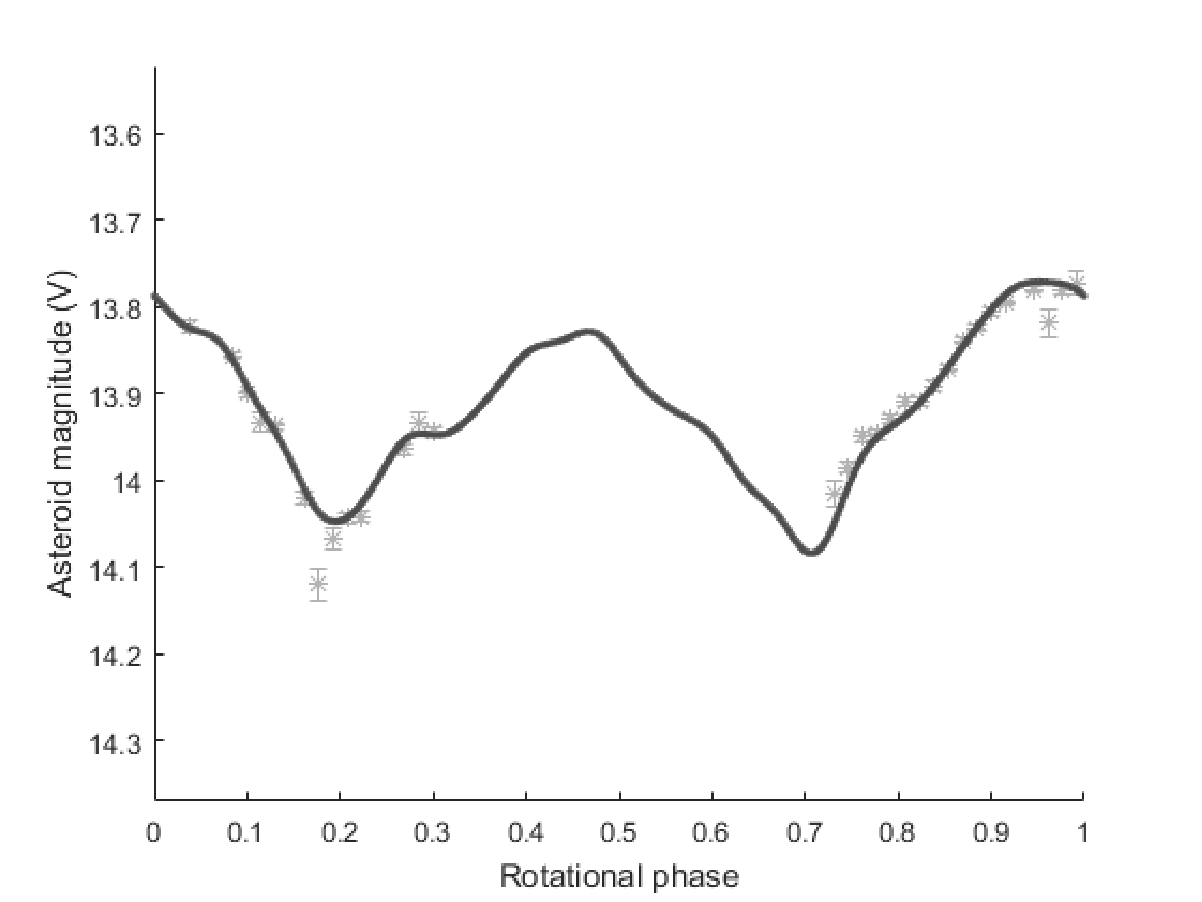}
\includegraphics[width=6.5cm]{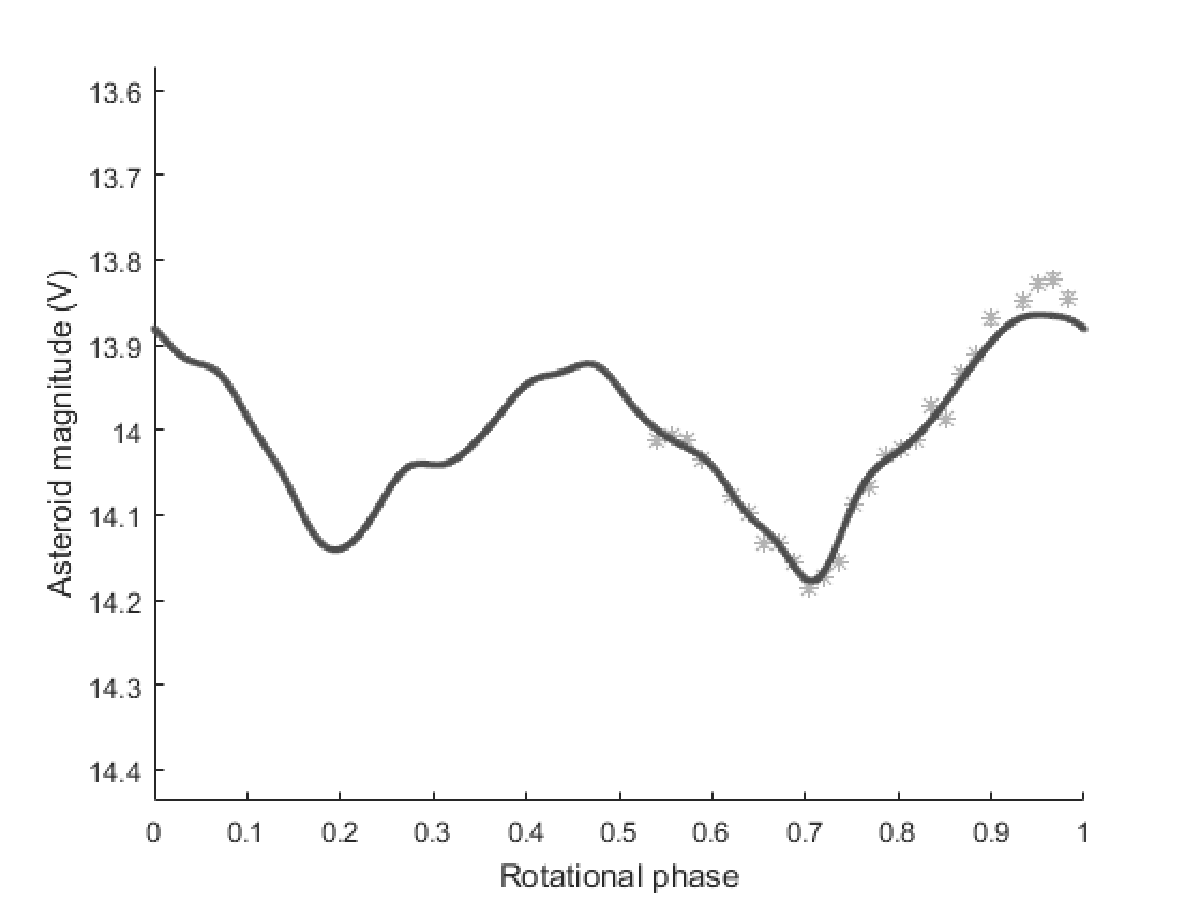}
\includegraphics[width=6.5cm]{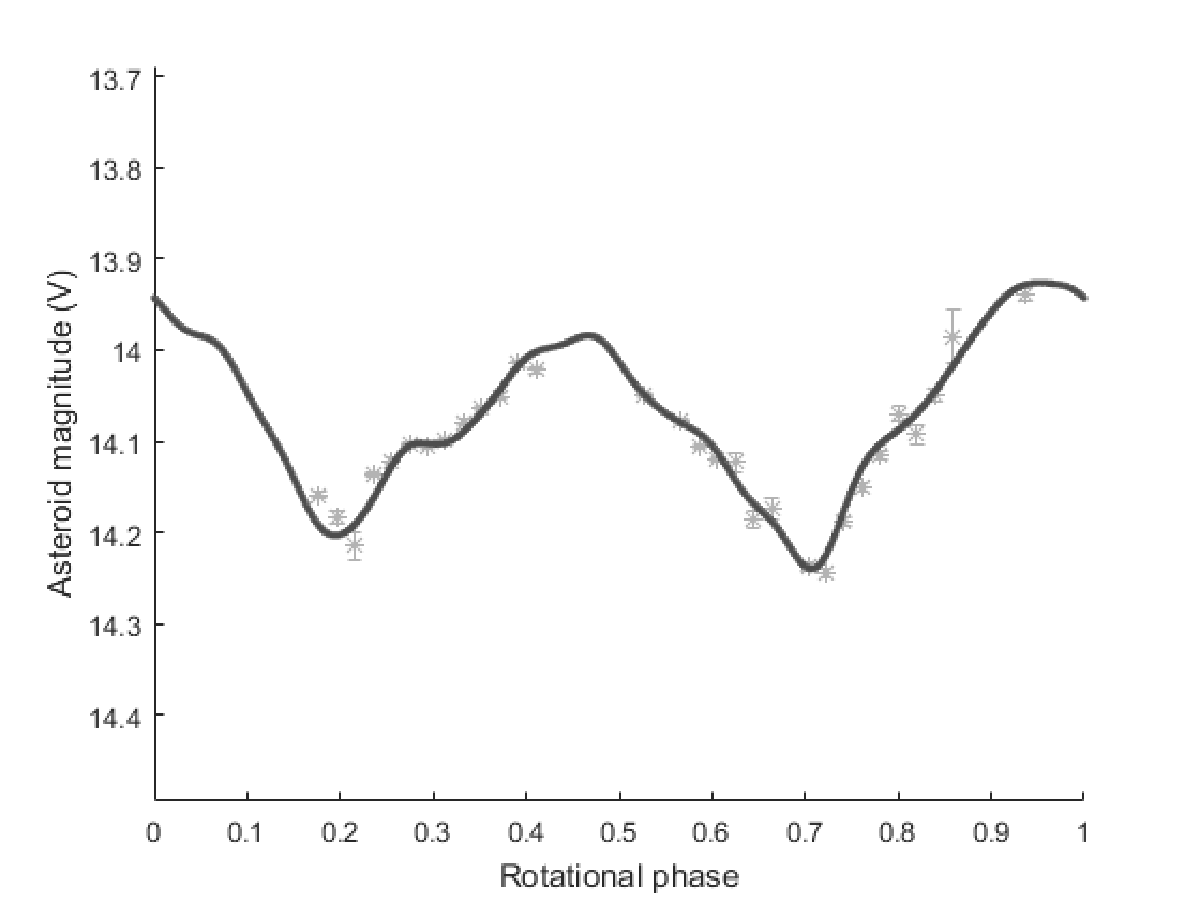}
\includegraphics[width=6.5cm]{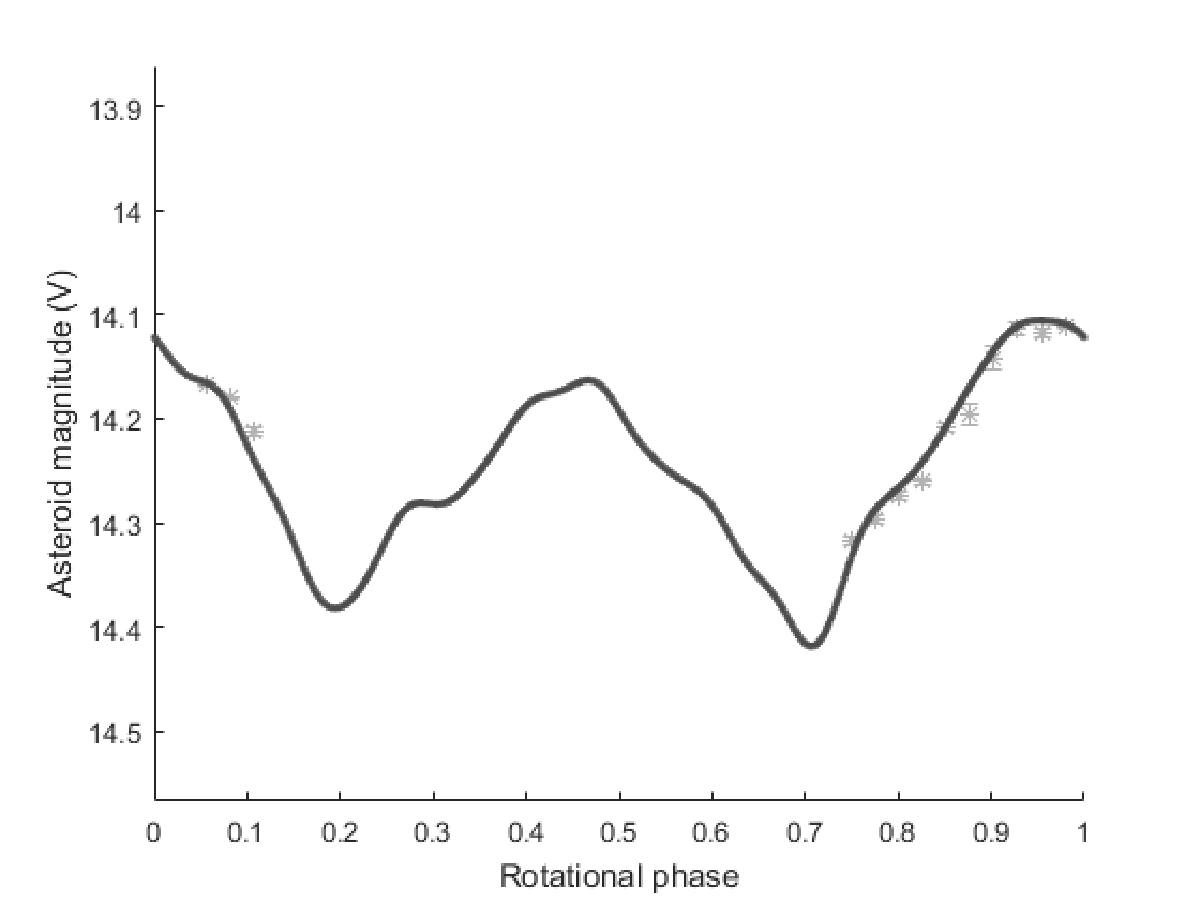}
\includegraphics[width=6.5cm]{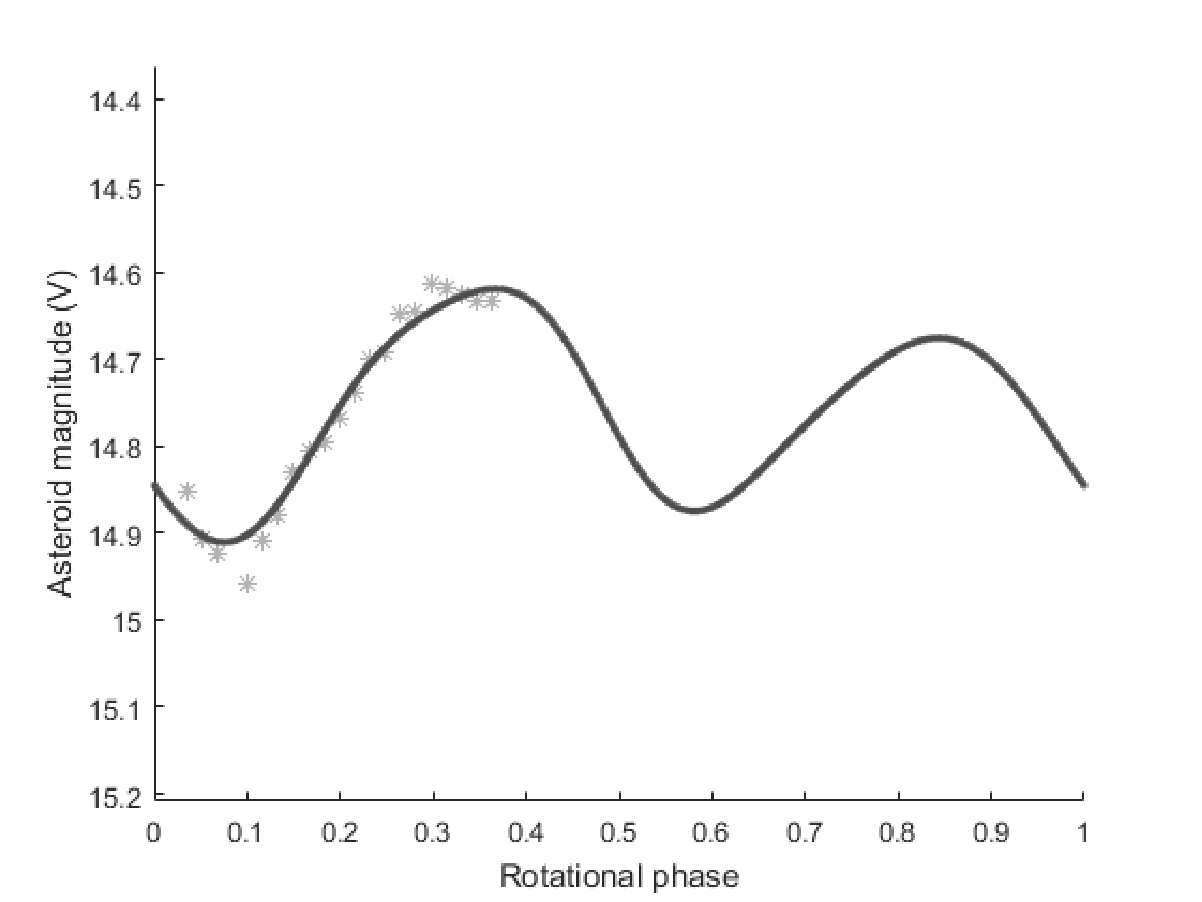}
\includegraphics[width=6.5cm]{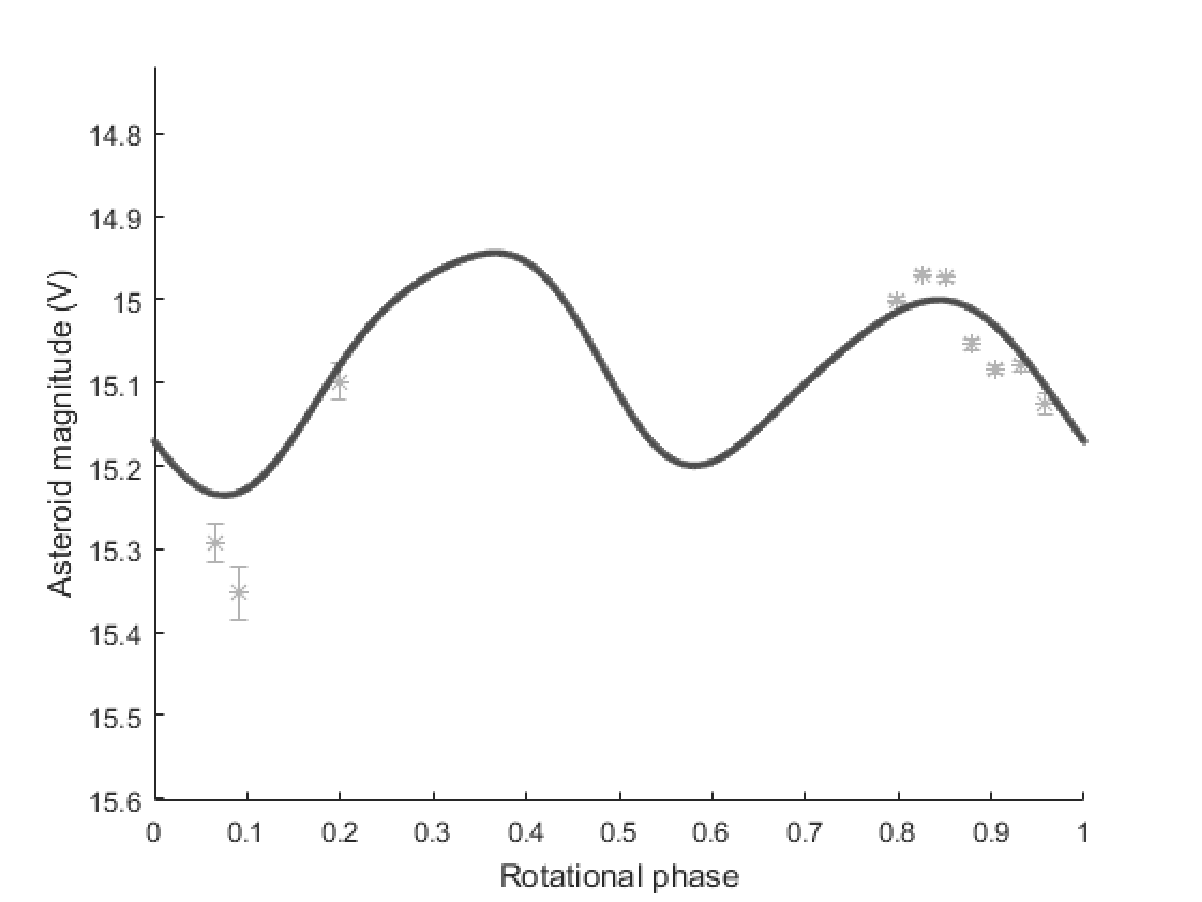}
\end{center}
\caption{The two Fourier fit models for 522 Helga superimposed on the observed data sessions (yyymmdd): 20121002, 20121003, 20121010, 20121013, 20121021, 20121116, 20121206, from left to right, top to bottom.} 
\label{FourierFit522}
\end{figure}

\begin{table*}
  \caption{Summary of the lightcurve Fourier models with the reference Julian days - 2456000.5 and the fit orders used for each asteroids.
           In this table the asteroid 444 Gyptis is missing, it has not been analyzed because it is possible that it is a binary system.}
\begin{center}
\begin{tabular}{lcccccc}
\hline
 Asteroids & $JD_{1}$ & $n_1$ & $JD_{2}$ & $n_2$ & $JD_{3}$ & $n_3$\\
\hline
085 Io	        &      214 	  &    4  &    253   &   3   &    -     &   -  \\
135 Herta 	    &      280 	  &    5  &    -     &   -   &    -     &   -  \\
208 Lacrimosa	&      264 	  &    5  &    -     &   -   &    -     &   -  \\
236 Honoria	    &      197 	  &    4  &    203   &   5   &    -     &   -  \\
306 Unitas	    &      255	  &    5  &    -     &   -   &    -     &   -  \\
308 Polyxo	    &      306 	  &    5  &    -     &   -   &    -     &   -  \\
313 Chaldaea	&      197 	  &    4  &    214   &   4   &    221   &   4  \\
338 Budrosa	    &      295 	  &    5  &    -     &   -   &    -     &   -  \\
522 Helga	    &      209 	  &   15  &    217   &   5   &    -     &   -  \\
925 Alphonsina	&      318 	  &    5  &    -     &   -   &    -     &   -  \\
\hline
\end{tabular}
\end{center}
\label{fourier_models}
\end{table*}

Using several nights differential photometry data, we have construct a complete rotational lightcurve, with full coverage of the asteroid rotation. Once the full lightcurve and period have been determined, we created a Fourier fit with sufficient orders to capture all of the essential features of the lightcurve. Next we use the Fourier fit model of the full lightcurve as a way of extrapolating to data points that are not measured on a given night, i.e. the lightcurve maximum used for the phase curves. The best fit between data session and model was obtained by plotting the data on the Fourier curve and minimizing the squared error between the Fourier fit and the observed data, in this way the maximum brightness is determined with the contribution of the points of the whole session. In this appendix there are not all Fourier models but the selection shown is sufficient to understand what method has been adopted.\\



\end{document}